\documentclass[12pt]{article}

\usepackage{float}

\usepackage{graphicx}
\usepackage{epsfig}
\usepackage{amssymb}
\usepackage{xcolor}

\def\gtwid{\mathrel{\raise.3ex\hbox{$>$\kern-.75em\lower1ex\hbox{$\sim$}}}}
\def\ltwid{\mathrel{\raise.3ex\hbox{$<$\kern-.75em\lower1ex\hbox{$\sim$}}}}
\def\square{\kern1pt\vbox{\hrule height 1.2pt\hbox{\vrule width 1.2pt\hskip 3pt
   \vbox{\vskip 6pt}\hskip 3pt\vrule width 0.6pt}\hrule height 0.6pt}\kern1pt}
   
\usepackage{amsmath}
\usepackage{hyperref}

\begin{document}

\begin{titlepage}

\begin{flushright}
\today
\end{flushright}

\vskip 0.7cm

\begin{center}
{\bf\Large Field-theoretic derivation of bubble-wall force}
\end{center}

\vskip .3cm

\begin{center}
Marc Barroso Mancha$^{1\star}$, Tomislav Prokopec$^{2\diamond}$ 
 and Bogumi{\l}a {\'S}wie{\.z}ewska$^{2,3\ddagger}$

\end{center}

\vskip .3cm

\begin{center}
\it{$^{1}$ Institut fuer Kernphysik, Technische Universitaet Darmstadt,
Schlossgartenstraße 2, D-64289 Darmstadt, Germany}
\end{center}

\begin{center}
\it{$^{2}$ Institute for Theoretical Physics, Utrecht University,\\ Princetonplein 5, 3584 CC Utrecht, the Netherlands}
\end{center}

\begin{center}
\it{$^{3}$ Faculty of Physics, University of Warsaw,\\Pasteura 5, 02-093 Warsaw, Poland}
\end{center}

\vspace{.5cm}

\begin{center}
ABSTRACT
\end{center}
We derive a general quantum field theoretic formula 
for the force acting on expanding bubbles 
of a first order phase transition in the early Universe setting.
In the thermodynamic limit the force is proportional 
to the entropy increase across the bubble
 of 
active species that exert a force on the bubble interface. 
When local thermal equilibrium is attained,
we find a strong friction force which grows as the Lorentz factor squared,
such that the bubbles quickly reach stationary state and {\it cannot run away}. We also study an opposite case when scatterings are negligible across the wall
(ballistic limit), finding that the force saturates for moderate Lorentz factors thus allowing for a runaway behavior.
We apply our formalism to a  massive 
real scalar field,
the standard model and its simple portal extension.
For completeness, we also present a derivation of the 
renormalized, one-loop, thermal energy-momentum tensor
for the standard model and demonstrate its gauge independence. 

\begin{flushleft}
\end{flushleft}

\vskip .5cm

\begin{flushleft}
$^{\star}$ e-mail: mbarroso@theorie.ikp.physik.tu-darmstadt.de,  
\\
$^{\diamond}$ e-mail: T.Prokopec@uu.nl \\
$^{\ddagger}$ e-mail: bogumila.swiezewska@fuw.edu.pl
\end{flushleft}

\end{titlepage}

\section{Introduction} 
\label{Introduction}

 Modeling the bubble-wall velocity of expanding bubbles at a 
 first order electroweak transition is of an essential importance for 
an accurate modeling of the gravitational wave 
production~\cite{Caprini:2019egz}
and for baryogenesis at the electroweak 
scale~\cite{Morrissey:2012db,Konstandin:2013caa}.
In spite of a lot of progress, we still lack a reliable first principle calculation which -- based on an (out-of-equilibrium) 
quantum field theoretic framework -- provides a formula 
which can be used to get a quantitatively reliable information 
about the phase transition dynamics.
In this paper we present a first principle (quantum field
theoretic) derivation of
the {\it force} acting on {\it expanding bubbles} that reached stationary state during a strong first order transition, 
providing thus an important step towards that noble goal.

To determine the dynamics  of  expanding bubbles one ought to know the friction force exerted on them by the plasma. 
On the other hand, the bubbles back-react on the plasma, thereby changing its properties. One way to find a general solution of this complex problem is to solve the Boltzmann 
equations for the relevant species in the presence of expanding bubbles~\cite{Dine:1992, Liu:1992, Moore:1995ua,Moore:1995si}, which is a formidable task. For that reason, in many papers the bubble velocity is treated as a free parameter whose value is assumed or
roughly estimated.

An important question from the point of view of gravitational wave production is whether the bubble can run away, \textit{i.e.}\ permanently accelerate, asymptotically reaching the speed of light. If such a situation is possible the latent heat of the transition is pumped into the scalar field, resulting in a characteristic gravitational wave spectrum with a very strong peak amplitude. 
An important step towards solving this puzzle 
was made by B{\"o}deker and Moore in Ref.~\cite{Bodeker:2009qy}, 
where they considered the bubble force in the relativistic limit and at the leading order 
in the relevant coupling constants, 
following an earlier work by Arnold~\cite{Arnold:1993wc}. 
They found that, in the limit of a large Lorentz factor $\gamma$,
the force does not depend on $\gamma$ and thus concluded that the bubbles can run away if there is enough latent heat released. In their follow-up paper~\cite{Bodeker:2017cim}, 
the authors considered the next-to-leading order effects, 
finding a $\gamma$ dependence in the friction force, 
which in principle precludes the runaway scenario, 
but still allows for highly relativistic walls.
Some of the relevant works that discuss how fast 
the bubbles can expand 
include~\cite{Giese:2020rtr,Dorsch:2018pat,Konstandin:2014zta,Kozaczuk:2015owa}.~\footnote{After submission of the first version of this paper, two related works appeared~\cite{Hoeche:2020, Balaji:2020}.}
In particular, Ref.~\cite{Dorsch:2018pat} refines some of the 
arguments presented in~\cite{Bodeker:2009qy,Bodeker:2017cim}.

The approach of the present paper is based on covariant conservation of the energy-momentum tensor of the bubble-plasma system.   
When applied to the limit when 
an approximate Lorentz 
symmetry~\footnote{Even though
an expanding bubble explicitly breaks Lorentz symmetry, one can speak of an approximate Lorentz symmetry maintained by the state,  
if it can be characterized by a Lorentz covariant distribution function. A notable example 
of such a state is local thermal equilibrium.} 
holds, we find that 
the friction force scales as $(\gamma^2-1)$.   Therefore, for the most of physically interesting cases, the wall will reach stationary state for moderate velocities and thus will not run away.  
This is the case when 
scatterings are efficient and 
local thermal equilibrium is enforced. 
If, on the other hand, 
scatterings are inefficient, 
a ballistic approximation~\cite{Moore:1995si} better describes the actual situation. 
In that case the bubble force saturates and there is no obstacle for the bubbles to
run away. In this work our formalism 
is applied to a toy model with one real scalar field in a heat bath,
as well as to a model with the standard-model field content featuring a first order phase transition and to 
its simple extension.

The paper is organized as follows. In 
section~\ref{The force on bubbles from the renormalized stress energy tensor} we derive our main formula for the bubble force 
and discuss its applicability and possible generalizations. 
In sections~\ref{Real scalar field} 
and~\ref{Standard model and its extensions}
we show how to 
apply our formalism to a real scalar field
and to the standard model and its simple portal model extension.
In section~\ref{Comparison with existing results} 
we compare our formalism with the literature, 
and in section~\ref{Summary and discussion} we conclude.
In an extensive appendix we calculate 
the one-loop energy-momentum tensor of the standard model
in local thermal equilibrium.  
A particular attention is devoted to renormalization
and to showing gauge 
independence of the renormalized energy-momentum tensor.


\section{The bubble force from the renormalized energy-momentum tensor} 
\label{The force on bubbles from the renormalized stress energy tensor}

When taken together, the energy-momentum tensor of the plasma $T_{\mu\nu}^p$ and
expanding bubbles $T_{\mu\nu}^b$ must be covariantly conserved, 
\begin{equation}
\nabla^\mu \langle \hat T_{\mu\nu}^p + \hat T_{\mu\nu}^b\rangle =0
\,.
\label{conservation of Tmn}
\end{equation}
For simplicity we shall assume that the bubbles are large and nearly spherical,
such that their front can be approximated by nearly planar walls propagating through
the plasma. Because the surface tension of the bubbles is typically 
quite large, this approximation is justified.
Moreover, we shall assume that the following hierarchy of scales holds,
$L,D,\tau_{\rm int}=1/\Gamma \ll R_H$, where $L$ is the bubble wall thickness,
$D$ is the relevant diffusion time (recall that $c=1$), 
$\Gamma$ is the scattering rate of the relevant plasma species,
and $R_H=1/H$ is the Hubble radius characterizing the expansion of the Universe. Then the covariant derivative 
in~(\ref{conservation of Tmn}) can be approximated by an ordinary derivative and the expansion of the Universe can be, to the leading order in adiabatic expansion, encoded by the temperature dependence on
the scale factor of the Universe $a(t)$. 
Taking account of these, 
equation~(\ref{conservation of Tmn}) simplifies to,
\begin{eqnarray}
-\partial_t\langle \hat T_{00}^p\rangle +\partial_z\langle \hat T_{z0}^p\rangle
-\partial_t\langle \hat T_{00}^b\rangle +\partial_z\langle \hat T_{z0}^b\rangle &=&0,
\label{conservation of Tmn:2a}\\
-\partial_t\langle \hat T_{0z}^p\rangle +\partial_z\langle \hat T_{zz}^p\rangle
-\partial_t\langle \hat T_{0z}^b\rangle +\partial_z\langle \hat T_{zz}^b\rangle &=&0
\,.
\label{conservation of Tmn:2b}
\end{eqnarray}
Several remarks are in order:  
\begin{enumerate}
\item[$\bullet$] The energy-momentum
tensor in Eqs.~(\ref{conservation of Tmn:2a}--\ref{conservation of Tmn:2b}) is a composite operator which diverges, 
so to make it finite it ought to be regularized and renormalized;
\item[$\bullet$] The expectation values in~(\ref{conservation of Tmn:2b}) ought to be calculated in a state that includes 
the bubble, or which approximates it well enough;
\item[$\bullet$] 
Eqs.~(\ref{conservation of Tmn:2a}--\ref{conservation of Tmn:2b})
simplify further in the {\it bubble frame}, in which the terms containing time derivatives drop out.    This applies when the bubble reaches stationary state and that is where our analysis holds. The regime of nonstationary bubbles is worth investigating in a separate work.
\end{enumerate}

In the bubble frame, one can then integrate
equations~(\ref{conservation of Tmn:2a}--\ref{conservation of Tmn:2b}) 
across the bubble to obtain, 
\begin{eqnarray}
\Delta \langle \hat T_{z0}^p\rangle+\Delta \langle \hat T_{z0}^b\rangle &=&0,
\label{conservation of Tmn:3a}\\
\Delta\langle \hat T_{zz}^p\rangle+\Delta\langle \hat T_{zz}^b\rangle &=&0
\,,
\label{conservation of Tmn:3b}
\end{eqnarray}
where $\Delta \langle \hat T_{\mu\nu}^{p,b}\rangle$ denotes the change of the $\mu\nu$ 
components of the energy-momentum tensor of the plasma or bubble across the bubble.
In what follows we focus on Eq.~(\ref{conservation of Tmn:3b}) since the bubble-wall speed is determined by the balance 
of the two terms in~(\ref{conservation of Tmn:3b}) which encapsulate the driving force of the vacuum energy and the friction force from interactions with the plasma.

When calculating the plasma contribution in~(\ref{conservation of Tmn:3b}), 
it is convenient to calculate it in the plasma frame, in which 
$\langle (\hat T_{00})_{\rm plasma}\rangle\equiv\rho_p$ and 
$\langle (\hat T_{zz})_{\rm plasma}\rangle\equiv{\cal P}_p$. As $T_{\mu\nu}$ is a tensor, Lorentz boosting it to the bubble frame results in, 
$\langle \hat T_{zz}^p\rangle = (\gamma^2-1)(\rho_p+{\cal P}_p)+{\cal P}_p$,
$\langle \hat T_{zz}^b\rangle = {\cal P}_b$,~\footnote{
The 
$\gamma$ dependence in this relation is exact if 
in the plasma frame there is 
no significant violation of Lorentz symmetry. For now 
it suffices to note that this is
the case if {\it e.g.} local thermal equilibrium is maintained
across the moving bubble.
} 
where in the last relation we used
$\rho_b+{\cal P}_b=0$ (since the bubble possesses no entropy).  
Taking account of the thermodynamic relation for the entropy density $s$, 
\begin{equation}
s_p=\frac{\rho_p+{\cal P}_p}{T} = s
\,,
\label{entropy density}
\end{equation}
we arrive at the following expression for the  friction force $F$ per 
volume $V$ on an expanding bubble, 
\begin{equation}
 \boxed{\frac{F}{V}
        \equiv -\Delta {\cal P} =(\gamma^2-1)T\Delta s}
\,,
\label{conservation of Tmn:4}
\end{equation}
where we used, ${\cal P}={\cal P}_p+{\cal P}_b$ and 
$\Delta{\cal P}_b=\Delta\langle \hat T_{zz}^b\rangle$.
The relation~(\ref{conservation of Tmn:4}) tells us that the {\it change in the pressure} 
is balanced by the {\it change in the entropy density} of the plasma 
across the bubble and that the effect grows quadratically with the Lorentz factor 
$\gamma=1/\sqrt{1-v^2}$, unless some compensating $\gamma$ dependent terms arise from the plasma frame $\Delta s$, 
which we discuss later. Since in the derivation of Eq. ~(\ref{conservation of Tmn:4}) we assumed stationary bubbles, making use of~(\ref{conservation of Tmn:4}) one can determine 
the terminal bubble speed. This simple observation constitutes on of the principal results of this work. 

Equation~(\ref{conservation of Tmn:3a}) in the plasma frame reads $\Delta(\gamma^2 v T s)=0$ and thus implies that (with nonzero variation of entropy across the bubble) the temperature or velocity of the plasma must change across the bubble. Here we focus on the bubble-wall force, Eq.~(\ref{conservation of Tmn:4}), therefore studying the effect of Eq.~(\ref{conservation of Tmn:3a}) is beyond the scope of the present work.\footnote{Variation of temperature and velocity across the bubble wall has been studied in ref.~\cite{Balaji:2020} which appeared after the submission of the first version of this article.}

Before we proceed to applications, in what follows we discuss applicability 
of formula~(\ref{conservation of Tmn:4}).

\medskip

{\bf 1.} To arrive at~(\ref{conservation of Tmn:4}) we took an
expectation value of the energy-momentum tensor.
This can be exacted by making use of 
the full quantum formalism, which gives accurate answers, 
but it is hard to implement because it involves 
quantum field theory 
in an out-of-equilibrium setting. 
A much simpler procedure is to take a semiclassical limit, 
according to which the plasma can be described as a collection of 
quasiparticles that remain approximately {\it on-shell} across
the bubble interface. To estimate when 
the quasiparticle approximation is reliable, recall that according to 
the uncertainity principle particles can be off-shell
for short periods of time satisfying,
$\Delta t\lesssim 1/\Delta E\sim 1/E$, 
where $E=\sqrt{p^2+m^2}$ is the energy of the particle. 
Conversely, when the wall passage time $\sim L/(\gamma v)$ is
longer than the bubble wall thickness, $L/\gamma$ (which gets Lorentz-contracted in the plasma frame), then 
particles will be approximately on-shell, {\it i.e.}
\begin{equation}
 \gamma v < \gamma   < LT
  \qquad {\rm (on\!\!-\!\!shell\; condition)}
\,,
\label{on-shell condition}
\end{equation}
where we assumed that $E\sim T$.~\footnote{ 
The condition~(\ref{on-shell condition}) 
is not the most general one. Namely, in view of Einstein's relation, $E=\sqrt{p^2+m^2}$, 
the on-shell condition is most restrictive for 
highly infrared particles, for which $p^2<m^2$, such that $E\approx m$ and
the on-shell condition~(\ref{on-shell condition})  
becomes $\gamma v < \gamma   < Lm$. Since typically the most massive particles dominate
the bubble friction, it is often the case that $m\sim T$ and~(\ref{on-shell condition})  applies.
If, however, the heaviest particle's mass is much smaller than $T$, then ~(\ref{on-shell condition}) 
ought to be replaced by the more stringent condition, 
$\gamma v < \gamma   < Lm$.}
A typical bubble wall thickness $L$ at the electroweak transition is of the order
$L\sim 10/T$~\cite{Moore:1995si},
implying that when 
$\gamma\lesssim 10$ the plasma quasiparticles will be
on-shell and the semiclassical kinetic  
formalism applies.
When the on-shell condition~(\ref{on-shell condition}) 
is satisfied, that does not yet mean that local thermal equilibrium is
reached. In fact, an additional condition -- efficient scatterings
on the bubble interface -- must be met for a local thermal 
equilibrium to be reached. We elaborate more on that below.

\begin{figure}[ht]
\vskip-0.6cm
\begin{center}
\includegraphics[width=6.7cm]{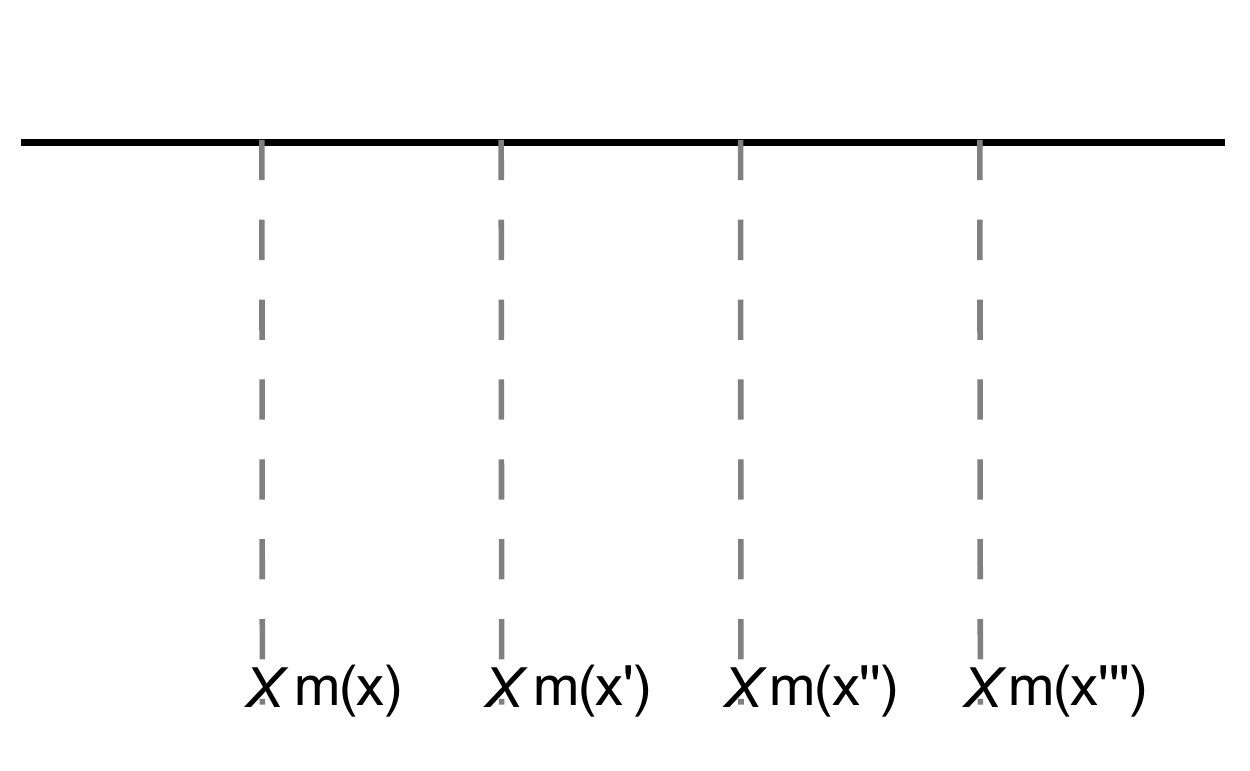}
\end{center}
\vskip-0.8cm
\caption{\small The propagation of a particle (horizontal solid black line) in presence of 
field dependent mass insertions, $m(x)=\sqrt{\lambda/2}\phi_0(x)$ (vertical dashed lines).}
\label{figure mass insertions}
\end{figure}

The adiabatic approximation we employ here can be regarded as 
a semiclassical approximation, in which the effects of 
varying backgrounds are modeled by 
{\it mass insertions} along a particle's trajectory, as illustrated in figure~\ref{figure mass insertions}.
This is a kinematic effect enforced by the energy conservation in the bubble frame,
and the approximation holds as long as the quantum off-shell effects are small. On top of this, particles can 
interact and scatter off each other. Even though scattering effects can be important for a complete understanding 
of the phase transition dynamics, we postpone their study for future work since they require 
the inclusion of higher-loop effects. In this work we 
investigate only two limits, namely very rapid and very slow scatterings.

When bubbles are very fast (or very thin)  and the criterion~(\ref{on-shell condition}) is violated, 
quantum off-shell effects can become significant. To capture that, 
when constructing the energy-momentum tensor,
one ought to use exact mode functions in the moving bubble background, which is a much harder endeavor. For an example of how 
that can be done for fermionic fields and without loop effects, see {\it e.g.}\ Ref.~\cite{Prokopec:2013ax}.
To consistently include the quantum loop effects 
on top of such a tree level treatment, one would 
have to solve the corresponding out-of-equilibrium problem
using a perturbative quantum field theoretic framework such as 
the Schwinger-Keldysh 
formalism~\cite{Schwinger:1960qe,Keldysh:1964ud,Chou:1984es}. 
While progress has been made 
in applying such a formalism in 
baryogenesis/leptogenesis 
scenarios~\cite{Kainulainen:2001cn,Kainulainen:2002th,Prokopec:2003pj,Prokopec:2004ic}
(for reviews see~\cite{Konstandin:2013caa,Garbrecht:2018mrp})
and in cosmic inflation~\cite{Brunier:2004sb,Miao:2005am,Weinberg:2005vy,Prokopec:2008gw,Friedrich:2019hev}, 
little or no progress 
has been made in studying cosmological phase transitions.

\medskip

{\bf 2.} While our conclusions based on the consideration of 
the energy-momentum tensor operator are general,
the form of Eq.~(\ref{conservation of Tmn:4}) is based on the assumption that the (expectation value of the) 
energy-momentum tensor is well approximated by a perfect fluid form, 
$\langle \hat T_{\mu\nu}\rangle = (\rho+{\cal P})u_\mu u_\nu + g_{\mu\nu}{\cal P}$. To get a better 
understanding of the limitations of this approximation, recall that the perfect fluid form 
can be viewed as the leading order approximation in a gradient expansion. 
Including the first order gradient corrections yields the well known expression,
\begin{equation}
\langle \hat T_{\mu\nu}\rangle = (\rho+{\cal P})u_\mu u_\nu + g_{\mu\nu}{\cal P} + {\tau}_{\mu\nu}
\,,
\label{energy momentum gradient corrections}
\end{equation}
where 
\begin{equation}
{\tau}_{\mu\nu} = 2\eta\left[\nabla_{(\mu} u_{\nu)}+u_{(\mu} u\cdot \nabla u_{\nu)}\right]
 - \left(\frac23\eta-\zeta\right)(g_{\mu\nu}+u_\mu u_\nu)\nabla\cdot u
\,.
\label{viscosity part}
\end{equation}
Here 
 $\eta$ and $\zeta$ denote the {\it shear and the bulk viscosity},
respectively, and $(\alpha\beta)$ means that the indices 
are symmetrized. 
Notice that $\tau_{\mu\nu}$ is orthogonal to $u^\mu$, 
$u^\mu \tau_{\mu\nu}=0=\tau_{\mu\nu}u^\nu $, and that its trace is proportional to the bulk viscosity,
${\rm Tr}[\tau_{\mu\nu} ]= g^{\mu\nu} \tau_{\mu\nu} = 3\zeta \nabla\cdot u$, where we used,
$u^2=-1$ and $u^\mu (u\cdot \nabla) u_\mu = 0$. 
If the plasma is in thermal equilibrium, then (\ref{viscosity part}) vanishes. To see that recall that  
in the plasma (bubble) frame, $u^\mu = (1,0,0,0)$ ($u^\mu=(\gamma v,0,0,v$)), such that 
(if one neglects the expansion of the Universe) the covariant derivatives acting on $u^\mu$ give zero.
Therefore, one gets a nonvanishing contribution from the viscosity part of the energy-momentum 
tensor~(\ref{viscosity part})
only if both (a) plasma velocity is perturbed from its thermal equilibrium form and 
(b) the viscosity coefficients do not vanish. 
In a perturbative treatment the leading order contribution to the energy-momentum tensor comes from the 
one-loop approximation.  Since the one-loop contribution
is non-dissipative, the viscosities (whose nature is dissipative) 
acquire nonvanishing contributions only at two and higher 
loops. Therefore in weakly coupled theories 
the dominant contribution to the bubble force is captured by the one-loop calculation 
and the bubble dynamics can be obtained from Eq.~(\ref{conservation of Tmn:4}). 
That does not mean that an accurate answer for the bubble dynamics 
can be obtained just from a one-loop analysis, 
as higher loops may be essential for determining the accurate form of 
the state with respect to which one takes the expectation value of the energy-momentum tensor.

\medskip

{\bf 3.} Even though formula~(\ref{conservation of Tmn:4}) was obtained based on non-equilibrium 
considerations, it has a deceptively similar form to the fundamental thermodynamic law, 
\begin{equation}
dE= TdS - {\cal P}dV +\mu dN
\,,
\label{TD relation}
\end{equation}
where $E$ denotes the energy, $S$ is the entropy, ${\cal P}$ is the pressure, $V$ is the volume, $\mu$ 
the chemical potential and $N$ the particle number of the system. To see that let us divide~(\ref{TD relation})
 by $dV$ to obtain its local form, 
\begin{equation}
 \rho = Ts - {\cal P} + \mu n
\,,
\label{TD relation:2}
\end{equation}
where $\rho=dE/dV$ is the energy density, $s=dS/dV$ is the entropy density and $n=dN/dV$ is the number density.
Next,
close enough to thermal equilibrium 
$\mu\simeq 0$~\footnote{The condition 
$\mu\simeq 0$ does not 
mean that Eq.~(\ref{conservation of Tmn:4}) does not apply away
from thermal equilibrium in which chemical potentials are 
appreciable. It just means that we have subsumed 
all the relevant effects that 
contribute to the bubble friction into the entropy increase.}  
and the contribution of $\mu n$ 
can be neglected and one obtains
a standard thermodynamic relation,
that the change in energy density plus pressure
across the bubble equals to the 
change in the entropy density times the 
temperature, $\Delta (\rho+{\cal P})= T \Delta s$,
which was used in the derivation
of Eq.~(\ref{conservation of Tmn:4}).

Let us now look more closely at
how our approach compares
with the usual description 
according to which the transition dynamics 
is governed by 
the latent heat release $\ell=\Delta \rho$ and by  the change in the effective potential
across the bubble, see {\it e.g.}\ 
Ref.~\cite{Moore:1995si}. 
Our approach here is instead based on the change in the entropy density 
across the bubble, $\Delta s = \Delta\rho + \Delta {\cal P}=\ell + \Delta {\cal P}$, 
which naturally arises from the energy-momentum conservation. 
In the one-loop approximation and in local thermal
equilibrium, $\Delta{\cal P}=-\Delta V_{\rm eff}$, where 
$V_{\rm eff}$ is the one-loop effective potential.~\footnote{To see that,
note that from Eq.~(\ref{part of Tmn: real scalar: ij thermal contribution}) 
we can read off the thermal contribution to the pressure,
${\cal P}=-(6\pi^2\beta^4)^{-1}[z^{-1}\partial_{z}J_B(6,z)]_{z=\beta m}$, where $J_B(6,z)$ is given in Eq.~(\ref{JB n integral}). 
By a suitable partial integration one can show,
$z^{-1}\partial_z J_B(6,z)=-3J_B(4,z)$, from which we conclude,
${\cal P}=(2\pi^2\beta^4)^{-1}J_B(4,\beta m)=-V_{\rm eff}$,
which completes the proof.} Therefore, we have $\Delta s = \ell -\Delta V_{\rm eff}$.
It would be worth investigating whether 
the two approaches are equivalent in more general situations.

 The usual intuition from statistical physics
tells us that, in the limit of a thick bubble interface, 
entropy does not increase across the interface if local thermal equilibrium is maintained, {\it i.e.}\ 
plasma particles crossing a thick interface 
constitutes an isoentropic process. This is true 
only if a single particle species forms the plasma.
However, in realistic applications many particle species are present. Then the system naturally 
splits into {\it active species}, which
exert a significant force on the interface, and {\it passive}
species, whose mass remains 
approximately constant or zero across the interface, such 
that they do not exert a significant force on the interface.
The passive species form a heat reservoir. 
The standard thermodynamic picture then applies. 
The entropy density in the active part 
of the system reduces significantly, thus exerting 
the force on the bubble. The heat reservoir absorbs heat,
thus heating up. If the heat capacity of the reservoir is 
large enough, the total temperature of the system
plus reservoir does not change much, and 
the local thermal equilibrium description, which 
assumes equal temperature on both sides of the interface, 
can be considered as the leading order approximation.
In realistic situations,
in which the standard model contains most of the degrees 
of freedom, the active particles are the top quark,
the weak gauge bosons and the Higgs particle and they
comprise about $20\%$ of the relativistic degrees 
of freedom; the remaining $80\%$ constitute
a large heat reservoir.

\medskip

In order to illustrate how to use~(\ref{conservation of Tmn:4}) for 
the phase transition 
dynamics, in what follows we first consider a simple real scalar field model and then more general models.  For the scalar theory we will analyze two opposite cases: the local thermal equilibrium, where we assume that the interactions in the wall are very efficient; and the ballistic approximation in which the interactions in the wall are inefficient. The force that we find in both cases will display a very distinct behavior.


\section{Real scalar field} 
\label{Real scalar field}

In this section we calculate the one-loop energy momentum tensor of a 
self-interacting, massless scalar field in the presence of an expanding spherical bubble at 
a first order phase transition. 
For simplicity we shall first consider the case in which the scalar is in a {\it local thermal equilibrium} (lte),
which approximates well the scalar field state if the thermalization time scale,
$\tau_{\rm th} =1/\Gamma_{\rm th}$ 
is smaller than the time $\Delta t_b = L/(\gamma v)$ it takes the bubble to pass by an observer in 
the plasma frame, {\it i.e.}
\begin{equation}
 \tau_{\rm th} =\frac{1}{\Gamma_{\rm th}} < \frac{L}{\gamma v} \qquad 
 {\rm (local\;thermal\;equilibrium)}
 \,.
\label{local thermalization}
\end{equation}
Since $\Gamma_{\rm th}$ 
is controlled by a coupling constant, which is typically smaller than one, 
$\tau_{\rm th}> 1/T$, the condition~(\ref{local thermalization}) is more stringent 
than the on-shell condition~(\ref{on-shell condition}).~\footnote{We do not verify whether this condition is met for the toy model that we consider here, since in this model no phase transition is present and we treat it only as an illustration of the method. More realistic models are considered in Section~\ref{Standard model and its extensions}. }
If~(\ref{local thermalization})
is not met, then particles that move across the bubble partially thermalize, and in the extreme case do not 
thermalize at all, in which case a {\it ballistic} approximation applies. In what follows we firstly 
calculate the energy-momentum tensor by assuming local thermal equilibrium and then 
discuss how our results are affected if ballistic approximation represents a more appropriate 
description.

The free scalar field action is given by, 
\begin{equation}
S_0\left[\phi,g_{\mu\nu}\right]  = \int d^Dx \sqrt{-g}{\cal L}_\phi
\,,\qquad
{\cal L}_\phi=-\frac12g^{\mu\nu}(\partial_\mu\phi)(\partial_\nu\phi)-\frac12 m^2\phi^2
\,,
\label{action: real scalar}
\end{equation}
where $g^{\mu\nu}$ is the inverse of the metric tensor $g_{\mu\nu}$, the signature of the metric is mostly plus,
$g={\rm det}\left[g_{\mu\nu}\right]$ and $m$ is the scalar mass. 
The origin of the mass can be either a tree level-mass, $m_0$, or it can be generated by 
 a field condensate. For example, adding a scalar self-interaction, 
\begin{equation}
{\cal L}_{\rm int}=-\frac{\lambda}{4!}\phi^4
\label{self interaction}
\end{equation}
to the Lagrangian in~(\ref{action: real scalar}) will generate in the presence of a condensate 
$\langle\hat \phi\rangle =\phi_0$ a field dependent mass,~\footnote{Since no condensate can be 
generated in a pure massless $\lambda \phi^4$ theory~\cite{Coleman:1973}, realistic 
models of spontaneous symmetry breaking must include extra fields or a negative mass term.
For simplicity in this section we  neglect this complication.}
\begin{equation}
 m^2(\phi_0) = \frac{\lambda}2\phi_0^2
 \,.
 \label{field dependent mass}
\end{equation}
If the background field $\phi_0=\phi_0(x)$ varies in spacetime,
the mass~(\ref{field dependent mass}) will follow the suit. As long as the variation is 
slow, it can be treated as an adiabatically varying quantity. 

\begin{figure}[h! ]
\vskip -0.4cm
\centering
\includegraphics[scale=0.6]{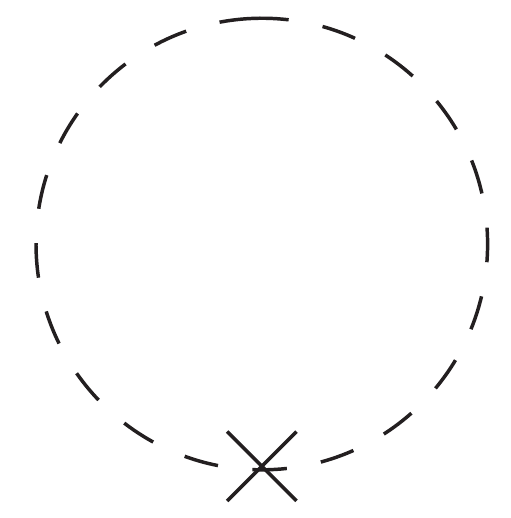}
\vskip -0.2cm
\caption{\small The Feynman diagram that contributes at one-loop level to 
the energy-momentum tensor of a real scalar field. The dashed circle denotes the scalar propagator
and the cross denotes the energy-momentum tensor insertion.}
\label{figure scalar loop Tmn}
\vskip -0.3cm
\end{figure}

Varying~(\ref{action: real scalar})
with respect to $g^{\mu\nu}$ results in the energy-momentum tensor, 
\begin{equation}
T_{\mu\nu}^\phi  \equiv -\frac{2}{\sqrt{-g}}
\frac{\delta S_0}{\delta g^{\mu\nu}}
 =(\partial_\mu\phi)(\partial_\nu\phi)+g_{\mu\nu}{\cal L}_\phi
\,.
\label{energy momentum: real scalar}
\end{equation}
Here we are interested in evaluating the expectation value  of the one-loop
energy-momentum tensor, 
\begin{equation}
\langle \hat T_{\mu\nu}^\phi\rangle 
 = \langle T[ (\partial_\mu\hat \phi)(\partial_\nu\hat \phi)]\rangle 
 +g_{\mu\nu}\langle T[\hat{\cal L}_\phi]\rangle 
\,. 
\label{expectation value Tmn: real scalar}
\end{equation}
The contributing one-loop diagram is shown in figure~\ref{figure scalar loop Tmn}, 
where the dashed line denotes the scalar propagator and the cross ($\times$) indicates
the energy-momentum tensor insertion.

\subsection{Local thermal equilibrium approximation}
\label{Local thermal equilibrium approximation}

Since at one-loop order the energy-momentum tensor~(\ref{expectation value Tmn: real scalar}) can 
be evaluated by making use of the field propagator, 
the next natural step is to construct the scalar field propagator, which is in general defined by, 
\begin{equation}
i\Delta_m(x;x') ={\rm Tr}\Big[\hat\rho(t) T[\hat\phi(x)\hat\phi(x')]\Big]
\,.
\label{scalar propagator}
\end{equation}
Here $\hat\rho$ denotes the density operator, which contains the information about the scalar field state 
and $T$ denotes the time ordering operator.
In local thermal equilibrium~(\ref{local thermalization}) holds and 
the density operator can be approximated by its local 
equilibrium form, 
$\hat\rho\rightarrow\hat\rho_{\rm th}= {\rm e}^{-\beta \hat H_\phi}/{\rm Tr}[{\rm e}^{-\beta \hat H_\phi}]$,
where $\hat H_\phi$ is the Hamiltonian operator.

In general the free scalar propagator~(\ref{scalar propagator}) obeys the equation of motion, 
\begin{equation}
\sqrt{-g}\left(\Box-m^2\right)i\Delta_m(x;x') = i\hbar\delta^D(x\!-\!x')
\,,
\label{thermal scalar propagator: EOM}
\end{equation}
where $m=m(\phi_0)$ is the field-dependent scalar field mass.  
The thermal propagator in~(\ref{thermal scalar propagator: EOM}) 
can be thought of as the inverse of the operator,
$\sqrt{-g}\left(\Box-m^2\right)\times\delta^D(x\!-\!y)$, where 
$\Box=g^{\mu\nu}\nabla_\mu\nabla_\nu$ denotes the
d'Alembertian operator with thermal boundary conditions imposed. In practice this can be done by 
inserting the thermal density operator $\hat\rho_{\rm th}$ in~(\ref{expectation value Tmn: real scalar}). 
Furthermore, since time scales that govern the phase transition dynamics are typically much shorter than
the Hubble time, the expansion of the Universe can be considered as adiabatic
and in the leading order adiabatic approximation 
the d'Alembertian reduces to the usual wave operator, 
$\Box\rightarrow \partial^2=\eta^{\mu\nu}\partial_\mu\partial_\nu$
(the expansion of the Universe is then captured by taking $T=T(t)$).
That means that for our purposes it suffices to determine the thermal propagator for a massive scalar in Minkowski space 
in {\it slowly varying background fields} (the metric tensor and the scalar field). 
That propagator is well known and in the plasma frame it reads,~\footnote{ 
As we argue at the end of 
section~\ref{The force on bubbles from the renormalized stress energy tensor}, 
the local thermal equilibrium assumed when constructing 
the propagator~(\ref{thermal scalar propagator}) 
approximates well the state if scatterings are efficient
and if the massive scalar is supplemented by a 
sufficiently 
large heat reservoir.
} 
\begin{equation}
 i\Delta_m(x;x') 
 = \frac{m^{D-2}}{(2\pi)^{D/2}}\frac{K_{\frac{D-2}{2}}\left(m\sqrt{\Delta x^2}\right)}
                                                        {(m\sqrt{\Delta x^2})^{\frac{D-2}{2}}}
\!+\! \int\frac{d^{D-1}p}{(2\pi)^{D-1}}
 \frac{{\rm e}^{i \vec p\cdot(\vec x-\vec x^{\prime})}}{E_p}
              \frac{\cos[E_p(t-t^\prime)]}{{\rm e}^{\beta E_p}\!-\!1}
\,,
\label{thermal scalar propagator}
\end{equation}
where $K_\nu(z)$ is the Macdonald function (the modified Bessel function of the second kind), 
$E_p=\sqrt{\vec p^{\,\,2} + m^2}$ and,
\begin{equation}
 \Delta x^2(x;x') = -(|t\!-\!t'|\!-\!\imath \epsilon)^2 + \|\vec x\!-\!\vec x^{\,\prime}\|^2
\label{invariant distance}
\end{equation}
is the invariant distance function on Minkowski space where 
the appropriate $\imath \epsilon$ prescription for the Feynman propagator is also indicated.

Note that in the free propagator~(\ref{thermal scalar propagator})
the vacuum ($\propto K_{(D-2)/2}$) and thermal contributions neatly split and that 
the thermal contribution is finite, and thus can be evaluated in $D=4$. While it is not 
generally possible to write in a closed form the thermal part of the 
propagator in~(\ref{thermal scalar propagator}), its coincident and near coincident
limits are possible to express in terms of the bosonic thermal integral, 
\begin{equation}
 J_B(n,z) \equiv \int_0^\infty dx x^{n-2}\ln\left(1-{\rm e}^{-\sqrt{x^2+z^2}}\right)  
\,,
\label{JB n integral}
\end{equation}
based on which one can obtain the thermal contribution to the one-loop effective potential
of a bosonic degree of freedom in $n$ space-time dimensions. For example, in
$n=4$ and for a particle of mass $m$, the one-loop thermal effective potential is 
$V_T^{(1)}=-[2\pi^2\beta^4]^{-1}J_B(4,\beta m)$. 

With these remarks in mind we can write a closed form expression for the coincident 
propagator~(\ref{thermal scalar propagator}), 
\begin{eqnarray}
 i\Delta_m(x;x) 
 &=& \frac{m^{D-2}\Gamma\big(1-\frac{D}{2}\big)}{(4\pi)^{D/2}}
\!+\!\frac{1}{2\pi^2\beta^2} \int_0^\infty dx\frac{x^2}{\sqrt{x^2+(\beta m)^2}}
              \frac{1}{{\rm e}^{\sqrt{x^2+(\beta m)^2}}\!-\!1}
\nonumber\\
 &=& \frac{m^{D-2}\Gamma\big(1-\frac{D}{2}\big)}{(4\pi)^{D/2}}
+\frac{1}{2\pi^2\beta^3m}\left[\partial_z J_B(4,z)\right]_{z=\beta m}
\,,
\label{thermal scalar propagator: coincidence}
\end{eqnarray}
such that the vacuum part is divergent in $D=4$ and ought to be regularized.
The vacuum contribution in~(\ref{thermal scalar propagator: coincidence}) was 
evaluated by noting that the Bessel function $K_\nu(z)$ in 
Eq.~(\ref{thermal scalar propagator}) can be expanded around the lightcone 
as a sum of two series ($\nu = (D-2)/2$),
\begin{equation}
\frac{K_\nu(z)}{z^\nu} = \frac{\Gamma\left(1\!-\!\frac{D}{2}\right)}{2^\frac{D}{2}} \sum_{n=0}^\infty\frac{(z/2)^{2n}}{\left(\frac{D}{2}\right)_n n!}
+\frac{\Gamma\left(\frac{D}{2}\!-\!1\right)}{2^\frac{D}{2}}
            \sum_{n=0}^\infty\frac{(z/2)^{2n+2-D}}{\left(2-\frac{D}{2}\right)_n n!}
\,,\quad
z=m\sqrt{\Delta x^2}
\,,\;
\label{expansion of Knu}
\end{equation}
and then used the fact that in dimensional regularization, by a clever use 
of the suitable analytic extension, one finds that the series with 
$D$ dependent powers in~(\ref{expansion of Knu}) 
does not contribute at coincidence.
Since all power-law divergences get subtracted in this way, this feature became known as the automatic subtraction.

 In order to evaluate~(\ref{expectation value Tmn: real scalar}) we need,
 \begin{equation}
 \langle T[ (\partial_\mu\hat \phi)(\partial_\nu\hat \phi)]\rangle
   =\left[\partial_\mu\partial_\nu^\prime  
    \langle T^*[\hat \phi(x)\hat \phi(x')]\rangle\right]_{x'\rightarrow x}
\,,
 \label{part of Tmn: real scalar}
 \end{equation}
where we introduced the usual $T^*$ time ordering which is defined to commute with the 
two external derivatives in~(\ref{part of Tmn: real scalar}). 
The vacuum part of~(\ref{part of Tmn: real scalar}) 
is $-2\eta_{\mu\nu}$ times the linear coefficient in $\Delta x^2$ of 
the propagator in~(\ref{thermal scalar propagator}), which is easily extracted 
from the integer series in~(\ref{expansion of Knu}),
\begin{equation}
\langle T[ (\partial_\mu\hat \phi)(\partial_\nu\hat \phi)]\rangle_{\rm vac}
 = \eta_{\mu\nu} \frac{m^{D}\Gamma\big(\!-\frac{D}{2}\big)}{2(4\pi)^{D/2}}
 \,.
 \label{part of Tmn: real scalar: vac}
 \end{equation}
 When taken together with the vacuum part in~(\ref{thermal scalar propagator: coincidence})
 this then implies, $\langle T[\hat{\cal L}]\rangle_{\rm vac}=0$, such that,
\begin{equation}
\langle \hat T_{\mu\nu}\rangle_{\rm vac}
 = \eta_{\mu\nu} \frac{m^{D}\Gamma\big(\!-\frac{D}{2}\big)}{2(4\pi)^{D/2}}
 \,,
 \label{Tmn: real scalar: vac}
 \end{equation}
 which has the form of a cosmological constant. 
 
The thermal contribution to~(\ref{part of Tmn: real scalar})
 is given by the integral over $p_\mu p_\nu$ of the coincident thermal integral
in~(\ref{thermal scalar propagator}). Recalling that $p^0=E_p$, the $00$ and 
$ij$ contributions ought to be evaluated separately (the $0i$ contribution vanishes),
\begin{eqnarray}
\langle T[ (\partial_0\hat \phi)(\partial_0\hat \phi)]\rangle_{\rm th}
 \!\!&=&\!\!\frac{1}{2\pi^2\beta^5m}\left[\partial_z J_B(6,z)\right]_{z=\beta m}
 \!+\!\frac{m}{2\pi^2\beta^3}\left[\partial_z J_B(4,z)\right]_{z=\beta m}
 \quad\;
 \label{part of Tmn: real scalar: 00 thermal contribution}\\
 \langle T[ (\partial_i\hat \phi)(\partial_j\hat \phi)]\rangle_{\rm th}
 \!\!&=&\!\!\delta_{ij}\frac{1}{6\pi^2\beta^5m}
                   \left[\partial_z J_B(6,z)\right]_{z=\beta m}
\,,
 \label{part of Tmn: real scalar: ij thermal contribution}
\end{eqnarray}
such that $\langle T[\hat {\cal L}]\rangle_{\rm th}=0$, and the contributions 
in~(\ref{part of Tmn: real scalar: ij thermal contribution}) are the thermal contributions
to the one-loop stress-energy tensor. 

To complete the calculation, we still ought to renormalize the vacuum
contribution~(\ref{Tmn: real scalar: vac}), which for a 
constant bare mass can be done 
by adding a cosmological constant counterterm. However, here we are primarily interested 
in a mass generated by a field condensate~(\ref{field dependent mass}), 
and the suitable counterterm action is of the form, 
\begin{equation}
S_{\rm ct}=\int d^Dx\sqrt{-g}\left[-\frac{\delta\lambda}{4!}\phi^4\right]
\,,
\label{counter term action: scalar}
\end{equation}
which contributes to the energy-momentum tensor as, 
\begin{equation}
 T_{\mu\nu}^{\rm ct} = -\frac{\delta\lambda}{4!}\phi^4 g_{\mu\nu}
 \,.
 \label{counter term Tmn: scalar}
\end{equation}
We shall use the {\it minimal subtraction scheme}, and to that purpose 
expand~(\ref{Tmn: real scalar: vac}) around $D=4$, 
\begin{equation}
\langle \hat T_{\mu\nu}\rangle_{\rm vac}
 =-\eta_{\mu\nu} \frac{m^4}{32\pi^2}\left[\frac{\mu^{D-4}}{D\!-\!4}
  +\frac{1}{2}\ln\left(\frac{m^2}{4\pi\mu^2}\right)+\frac{\gamma_E}{2}-\frac34\right]
 \,,
 \label{Tmn: real scalar: vac 2}
 \end{equation}
where $\mu$ is an arbitrary scale and $m^2=\lambda\phi_0^2/2$. 
Comparing~(\ref{Tmn: real scalar: vac 2}) with~(\ref{counter term Tmn: scalar}) we see that, 
\begin{equation}
\delta\lambda = -\frac{3\lambda^2}{16\pi^2}\frac{\mu^{D-4}}{D\!-\!4}
\,,
 \label{counter term lambda: scalar}
\end{equation}
removes the divergence from~(\ref{Tmn: real scalar: vac 2}),
resulting in the following renormalized energy-momentum tensor, 
\begin{eqnarray}
\langle\hat T_{\mu\nu}\rangle_{\rm ren} 
&=& -\eta_{\mu\nu} \frac{m^4}{64\pi^2}\left[\ln\left(\frac{m^2}{4\pi\mu^2}\right)
\!+\!\gamma_E\!-\!\frac32\right]
   \!+\!\frac{\eta_{\mu\nu}}{6\pi^2\beta^5m}
                   \left[\partial_z J_B(6,z)\right]_{z=\beta m}
\quad
\nonumber\\
 &+&\!\! \delta_\mu^0\delta_\nu^0
        \left\{
        \frac{2}{3\pi^2\beta^5m}\left[\partial_z J_B(6,z)\right]_{z=\beta m}
         \!+\!\frac{m}{2\pi^2\beta^3}\left[\partial_z J_B(4,z)\right]_{z=\beta m}
        \right\}
.
 \quad
\label{renormalized energy-momentum tensor: scalar}
\end{eqnarray}

From the point of view of the bubble force calculation, the change in 
the entropy density~(\ref{entropy density}) 
across the bubble is what determines the  friction force on the 
bubble interface, 
which is determined by the second line in~(\ref{renormalized energy-momentum tensor: scalar}) 
(Lorentz covariant contributions have a 
vanishing entropy density),
\begin{equation}
T\Delta s = \frac{2\pi^2}{45\beta^4} 
    \!-\! \frac{2}{3\pi^2\beta^5m}\left[\partial_z J_B(6,z)\right]_{z=\beta m}
         \!-\!\frac{m}{2\pi^2\beta^3}\left[\partial_z J_B(4,z)\right]_{z=\beta m}
\,.
\label{change in entropy density: scalar}
\end{equation}
Inserting this into Eq.~(\ref{conservation of Tmn:4}) gives 
an expression for the bubble speed as a function of the change in
the pressure across the bubble $-\Delta {\cal P}$
 due to the bubble nucleation and the change of the plasma 
 entropy density. 
Since $-\Delta {\cal P}$ depends on the amount of supercooling 
before bubbles start nucleating and on the detailed form of the effective potential, 
the toy model Lagrangian considered in this 
section cannot be used for a meaningful estimation 
of $\Delta {\cal P}$. For that reason
we shall not attempt to estimate it here,
but instead we shall treat it as a free parameter 
of the transition. 

\begin{figure}[h!]
\vskip-0.1cm
\begin{center}
\includegraphics[width=7.8cm]{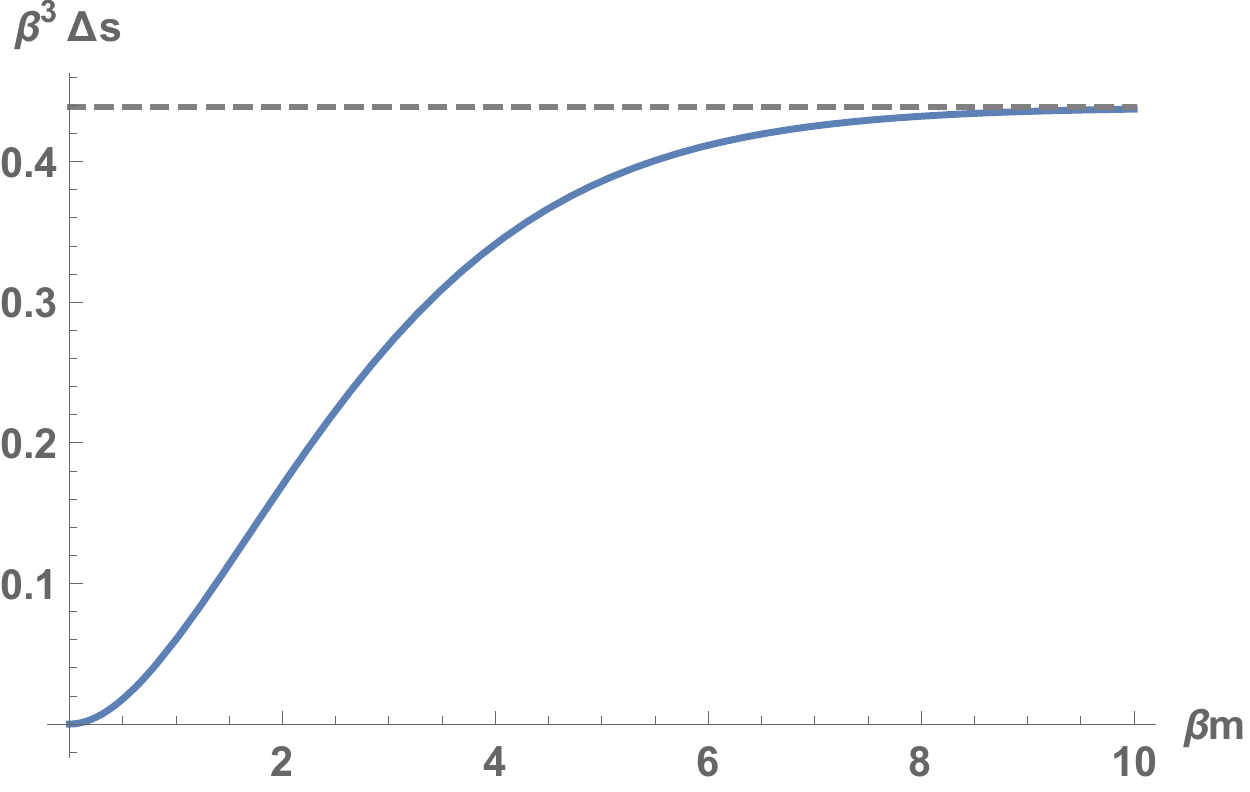}
\end{center}
\vskip-0.5cm
\caption{\small The dimensionless change in the entropy density $\Delta s/T^3$ (solid blue) 
of a real scalar field thermal plasma across the nucleated bubble as a function of the 
scalar mass $m/T$.
Since the scalar mass $m=\sqrt{\lambda/2}\phi_0(x)$ increases across the bubble
as the scalar condensate increases, the entropy in the broken phase decreases.
The maximum amount by which the entropy density can change is $2\pi^2T^3/45$,
which is formally reached when $m\rightarrow \infty$, when the entropy density inside the bubble tends to 
zero (horizontal dashed).}
\label{entropy production}
\end{figure}

In figure~\ref{entropy production} we show how the entropy density changes across
the bubble as a function of the scalar mass. The relativistic plasma limit, 
$s=2\pi^2 T^3/45$ (dashed) is reached in the limit when the mass in the broken phase 
$m\rightarrow \infty$, because then the entropy density inside the bubble tends to zero.
\begin{figure}[ht]
\vskip-0.1cm
\begin{center}
\includegraphics[width=7.8cm]{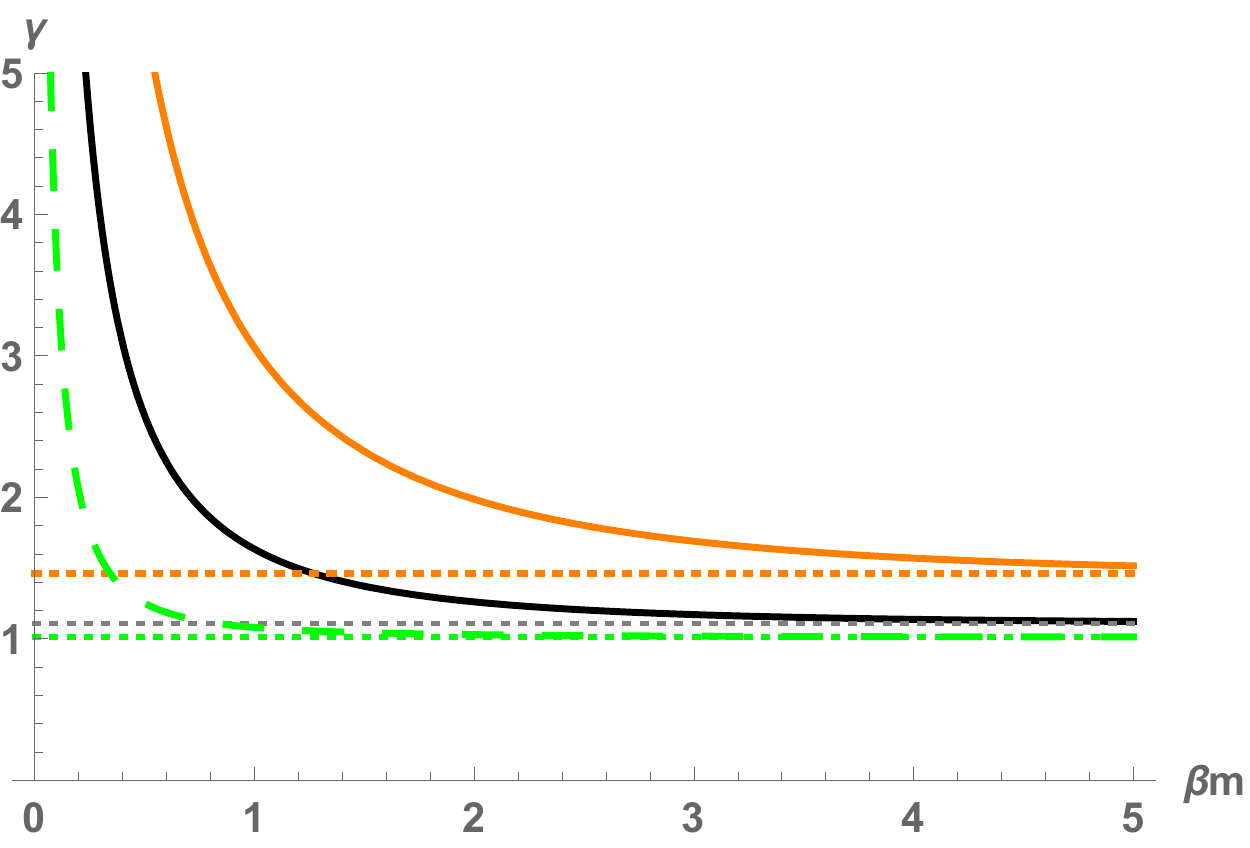}
\end{center}
\vskip-0.5cm
\caption{\small The bubble Lorentz factor $\gamma(v)$ 
as a function of the scalar mass $m/T$ for $\Delta {\cal P}=-0.01/\beta^4$ 
(green dashed), $\Delta {\cal P}=-0.1/\beta^4$ (solid black) and 
$\Delta {\cal P}=-0.5/\beta^4$ (solid orange).}
\label{bubble speed gamma}
\end{figure}
The main application of Eq.~\eqref{conservation of Tmn:4} is to determine the terminal velocity (or Lorentz factor)  of a bubble in stationary state. Figure~\ref{bubble speed gamma} shows the bubble's Lorentz factor $\gamma$ 
as a function of the scalar mass and the strength of the transition, expressed
as $\Delta {\cal P}$ across the bubble for a moderately strong transition, 
$\Delta {\cal P}=-0.1/\beta^4$ (solid black), for a strong transition 
$\Delta {\cal P}=-0.5/\beta^4$ 
(solid orange) and for a very weak transition $\Delta {\cal P}=-0.01/\beta^4$ 
(dashed green). For each choice of $\Delta {\cal P}$ there is a minimum
bubble Lorentz factor, $\gamma_{\rm min}=\sqrt{1+[45(-\Delta {\cal P})/(2\pi^2T^4)]}$,
reached when $m\rightarrow\infty$. The maximum $\gamma$ is reached when 
$m\rightarrow 0$, in which case the bubble runs away (since in that limit there is no force). 
However, for every $m^2>0$ and 
$\Delta {\cal P}<0$, no matter how small they may be, 
a finite $\gamma$ is reached. 

\begin{figure}[h]
\vskip-0.1cm
\begin{center}
\includegraphics[width=7.8cm]{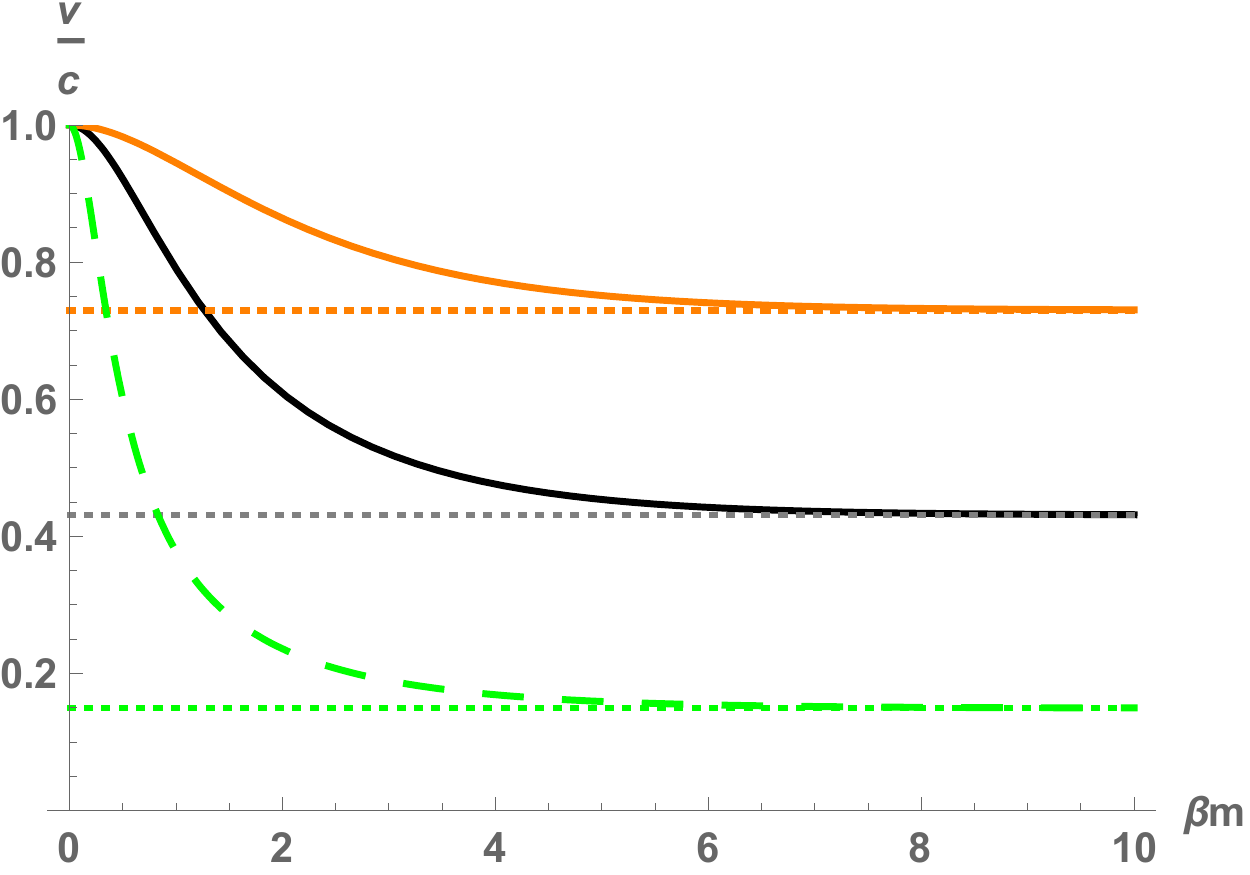}
\end{center}
\vskip-0.5cm
\caption{\small The bubble speed $v/c$ 
as a function of the scalar mass $m/T$ for the same choice of the parameters
as in figure~\ref{bubble speed gamma}:
$\Delta {\cal P}=-0.01/\beta^4$ 
(green dashed), $\Delta {\cal P}=-0.1/\beta^4$ (solid black) and 
$\Delta {\cal P}=-0.5/\beta^4$ (solid orange).}
\label{bubble speed v}
\end{figure}

Figure~\ref{bubble speed v} shows the bubble speed for the same choice of parameters
as in figure~\ref{bubble speed gamma}. Notice that, independently of $\Delta {\cal P}$,
all curves begin at $v=c$ for $m=0$.
The horizontal dashed lines indicate the lowest 
attainable bubble speed for a given phase transition strength $\Delta {\cal P}$.

We see that in the case of local thermal equilibrium, the bubble velocity stabilizes at relatively low values. Therefore, even based on our analysis which is limited to moderate velocities with $\gamma<10$, one can conclude that the bubbles cannot run away.

\subsection{Ballistic approximation}
\label{Ballistic approximation}

If the thermalization rate is not large 
enough to satisfy Eq.~(\ref{local thermalization}) then
the {\it ballistic approximation}~\cite{Moore:1995si} is the more appropriate one to model the state of the field. 
In this case, one assumes that
particles are in thermal equilibrium in the symmetric phase far in front of the expanding bubble,
and they move across the wall so fast that they interact semiclassicaly, but the time 
is so short that they do not reach a local thermal equilibrium. To solve for the force 
acting on a bubble in this case, one can solve the Liouville equation in the bubble frame, 
\begin{equation}
\left(\frac{p_z}{E}\partial_z - \frac{\partial_z(m^2)}{2E}\partial_{p_z}\right)f(p_z,z) =0
\,,
\label{Liouville equation}
\end{equation}
where $E=E(p_z,p_\perp,z)=\sqrt{p_\perp^2+p_z^2+m(z)^2}$. By observing that,
$\partial_z = (\partial_z m^2)\partial_{m^2}$, one sees that the general solution 
of~(\ref{Liouville equation}) can be written as a general function of $m^2+p_z^2$ and $p_\perp$
(or, equivalently, of $E$),
\begin{equation}
 f=f(p_z^2\!+\!m^2,\vec p_\perp)
 \,,
\label{Liouville equation: general solution}
\end{equation}
which is equivalent to saying that $E$ and $\vec p_\perp$ are conserved in the bubble frame.
On the other hand, in front of 
the bubble, where $m=0$ (inside the bubble 
where $m\neq 0$), 
the solution is given by,
\begin{eqnarray}
f_{-\infty} &=&\frac{1}{{\rm e}^{\beta\gamma(E-v p_z)}\!-\!1}
\quad (p_z>0, z\rightarrow -\infty, m=0)
\,,
\label{Liouville equation: boundary condition +}\\
f_{\infty} &=& \frac{1}{{\rm e}^{\beta\gamma(E-v p_z)}\!-\!1}
\quad (p_z<0, z\rightarrow +\infty, m\neq 0)
\,.
\label{Liouville equation: boundary condition -}
\end{eqnarray}
respectively. Notice that the negative $p_z$ branch of the 
distribution 
function~(\ref{Liouville equation: boundary condition -})
exists only if thermalisation inside the 
bubble takes place. Since the bubbles at the electroweak 
transition grow large before they start colliding,
this condition will be satisfied for typical bubbles. If not, 
the negative 
branch~(\ref{Liouville equation: boundary condition -})
will be absent.
When the boundary condition~(\ref{Liouville equation: boundary condition +}--\ref{Liouville equation: boundary condition -})
is imposed 
on~(\ref{Liouville equation: general solution}) one 
obtains the general ballistic solution, which is 
conveniently broken into three parts as follows.
\begin{itemize}
\item[\hskip 1cm {\bf Case A.}] {\it Transmission ($t_+$)
from the symmetric phase:}
\begin{eqnarray}
 f(z,p_z,\vec p_\perp) &=& \frac{1}
      {\exp\left[\beta\gamma\left(E\!-\!v \sqrt{p_z^2\!+\!m(z)^2}
         \right)\right]\!-\!1}
\nonumber\\
&& \hskip 4.1cm
\Bigl(p_z(z)\!>\!\sqrt{m_0^2\!-\!m(z)^2}\,\Bigr)
 \,.\qquad
\label{Liouville equation: particular solution A}
\end{eqnarray}
\item[{\bf Case B.}] {\it Reflection ($r$):}
\begin{eqnarray}
 f(z,p_z,\vec p_\perp) 
 \!\!&=&\!\! \frac{1}{\exp\left[\beta\gamma\left(E
  \!-\!v {\rm sign}[p_z]\sqrt{p_z^2\!+\!m(z)^2}\right)\right]
      \!-\!1}
\nonumber\\
&&\hskip 0.7cm
\Bigl(-\sqrt{m_0^2\!-\!m(z)^2}<p_z(z)
             <\sqrt{m_0^2\!-\!m(z)^2}\Bigr)
 \,. \qquad
\label{Liouville equation: particular solution B}
\end{eqnarray}
\item[{\bf Case C.}] {\it Transmission ($t_-$) 
           from inside the bubble:}
\begin{eqnarray}
 f(z,p_z,\vec p_\perp) 
 \!&=&\! \frac{1}{\exp\left[\beta\gamma
 \Bigl(E\!+\!v \sqrt{p_z^2\!+\!m(z)^2\!-\!m_0^2}\Bigr)\right]
    \!-\!1}
,
\nonumber\\
 &&\hskip 3.4cm
 \Bigl(p_z(z)<-\sqrt{m_0^2\!-\!m(z)^2}\,\Bigr)
\,,\qquad
\label{Liouville equation: particular solution C}
\end{eqnarray}
\end{itemize}
where 
we temporarily introduced an index $0$ on $m$ to 
denote the mass deep inside the bubble.
Note that in the reflected 
solution~(\ref{Liouville equation: particular solution B}) 
we included both the particle incoming on the wall 
 ($-m_0<p_z<0$) and the reflected particles ($0<p_z<m_0$),
 because they both contribute to the force on the wall.

The ballistic solution~(\ref{Liouville equation: particular solution A}--\ref{Liouville equation: particular solution C}) 
tells us that quasiparticles are still on-shell, 
but they are not thermally distributed. 
As a consequence of this departure from
thermal equilibrium Lorentz symmetry is violated.
This then introduces a dependence on the Lorentz factor 
$\gamma$ in the plasma frame quantities in 
Eq.~(\ref{conservation of Tmn:4}), making thus the $\gamma$ 
dependence of the bubble force more complex than what
Eq.~(\ref{conservation of Tmn:4}) suggests.

To obtain the bubble force one must calculate the change 
in the plasma pressure across the bubble interface.
The pressure in the bubble frame reads, 
\begin{equation}
 {\cal P}_z= \int \frac{d^3p}{(2\pi)^3}\frac{p_z^2}{E}f
 \,,
 \label{pressure inside bubble}
 \end{equation}
 where the index $z$ indicates that $\mathcal{P}_z$ is the pressure in the direction in which the bubble expands. 
For simplicity in Eq.~(\ref{pressure inside bubble}) 
we took account of 
the plasma contribution only (since the vacuum contribution
is unchanged).  
The form of the pressure in~(\ref{pressure inside bubble}) 
can be obtained from 
the $zz$ part of Eq.~(\ref{part of Tmn: real scalar}),
provided the thermal distribution function 
$1/({\rm e}^{\beta E_p}-1)$ 
in~(\ref{thermal scalar propagator})
gets replaced by the ballistic distribution 
function~(\ref{Liouville equation: particular solution A}--\ref{Liouville equation: particular solution C}).
This ballistic approximation 
is valid as long as the semiclassical on-shell 
condition~(\ref{on-shell condition}) holds true.
Since the pressures in the other two spatial directions do not change by the expanding bubble, 
we do not need to consider them here. The quantity ${\cal P}_z$ in~(\ref{pressure inside bubble})
represents the total pressure in the $z$ direction in the wall frame exerted by the plasma,
and it is therefore equal to $\langle\hat T^p_{zz}\rangle = (\gamma^2-1)Ts + {\cal P}_p$, where 
$s$ and ${\cal P}_p$ are the plasma frame quantities, and therefore -- up to an unimportant term ${\cal P}_p$ --
it equals the differential bubble force in Eq.~(\ref{conservation of Tmn:4}).
 To evaluate the integrals in~(\ref{pressure inside bubble}), it is convenient to first integrate over $E$, 
 in which case the ranges of integration are, $E\in [\sqrt{p_z^2+m^2},\infty)$, $p_z\in(-\infty,\infty)$.
By making use of $EdE = p_\perp dp_\perp$ and integrating {\it e.g.}\ the transmitted 
contribution~(\ref{Liouville equation: particular solution A}) 
over $E$ one gets, 
\begin{equation}
 {\cal P}^{t_+}_z=-\frac{1}{4\pi^2\beta\gamma}
  \int_{\sqrt{m_0^2-m(z)^2}}^\infty dp_z p_z^2  
 \ln\left[1-\exp\left(-\beta\gamma(1-v)\sqrt{p_z^2+m(z)^2}\right)\right]
\,.
 \label{pressure inside bubble:2}
 \end{equation}
The integral over $p_z$ cannot be done analytically. 
To simplify it, it is convenient to introduce dimensionless variables,
$M_\pm=\beta\gamma(1\pm v)m$, 
$x=\beta\gamma(1\pm v)p_z$, 
and the integral~(\ref{pressure inside bubble:2}) becomes,
\begin{eqnarray}
 {\cal P}_z^{t_+}
 \!&=&\!-\frac{1}{4\pi^2\beta^4\gamma^4(1\!-\!v)^3}
  \int_{\beta\gamma(1-v)\sqrt{m_0^2-m(z)^2}}^\infty\! dx x^2  
\ln\left(1\!-\!{\rm e}^{-\sqrt{x^2+M_-^2}}\right)
\,,\qquad
 \label{pressure inside bubble:3}
 \end{eqnarray}
 where $v\equiv v(\gamma)=\sqrt{1-(1/\gamma^2)}$. 
 An analogous procedure gives for the reflected 
 contribution~(\ref{Liouville equation: particular solution B}) 
\begin{eqnarray}
 {\cal P}_z^{r}
 \!&=&\!-\frac{1}{4\pi^2\beta^4\gamma^4}
 \Bigg\{\frac{1}{(1\!-\!v)^3}
  \int_0^{\beta\gamma(1-v)\sqrt{m_0^2-m(z)^2}}\! dx x^2  
\ln\left(1\!-\!{\rm e}^{-\sqrt{x^2+M_-^2}}\right)
\nonumber\\
&&+\,\frac{1}{(1\!+\!v)^3}
  \int_0^{\beta\gamma(1+v)\sqrt{m_0^2-m(z)^2}}\! dx x^2  
\ln\left(1\!-\!{\rm e}^{-\sqrt{x^2+M_+^2}}\right)
\Bigg\}
\,,\qquad
 \label{pressure inside bubble:3 B}
 \end{eqnarray}
 and for the transmitted 
 contribution~(\ref{Liouville equation: particular solution C}), 
 %
\begin{eqnarray}
 {\cal P}_z^{\rm t_-}
 \!\!&=&\!\!-\,\frac{1}{4\pi^2\beta^4\gamma^4}
  \int_{\beta\gamma\sqrt{m_0^2-m(z)^2}}^\infty\! dx x^2
\label{pressure inside bubble:3 C}
\\
&&\hskip -0.1cm
\times\,  
\ln\!\left[1\!-\!{\exp}\left(
    \!\!-\sqrt{x^2\!+\!\bigl(\beta\gamma m(z)\bigr)^{\!2}}
    \!-\!v\sqrt{x^2\!+\!(\beta\gamma)^2
              \Bigl(m(z)^2\!-\!m_0^2\Bigr)^{\!2}}\,
                         \right)\right]\!
.
\nonumber 
 \end{eqnarray}
 %
 
 The total pressure
 is then simply the sum of the three contributions,
 \begin{equation}
  {\cal P}_z={\cal P}^{t_+}_z+{\cal P}^r_z+{\cal P}^{t_-}_z 
 \,,
 \label{total ballistic pressure}
 \end{equation}
 and the pressure difference across the bubble reads
 \begin{equation}
 \Delta {\cal P}_z=\Delta {\cal P}^{t_+}_z
    +\Delta {\cal P}^r_z+\Delta{\cal P}^{t_-}_z 
 \,.
 \label{total ballistic pressure difference}
 \end{equation}
 There is a subtlety involved in 
 evaluating~(\ref{total ballistic pressure}) 
in the reflected 
contribution~(\ref{pressure inside bubble:3 B}),
which is simply equal to, $\Delta {\cal P}^r_z={\cal P}^r_z$.
This is because the contribution
of the reflected particles to $\Delta {\cal P}^r_z$ is given 
by  ${\cal P}^r_z$ in front of the wall, minus the pressure 
at the turning point. But this contribution is zero 
because the phase space at the turning point is zero
(all particles at the turning point have $p_z=0$, such that 
the integral over $p_z$ vanishes).
Upon combining all of the 
contributions~(\ref{pressure inside bubble:3}),
(\ref{pressure inside bubble:3 B}) 
and~(\ref{pressure inside bubble:3 C})
one obtains for the pressure difference,~\footnote{
To arrive 
at Eq.~(\ref{pressure difference ballistic}) we made use of 
the integral,
\begin{equation}
\int_0^M dxx^2\ln\left(1\!-\!e^{-x}\right)
 =-\frac{\pi^4}{45} + M^2{\rm Li}_2\left(e^{-M}\right)
     \!+\! 2M{\rm Li}_3\left(e^{-M}\right)
     \!+\!2{\rm Li}_4\left(e^{-M}\right)
\,.
\nonumber
\end{equation}
} 
\begin{eqnarray}
\Delta {\cal P}_z(\beta m,\gamma) \!
&=&\!\frac{\pi^2}{90\beta^4}\left[4(\gamma^2\!-\!1)\!
       +\!1\right]
-\,\frac{1}{4\pi^2\beta^4\gamma^4}
 \Bigg\{\!\!-\frac{J_B(4,\beta\gamma(1\!-\!v)m)}
                 {(1\!-\!v)^3}
\qquad\;
\label{pressure difference ballistic}
\\
&&\hskip -1.1cm
+\,\,\frac{\left[M_+^2{\rm Li}_2\left(e^{-M_+}\right)
         \!+\!2M_+{\rm Li}_3\left(e^{-M_+}\right)
         \!+\!2{\rm Li}_4\left(e^{-M_+}\right)\right]
                     _{M_+=\beta\gamma(1\!+\!v)m}}
         {(1\!+\!v)^3}
\nonumber\\
\!&&\hskip -1.8cm
 +\int_{0}^\infty\!\! dx x
   \left[\sqrt{x^2\!+\!(\beta\gamma m)^2}\!-\!x\right]
\ln\!\left[1\!-\!{\exp}\left(
    \!\!-\sqrt{x^2\!+\!\bigl(\beta\gamma m\bigr)^{\!2}}
    \!-\!vx\right)\right]\!
\Bigg\}
\,,\;
\nonumber
\end{eqnarray}
where ${\rm Li}_s(z)=\sum_{n=1}^\infty({z^n}/{n^s})$ 
denotes the polylogarithm function and we dropped 
the index $0$ on the mass. The first term 
in~(\ref{pressure difference ballistic}) is 
the pressure at the vanishing mass, 
 \begin{equation}
  {\cal P}_z(0,\gamma) = \frac{\pi^2}{90\beta^4}\left[4(\gamma^2\!-\!1)\!+\!1\right]
\,,
 \label{pressure inside bubble:4 M=0}
 \end{equation}
which is precisely of the form,
$(\gamma^2-1)T s_0 + {\cal P}_0$, 
with $s_0=(\rho_0+{\cal P}_0)/T$, 
 $\rho_0$ and ${\cal P}_0$ being the entropy density, energy density and pressure of 
 an ultrarelativistic plasma with one degree of freedom.
 The third line in~(\ref{pressure difference ballistic})
 comes from particles penetrating the interface from 
 inside the bubble. As noted above, this 
 contribution will be present only if thermalization inside 
 the bubble takes place.

\begin{figure}[ht]
\vskip-0.1cm
\begin{center}
\includegraphics[width=7.8cm]{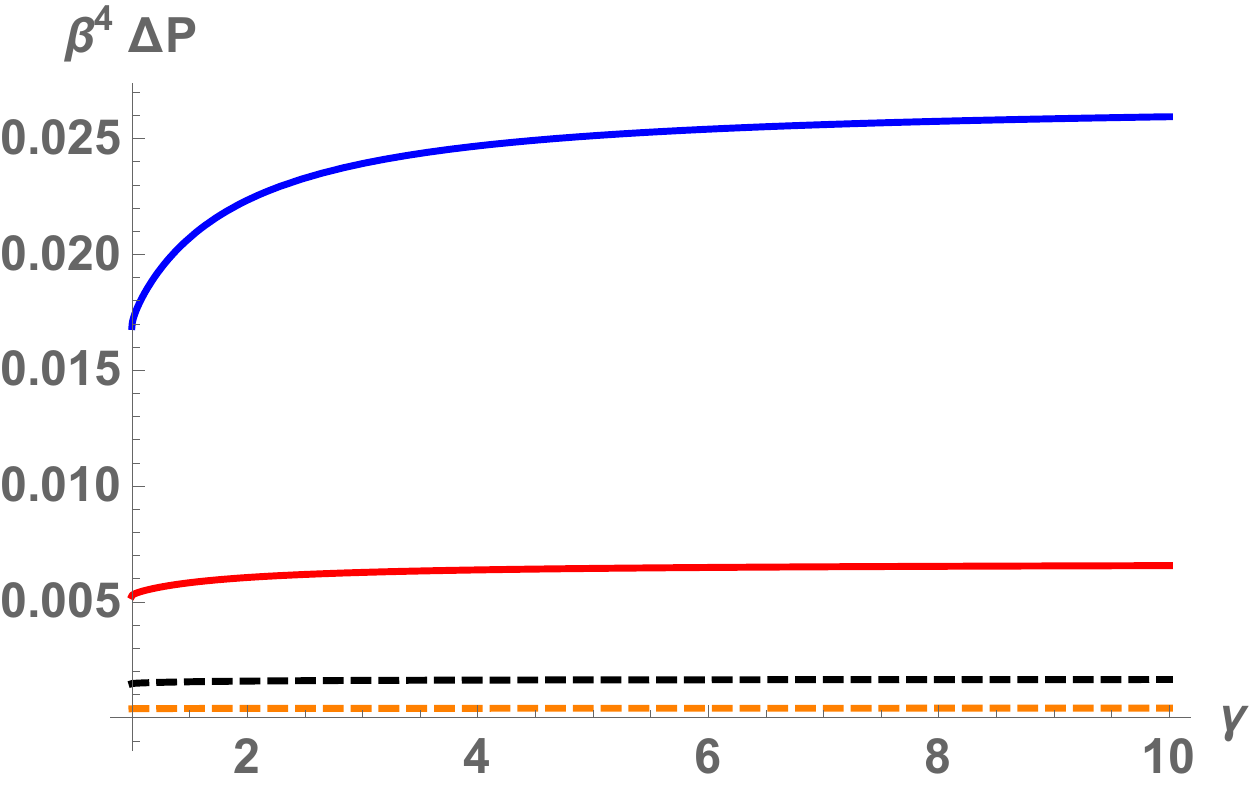}
\end{center}
\vskip-0.5cm
\caption{\small The force on the bubble $\Delta {\cal P}_z$ defined in~(\ref{pressure difference ballistic}) 
as a function of the Lorentz factor $\gamma$ 
for $m = 0.1T$ (orange dashed), $m=0.2T$ (black dashed), $m=0.4T$ (solid red), and $m=0.8T$ (solid blue).
As expected, the force is smaller for smaller masses. 
}
\label{delta P ballistic 1}
\end{figure}

Eq.~(\ref{pressure difference ballistic}) 
is to be compared with the pressure 
difference across the interface obtained assuming 
local thermal equilibrium (lte) close to the interface,
for which $f=1/[e^{\beta\gamma(E-v p_z)}-1]$ and hence, 
\begin{eqnarray}
 \Delta{\cal P}_{z,\rm lte}(\beta m,\gamma)
 &=&\!
 \frac{\pi^2}{90\beta^4}\left[4(\gamma^2\!-\!1)\!+\!1\right]
 \nonumber\\
&+&\!\frac{1}{4\pi^2\beta^4\gamma}
\int_{-\infty}^\infty dx x^2  
\ln\left(1\!-\!{\rm e}^{-\gamma(\sqrt{x^2+(\beta m)^2}-vx)}\right)
\,.\qquad
 \label{pressure inside bubble: local thermal equilibrium}
 \end{eqnarray}
 %
From Eqs.~(\ref{Liouville equation: particular solution A}--\ref{Liouville equation: particular solution C})
we see that, in the ballistic approximation, 
 all particles ascending onto the interface (Cases A and B)
slow down, such that  
$f>f_{\rm lte}$, implying that
 ${\cal P}_z>{\cal P}_{z,\rm lte}$, from which we infer 
$\Delta{\cal P}_z<\Delta {\cal P}_{z,\rm lte}$, meaning that the 
ballistic approximation yields a smaller bubble 
force 
and thus also faster bubbles.~\footnote{
This conclusion is valid 
provided the particles descending 
the wall (Case C), for which $f<f_{\rm lte}$ and  
hence $\Delta{\cal P}_z>\Delta {\cal P}_{z,\rm lte}$,
contribute sub-dominantly to the force. 
This must be the case because
in the wall frame the distribution function 
of the descending particles is always suppressed when 
compared with that of the ascending particles, implying that 
they also sub-dominantly contribute to the bubble force.} 
This should not come as a surprise, since
the ballistic approximation completely neglects particle interactions on the interface, it minimizes entropy production
across the bubble, thus {\it underestimating} the bubble force. 
To get an idea by how much, in figure~\ref{delta P ballistic 1} we plot 
 $\Delta {\cal P}_z$ defined in~(\ref{pressure difference ballistic}) as a function of the Lorentz factor
 $\gamma$ 
 for $m = 0.1T$ (orange dashed), $m=0.2T$ (black dashed), $m=0.4T$ (solid red), and $m=0.8T$ (solid blue).
As expected, the force is larger for larger masses. 
 Unlike in the case of the local thermal equilibrium force, the ballistic force {\it saturates} with $\gamma$. Interestingly, from the point of view of applicability range of our approach, the force saturates already for moderate values of $\gamma<10$, suggesting the possibility of runaway in the ballistic limit.
Motivated by that, we evaluate the  maximum force reached when $\gamma\rightarrow \infty$ 
from 
Eqs.~(\ref{pressure difference ballistic}--\ref{pressure inside bubble:4 M=0}) as follows.
In the limit when $\gamma\gg 1$,
\begin{equation}
\gamma(1-v)\approx
   \frac{1}{2\gamma}\left(1+\frac{1}{4\gamma^2}\right)
\,,\qquad 
\gamma(1+v)\approx
   2\gamma\left(1-\frac{1}{4\gamma^2}\right)
\,,
\label{large gamma estimate}
\end{equation}   
and the pressure difference 
$\Delta {\cal P}_z$ in Eq.~(\ref{pressure difference ballistic})
can be estimated as,
\begin{equation}
 \Delta {\cal P}_z 
                       = \frac{m^2}{24\beta^2} 
+ {\cal O}\left(\frac{1}{\gamma},
            \gamma^3 e^{-\gamma\beta m}\right)
\,.
\label{analytical estimate saturation force}
\end{equation}
In this limit our 
estimate~(\ref{analytical estimate saturation force})
agrees with the result of Ref.~\cite{Bodeker:2009qy}, Eq.~(2.4).
To get the result~(\ref{analytical estimate saturation force}) 
we have used the small argument expansion 
of the $J_B$ integral in the first line 
of Eq.~(\ref{pressure difference ballistic}). 
The terms in the second line can be 
neglected, since in that limit 
the polylogarithm functions are exponentially suppressed as 
$\sim {\rm e}^{-2\gamma \beta m}$,
which is easily seen from their small argument 
 expansion. 
Finally, the third line in~(\ref{pressure difference ballistic})
is exponentially suppressed at least as
$\sim {\rm e}^{-\gamma \beta m}$,~\footnote{
 To see this, let us rewrite the 
integral in the third line in~(\ref{pressure difference ballistic}) 
,
\begin{equation}
I_{t_-}= -\frac{1}{\gamma}\int_{0}^\infty\!\! dx' x'
   \left[\sqrt{x'^2\!+\!M^2}\!-\!x'\right]
\ln\!\left[1\!-\!{\exp}\left(
    \!\!-\gamma(\sqrt{x'^2\!+\!M^2}
    \!+\!vx')\right)\right]
\,,
\label{t- integral}
\end{equation}
where we used the variables $x'=\gamma x$, $M=\beta m$
and we included the prefactor $-1/\gamma^4$ in
the definition of the integral.
Next, it is convenient to introduce Lorentz boost variables,
$x=\gamma(x'+ve')$, $e=\gamma(e'+vx')$,
$e(x)=\sqrt{x^2+M^2}$, $e'=\sqrt{x'^2+M^2}$, 
upon which the integral naturally splits into two parts, 
$I_{t_-}=I_1+I_2$, where 
%
\begin{eqnarray}
I_1 &=& \gamma^2(1\!+\! v)^2\left[\left(
     \sqrt{1\!+\!(\gamma v)^2}M\!-\!M^2\right)
     {\rm Li}_2\left(e^{-\sqrt{1+(\gamma v)^2}M}\right)
     \!+\! {\rm Li}_3\left(e^{-\sqrt{1+(\gamma v)^2}M}\right)
     \right]
\nonumber\\
 &+&\gamma^2(1\!+\! v)
     {\rm Li}_2\left(e^{-\sqrt{1+(\gamma v)^2}M}\right)
     \sim 2\gamma^2
      \left(2\gamma M \!+\!3\!-\!2M^2\right)e^{-\gamma M}
\,,
\end{eqnarray}
where the last estimate holds in the limit 
$\gamma\rightarrow \infty$. 
The second integral cannot be evaluated exactly, 
but it can be 
bounded from above by the simpler integral
obtained by replacing 
$\sqrt{x^2+M^2}\rightarrow x$ 
in the exponent, which can be evaluated,
\begin{eqnarray}
-I_2&<&
2\gamma^2(1\!+\! v)^2
   {\rm Li}_4\left(e^{-\gamma v M}\right)
      +\gamma^2(1\!+\! v)vM^2
   {\rm Li}_2\left(e^{-\gamma v M}\right)
   \sim 2\gamma^2\left(4\!+\!M^2\right)e^{-\gamma M}
\,,
\nonumber
\end{eqnarray}
from where we conclude that 
$I_{t_-}$ is suppressed at least as 
$\sim e^{-\gamma \beta m}$.
} 
and hence
does not contribute at the leading order 
to~(\ref{analytical estimate saturation force}).

Notice that, if there are heavy particles in the plasma, 
$m\gg T$, the bubble force is $\propto m^2$ 
and thus gets saturated at 
much larger values and for large 
Lorentz factors, $\gamma\gtrsim m/T$.
 Namely,
even though the number of heavy particles 
is exponentially suppressed in 
the plasma frame, their energy in the wall frame gets 
boosted by the Lorentz factor $\gamma$, thereby 
reducing their suppression.
An important lesson to take from this analysis is that,
if a phase transition is strong and bubbles are relativistic, then 
the existence of very heavy particles (with a mass $m\gg T$)
can be of a crucial importance for the correct determination of 
the terminal bubble-wall velocity. 
In particular, heavy particles 
can determine whether the bubbles run away or not.


%

\begin{figure}[ht]
\vskip-0.1cm
\begin{center}
\includegraphics[width=6.75cm]{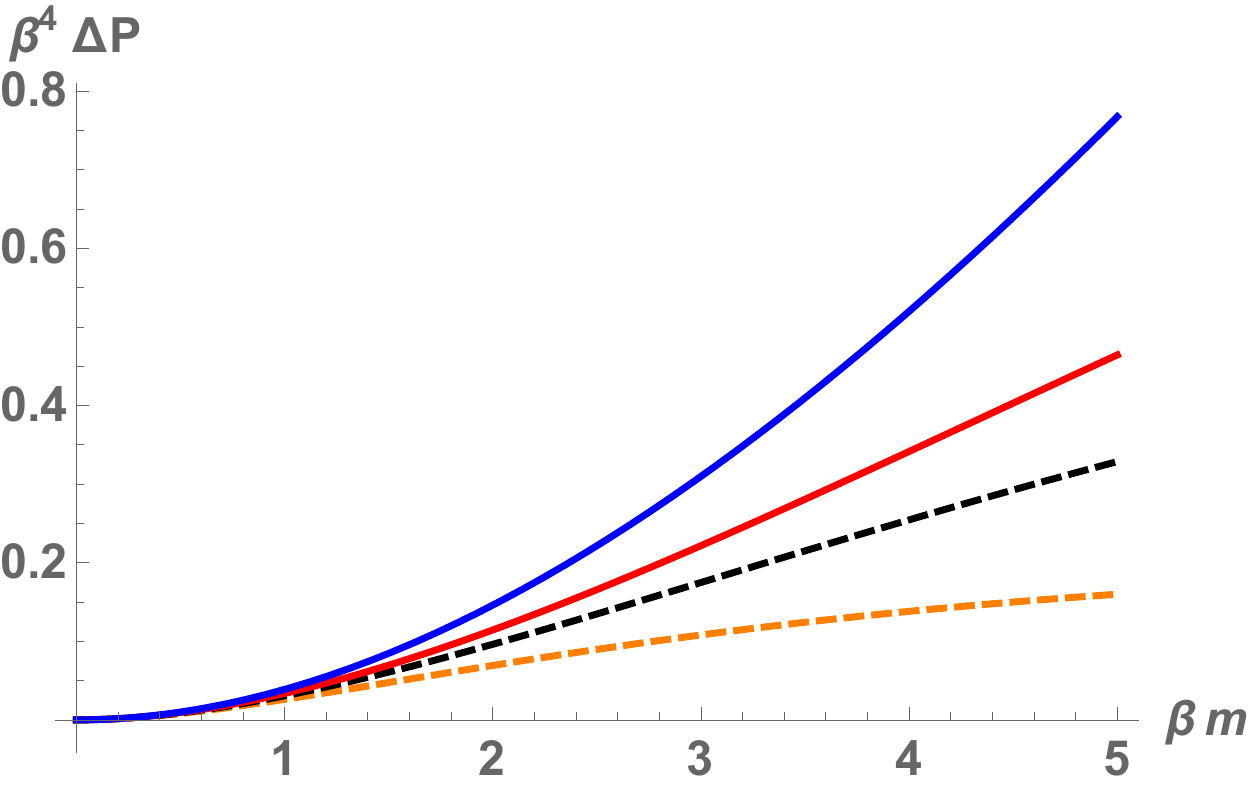}
\includegraphics[width=6.75cm]{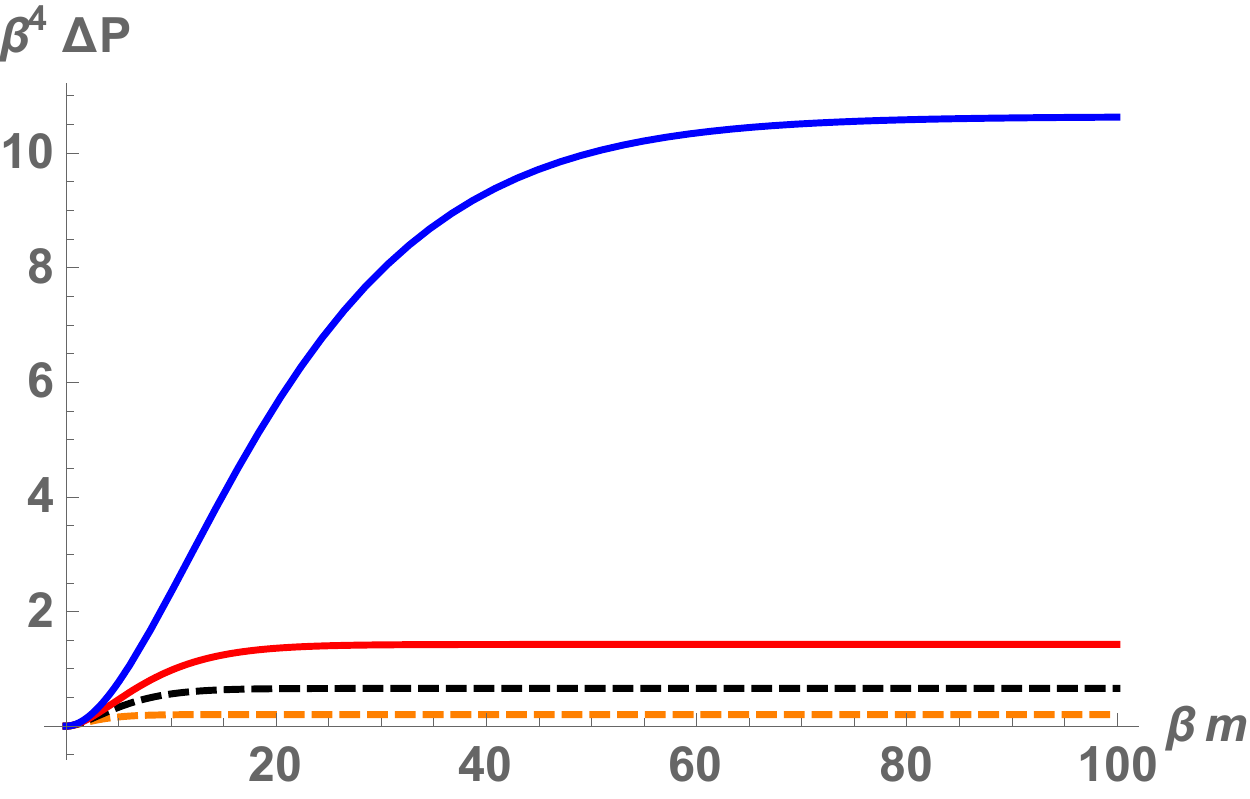}
\end{center}
\vskip-0.5cm
\caption{\small The force on the bubble $\Delta {\cal P}_z$ defined in~(\ref{pressure difference ballistic}) 
as a function of $m/T$ 
for $\gamma=1.1$ (orange dashed), $\gamma=1.5$ (black dashed), $\gamma=2$ (solid red), and $\gamma=5$ (solid blue).
}
\label{delta P ballistic 2}
\end{figure}

To illustrate how the bubble force depends on the particle mass, 
in figure~\ref{delta P ballistic 2}
we plot $\Delta {\cal P}_z$ in~(\ref{pressure difference ballistic}) as a function of $m/T$ for 
a fixed Lorentz factor $\gamma$. The values of $\gamma$ (starting from bottom up) are
$\gamma=1.1$ (orange dashed), $\gamma=1.5$ (black dashed), $\gamma=2$ (solid red), and $\gamma=5$
(solid blue), respectively. For small masses the force in 
the ballistic approximation increases rather slowly (left panel), but then it continues increasing 
and eventually saturates for very large masses. This is in contrast with the local thermal approximation result in 
figure~\ref{entropy production}, which shows that the lte force already saturates for quite modest masses. 

\begin{figure}[ht]
\vskip-0.1cm
\begin{center}
\includegraphics[width=6.75cm]{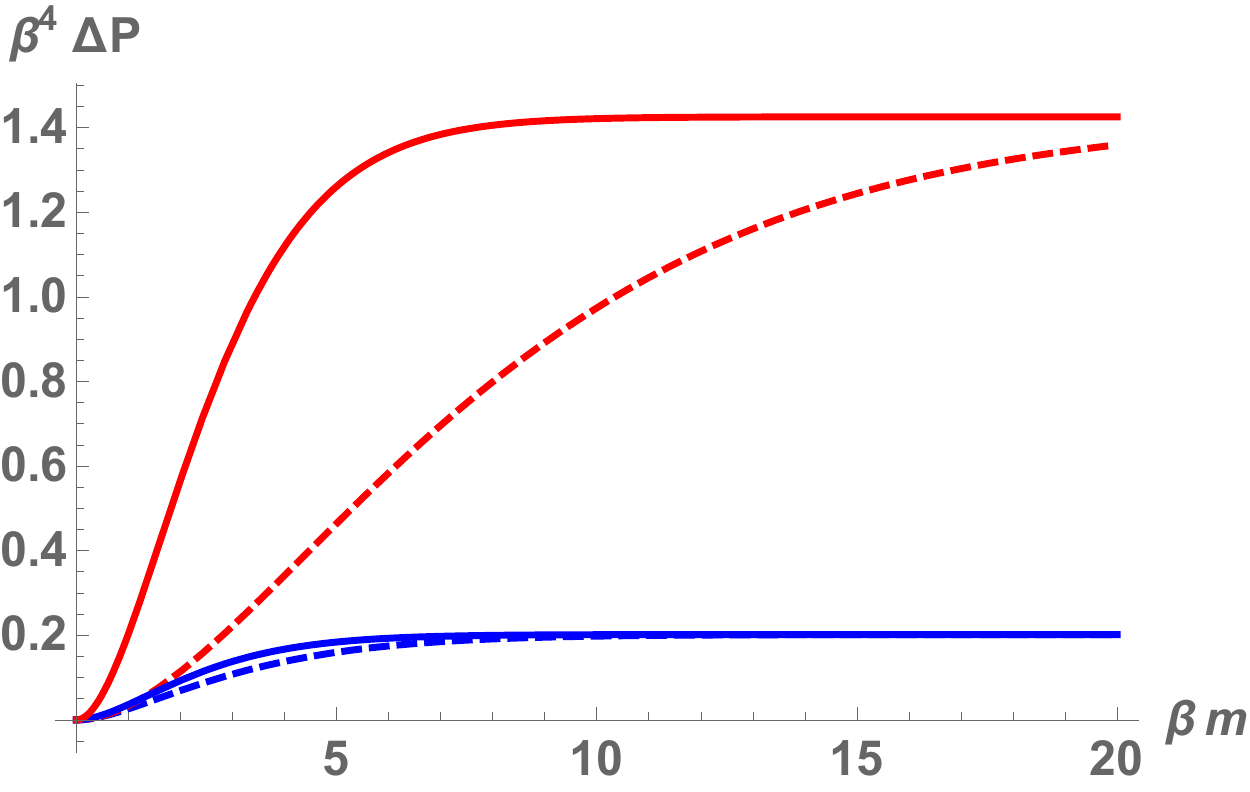}
\includegraphics[width=6.75cm]{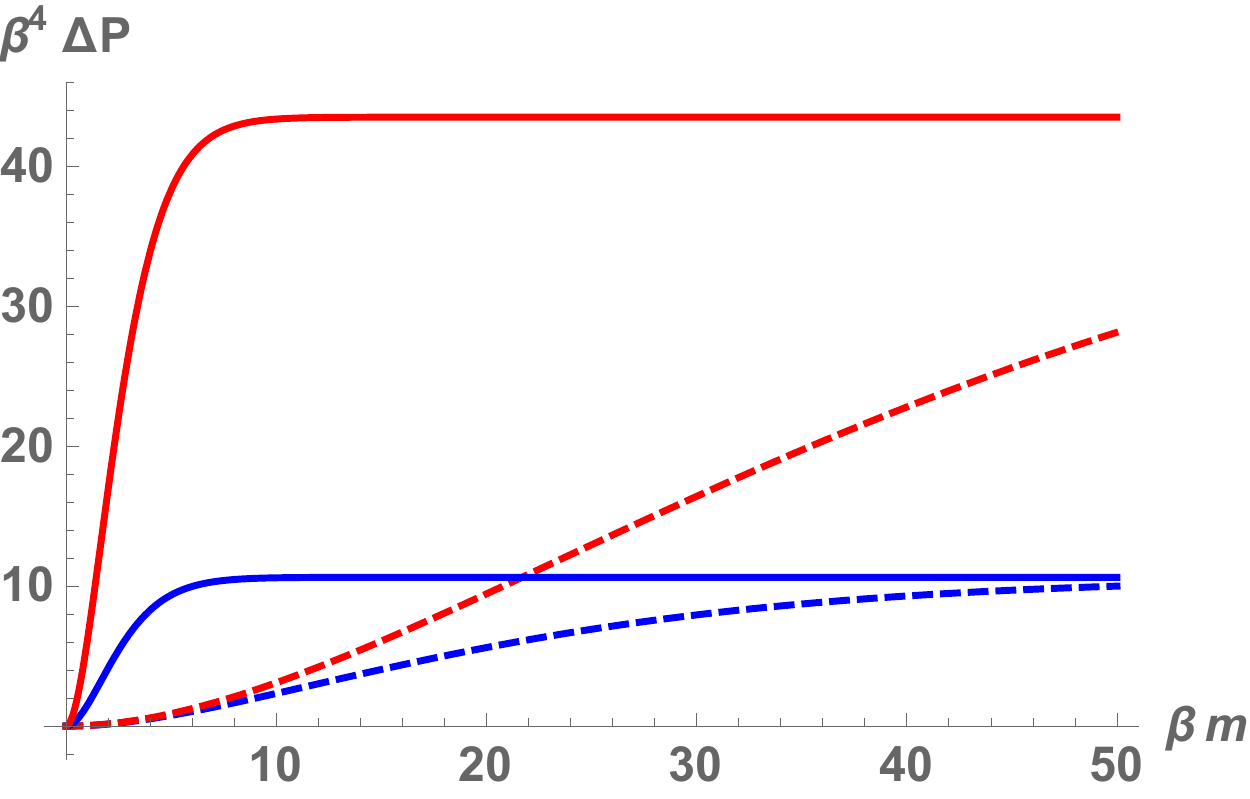}
\end{center}
\vskip-0.5cm
\caption{\small Comparison of the force on the bubble $\Delta {\cal P}_z$ given by the ballistic approximation~(\ref{pressure difference ballistic}) and the lte approach~\eqref{pressure inside bubble: local thermal equilibrium}
as a function of $m/T$. Left panel:  
$\gamma=1.1$ (lte: solid blue; ballistic: dashed blue), $\gamma=2$ (lte: solid red; ballistic: dashed red).
Right panel:  
$\gamma=5$ (lte: solid blue; ballistic: dashed blue), $\gamma=10$ (lte: solid red; ballistic: dashed red).
}
\label{delta P ballistic 3}
\end{figure}

\begin{figure}[ht]
\vskip-0.1cm
\begin{center}
\includegraphics[width=6.75cm]{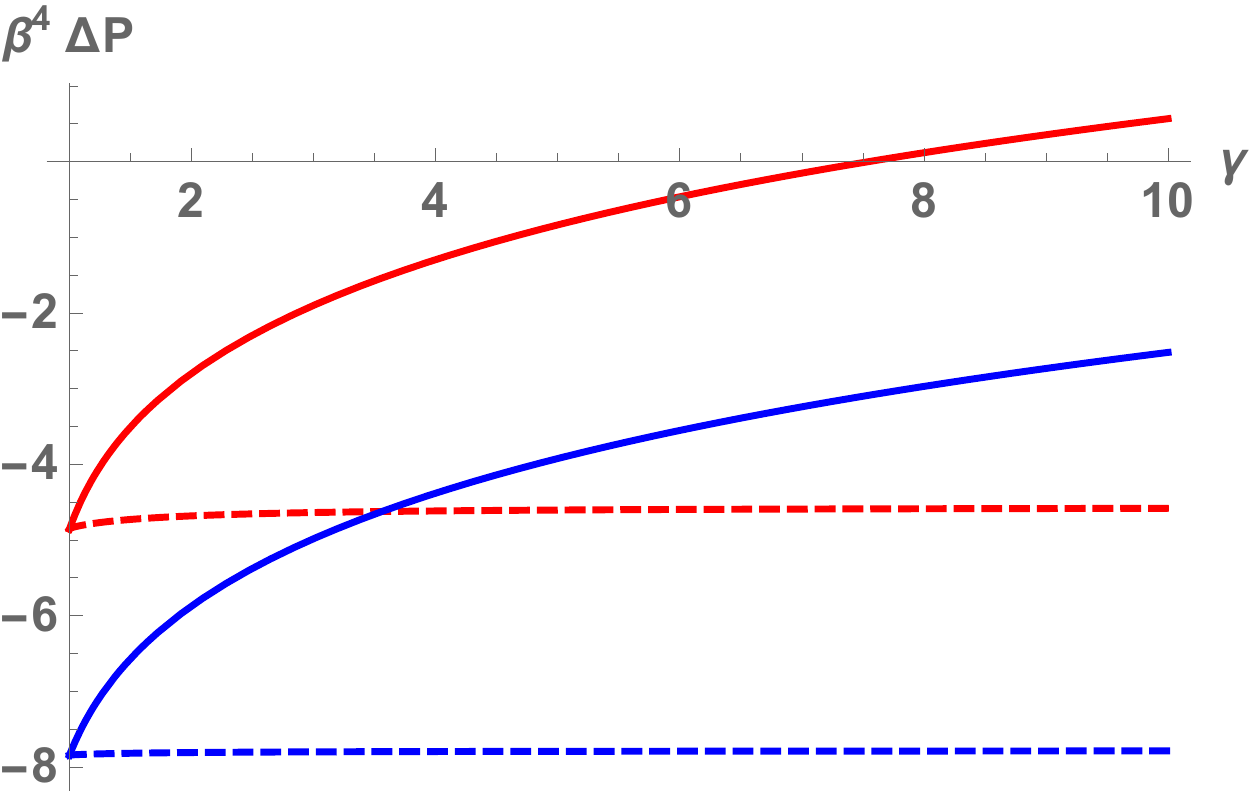}
\includegraphics[width=6.75cm]{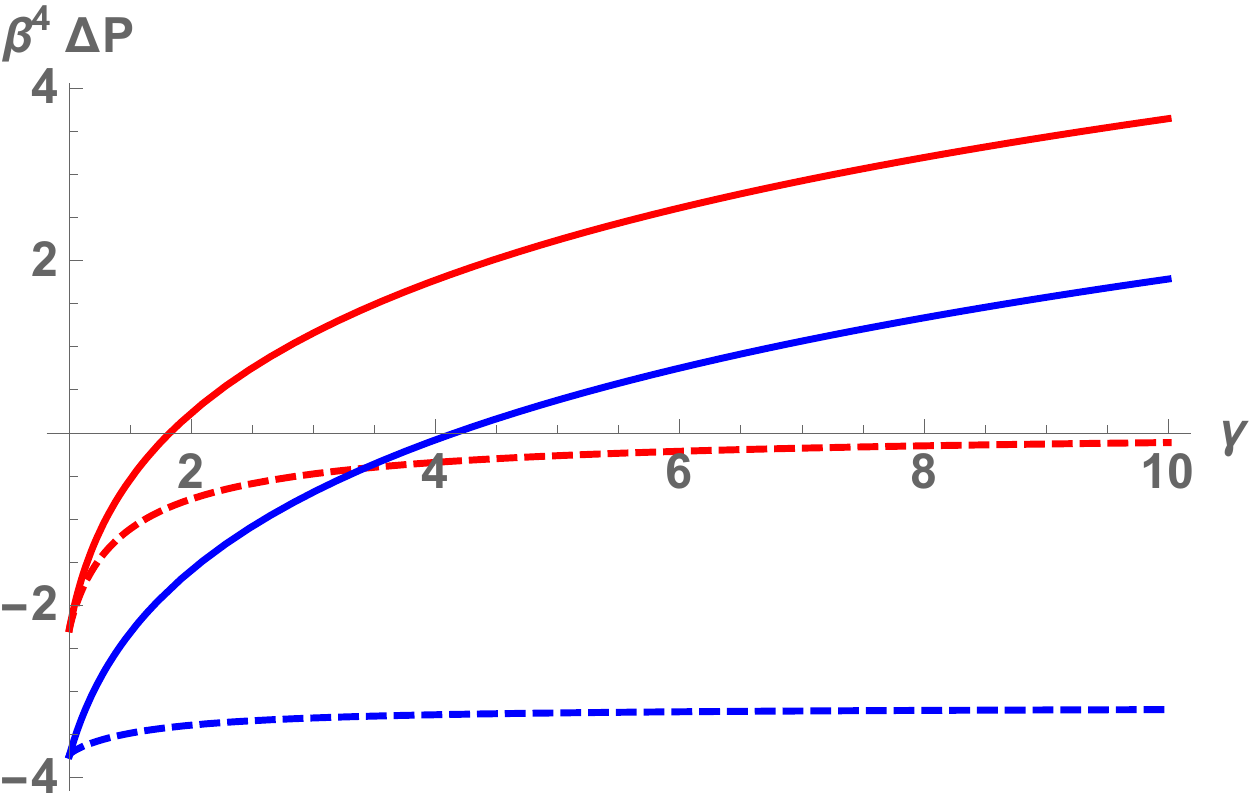}
\end{center}
\vskip-0.5cm
\caption{\small Natural logarithm of the force on the bubble $\Delta {\cal P}_z$ given by the ballistic approximation~(\ref{pressure difference ballistic}) and the lte approach~\eqref{pressure inside bubble: local thermal equilibrium} 
as a function of the Lorentz factor $\gamma$. Left panel:  
$m=0.1T$ (lte: solid blue; ballistic: dashed blue), $m=0.5T$ (lte: solid red; ballistic: dashed red).
Right panel:  
$m=1T$ (lte: solid blue; ballistic: dashed blue), $m=5T$ (lte: solid red; ballistic: dashed red).
}
\label{delta P ballistic 4}
\end{figure}

 In order to get a better insight into how much the bubble force calculated in the ballistic approximation
 is smaller than in the local thermal equilibrium (lte), 
in figure~\ref{delta P ballistic 3}
we plot $\Delta {\cal P}_z$~(\ref{pressure difference ballistic}) as a function of $m/T$ for 
a fixed Lorentz factor $\gamma$. In the left panel we show, from bottom up, $\gamma=1.1$ 
(lte is solid blue curve, ballistic is dashed blue) and $\gamma=2$ (lte is solid red curve, ballistic is dashed red),
while in the right panel the force for $\gamma=5$ 
(lte is solid blue curve, ballistic is dashed blue) and $\gamma=2$ (lte is solid red curve, ballistic is dashed red)
are shown. 
Notice that the force in the lte approximation is always larger, but for sufficiently large masses the two forces 
asymptote to the same value. This means that very massive particles ($m\gg T$) contribute more in the 
ballistic approximation than in the lte approximation. This observation can be important for the correct determination
of the phase transition dynamics, particularly in systems with very heavy degrees of freedom. 
Finally, in figure~\ref{delta P ballistic 4}
we plot the natural logarithm of 
$\Delta {\cal P}_z$~(\ref{pressure difference ballistic}) as a function of 
the Lorentz factor $\gamma$ for 
a fixed mass $m$. Left panel shows, from bottom up, $m=0.1T$ (lte: solid blue; ballistic: dashed blue), $m=0.5T$ (lte: solid red; ballistic: dashed red), while 
right panel:  
$m=1T$ (lte: solid blue; ballistic: dashed blue), $m=5T$ (lte: solid red; ballistic: dashed red).
Notice that the force calculated in the local thermal equilibrium approximation can be orders 
of magnitude larger than that calculated in the ballistic approximation. Notice further 
that, as $\gamma$ increases the ratio of the two forces increases, which can be explained by recalling
that, as $\gamma$ increases, the ballistic force saturates.

To conclude, the analysis in this section shows that the entropy production generated 
by a passing bubble (and thus also the bubble force) minimizes in the limit when particle 
interactions are negligible and can be treated in a ballistic approximation, in which case the bubbles' runaway is expected. 
In the opposite limit, when particles scatter efficiently
and can be consequently approximated by a local thermal equilibrium,
 the entropy production and the bubble force maximize. As long as the local thermal equilibrium approximation applies, 
entropy production grows as the Lorentz factor squared and quickly reach terminal velocity which can be estimated
from the one-loop energy-momentum tensor. We emphasize that formula~(\ref{conservation of Tmn:4}) 
derived in section~\ref{The force on bubbles from the renormalized stress energy tensor}
 applies in both of those limits as well as in more general situations,
 provided one takes proper care of the state of the field.
 Furthermore, both the ballistic 
and the local thermal equilibrium approximation can be understood 
as the leading order approximation in a more general 
treatment, in which they represent the leading 
(zeroth) order approximation to an expansion in powers 
of the perturbative parameter $\Gamma L/v$
for the ballistic approximation and
$v/(\Gamma L)$ for the local thermal equilibrium
approximation, where $L$ is the thickness of the bubble 
interface, $v$ its speed and $\Gamma$ the relevant thermalization rate.

\section{Comparison with existing results} 
\label{Comparison with existing results}

Since the problem of bubble dynamics at first order transitions is long-standing and there exists rich literature of the topic, it is useful to clarify where we differ when compared with the existing work.
Our principal statement is that the force on an expanding bubble can be extracted from 
the energy-momentum conservation law~(\ref{conservation of Tmn:2b}).
When written in the bubble-wall frame~(\ref{conservation of Tmn:3b}) it can be recast 
as~(\ref{conservation of Tmn:4}), 
from which we get a differential force on the bubble,  
\begin{eqnarray}
d{\cal P} = -\frac{dF}{V} =- (\gamma^2\!-\!1)Tds 
=- (\gamma^2\!-\!1)d(\rho_p\!+\!{\cal P}_p)
= -d\left[\langle\hat T_{zz}^p\rangle \!-\! {\cal P}_p\right]
,\quad
\label{comparison 1}
\end{eqnarray}
where $\rho_p$ and ${\cal P}_p$ are plasma frame quantities and $\langle\hat T_{zz}^p\rangle$
is calculated in the bubble frame. Note that ${\cal P}_p$ is just $\langle\hat T_{zz}^p\rangle$ in the 
plasma frame, such that Eq.~(\ref{comparison 1}) indicates that the force on the bubble
is present only when the bubble is moving.

The precise form of the energy-momentum tensor in~(\ref{comparison 1}) is complicated 
and it depends both on the state chosen and on the accuracy at which $\hat T^p_{zz}(x)$ is evaluated. For simplicity of the argument here we work in the one-loop approximation
and assume a state in which the bubble profile is an adiabatic function of $z$.
Then for a real scalar field in thermal equilibrium (outside the bubble) one can use the thermal part of 
the massive scalar propagator~(\ref{thermal scalar propagator}) 
to calculate the force in~(\ref{comparison 1}). When integrated across the bubble, 
Eq.~(\ref{comparison 1}) can then be recast as, 
\begin{equation}
\frac{F}{V}
= \int dz\frac{d\phi_0}{dz}\frac{dm^2}{d\phi_0}\int \frac{d^3p}{(2\pi)^3E}
 \left[\frac{1}{{\rm e}^{\beta\gamma (E-v p_z)}-1}-\frac{1}{{\rm e}^{\beta E}-1}\right]
,\quad
\label{comparison 2}
\end{equation}
where we made use of the bubble frame relations, 
\begin{equation}
d(p_z^2) = - d(m^2)\,,\qquad dE = 0
\label{bubble frame relations for pz and E}
\end{equation}
and of the differential chain rule, $d(p_z^2)=dz\partial_z(p_z^2) = -dz(d\phi_0/dz)(dm^2/d\phi_0)$.

If we compare Eq.~(3.3) from Ref.~\cite{Bodeker:2009qy} to the result in Eq.~\eqref{comparison 2} we see two unimportant differences. 
Firstly,  the force in Eq.~(\ref{comparison 2}) is two times larger,
and secondly, Eq.~(3.3) in Ref.~\cite{Bodeker:2009qy}
does not subtract the term needed to make the force vanish in the static limit.~\footnote{In Ref.~\cite{Bodeker:2009qy} it is mentioned that the force coming from the difference in vacuum potential on two sides of the wall should be taken into account. Then, a static limit is discussed and a conclusion is reached that a wall can be static only at the critical temperature.} Moreover, Eq.~(3.3) takes into account that the distribution function depends on the $z$ coordinate and is different at each point. Once a simplifying assumption of local
thermal equilibrium in front of the bubble is imposed as was done above, Eq.~(3.3) from Ref.~\cite{Bodeker:2009qy} and our result in Eq.~\eqref{comparison 2} do agree up to the minor differences already discussed.

The above analysis shows that, starting with our main result~(\ref{conservation of Tmn:4}),
and up to the well understood differences,
our general quantum expression for the bubble force reduces to the standard semi-classical 
expression in Refs.~\cite{Bodeker:2009qy,Moore:1995ua,Moore:1995si}. 
The question that then naturally arises is:  
why did we reach a different conclusion from~\cite{Bodeker:2009qy}  and from the result above in Eq.~\eqref{comparison 2}
when studying rapidly expanding bubbles in thermal equilibrium?

The answer lies in the use of the energy conservation across the wall as in Eq.~\eqref{bubble frame relations for pz and E}. By applying it, we impose a non-equilibrium distribution inside the bubble and thus we violate Lorentz symmetry which was crucial for obtaining the $(\gamma^2-1)$ scaling of the friction force in the lte case.

In tracing individual particles and applying energy-momentum conservation the approach of Ref.~\cite{Bodeker:2009qy} is closer to our classical treatment of the ballistic regime.  We derived the distribution inside the bubble, assuming thermal equilibrium outside and applying Liouville equation (see Eq.~\eqref{Liouville equation}).  If highly relativistic wall velocities are considered typical momenta in the wall frame are much larger than the masses involved. Then, the distribution inside and outside the bubble are approximately the same. This brings our 
analysis of the ballistic limit close to the analysis of 
Ref.~\cite{Bodeker:2009qy}
and, moreover, explains why in that limit the result -- 
that the force saturates for large $\gamma$ -- is similar. 
Nevertheless, it is important to emphasize that our treatment of the ballistic limit is not limited to the highly relativistic regime. Moreover, our general formula~(\ref{conservation of Tmn:4})   
allows one to calculate
the bubble force not just in local thermal equilibrium or in the ballistic 
approximation, but also in more general settings that go
 beyond the semiclassical treatment. The limited, one-loop computation of this work does not include the transition radiation which was treated in Refs.~\cite{Bodeker:2017cim}, however, higher-order computations should account also for that effect.

\section{Standard model and its extensions} 
\label{Standard model and its extensions}

In this section we utilize the results from 
sections~\ref{The force on bubbles from the renormalized stress energy tensor},
\ref{Real scalar field} and Appendix~A to study the phase transition dynamics
in the standard model and some of its popular extensions. 
To keep our analysis as simple as possible we do not analyze here the bubble nucleation
(from which one can infer the latent heat of the transition, 
bubble surface tension of the transition, {\it etc.}),  neither we address the non-equilibrium aspects 
of the transition. Instead, we compute the force on the expanding bubbles, from which 
one can then infer the dynamics of the transition if one knows 
the pressure difference across the bubble interface. Below we apply the local thermal equilibrium approximation. To verify whether it holds one should evaluate the condition given by Eq.~\eqref{local thermalization} using some estimates for thermalization rates, see e.g. Refs.~\cite{Moore:1995ua,Moore:1995si}.

\medskip

{\bf Standard model.} Even though it is known that the transition in the standard model 
is a {\it crossover}~\cite{Rummukainen:1998as}, it is useful to analyze it since its particle content is verified by 
experiments and the first microscopic analysis of the dynamics of the electroweak phase transition
was presented in~\cite{Moore:1995ua,Moore:1995si}, where the authors assumed that the transition 
is first order. It is well known that not all fields of the standard model are relevant for the dynamics of the electroweak transition. Namely, only those fields which significantly contribute to the thermal effective 
potential and which exert a large force on the bubble are important, and these are the fields whose 
mass is of the order of the temperature. For the standard model that selects:  
the top quark (12 relativistic degrees of freedom, $m_t=173~{\rm GeV}$), the Higgs boson (1 relativistic degree of freedom, $m_H=125~{\rm GeV}$), $W^{\pm}$ bosons (6 relativistic degrees of freedom, $m_Z=91.2~{\rm GeV}$) and the $Z$ boson (3 relativistic degrees of freedom, $m_Z=91.2~{\rm GeV}$).
All other particles are much lighter and do not significantly
contribute. For example, the next heaviest particle is the bottom quark, whose 
mass is about $5~{\rm GeV}$. Unless there is a large supercooling such that 
nucleation occurs at a very low temperature, 
comparable with or lower than $5~{\rm GeV}$, 
the bottom quark and 
other particles of the standard model are not important
for the phase transition dynamics. 

According to Eq.~(\ref{conservation of Tmn:4}) 
it is the change in the entropy density across the bubble, $T\Delta s = \rho_p+{\cal P}_p$ 
(see~(\ref{change in entropy density: scalar}))
that determines the force per unit area on the bubble. With this in mind and from 
Eq.~(\ref{renormalized energy-momentum tensor: fermion}) 
we can extract the top contribution,
\begin{equation}
 T\Delta s_t =\frac{7\pi^2}{15\beta^4}
                    + \frac{8}{\pi^2\beta^5m_t}\left[\partial_z J_F(6,z)\right]_{z=\beta m_t}
                    +\frac{6m_t}{\pi^2\beta^3}\left[\partial_z J_F(4,z)\right]_{z=\beta m_t}
\,.\quad
\label{change in entropy density: top}
\end{equation}
From the analysis of the Abelian Higgs model in its condensate phase in Appendix~A we infer that 
the Higgs boson after symmetry breaking contributes as one massive scalar field 
(see Eqs.~(\ref{energy momentum Abelian Higgs: total}) 
and~(\ref{renormalized energy-momentum tensor: scalar}--\ref{change in entropy density: scalar})),
\begin{equation}
T\Delta s_H = \frac{2\pi^2}{45\beta^4} 
    \!-\! \frac{2}{3\pi^2\beta^5m_H}\left[\partial_z J_B(6,z)\right]_{z=\beta m_H}
         \!-\!\frac{m_H}{2\pi^2\beta^3}\left[\partial_z J_B(4,z)\right]_{z=\beta m_H}
\,.\quad
\label{change in entropy density: Higgs}
\end{equation}
Based on the same Abelian Higgs model analysis (see Eq.~(\ref{energy momentum Abelian Higgs: total})) 
we can conclude that $W^\pm$ and $Z$ bosons contribute as two and one massive gauge bosons, 
respectively,  
\begin{eqnarray}
T\Delta s_W \!\!&=&\!\! \frac{4\pi^2}{15\beta^4} 
    \!-\! \frac{4}{\pi^2\beta^5m_W}\left[\partial_z J_B(6,z)\right]_{z=\beta m_W}
         \!-\!\frac{3m_W}{\pi^2\beta^3}\left[\partial_z J_B(4,z)\right]_{z=\beta m_W}
\!,\quad\;\;\;
\label{change in entropy density: W}\\
T\Delta s_Z \!\!&=&\!\! \frac{2\pi^2}{15\beta^4} 
    \!-\! \frac{2}{\pi^2\beta^5m_Z}\left[\partial_z J_B(6,z)\right]_{z=\beta m_Z}
         \!-\!\frac{3m_Z}{2\pi^2\beta^3}\left[\partial_z J_B(4,z)\right]_{z=\beta m_Z}
\,.\quad\;\;
\label{change in entropy density: Z}
\end{eqnarray}
Note that in Eqs.~(\ref{change in entropy density: Higgs}--\ref{change in entropy density: Z}) we have assumed 
that a local thermal equilibrium is attained.  The total change in the entropy density in a model with the field content of the standard model 
is then obtained by simply summing the four contributions in 
Eqs.~(\ref{change in entropy density: top}--\ref{change in entropy density: Z}),
%
\begin{equation}
 T\Delta s_{\rm SM} = T\big(\Delta s_t + \Delta s_H + \Delta s_W +\Delta s_Z\big)
\,.
 \label{entropy production SM}
 \end{equation}

\begin{figure}[ht]
\vskip-0.1cm
\begin{center}
\includegraphics[width=7.8cm]{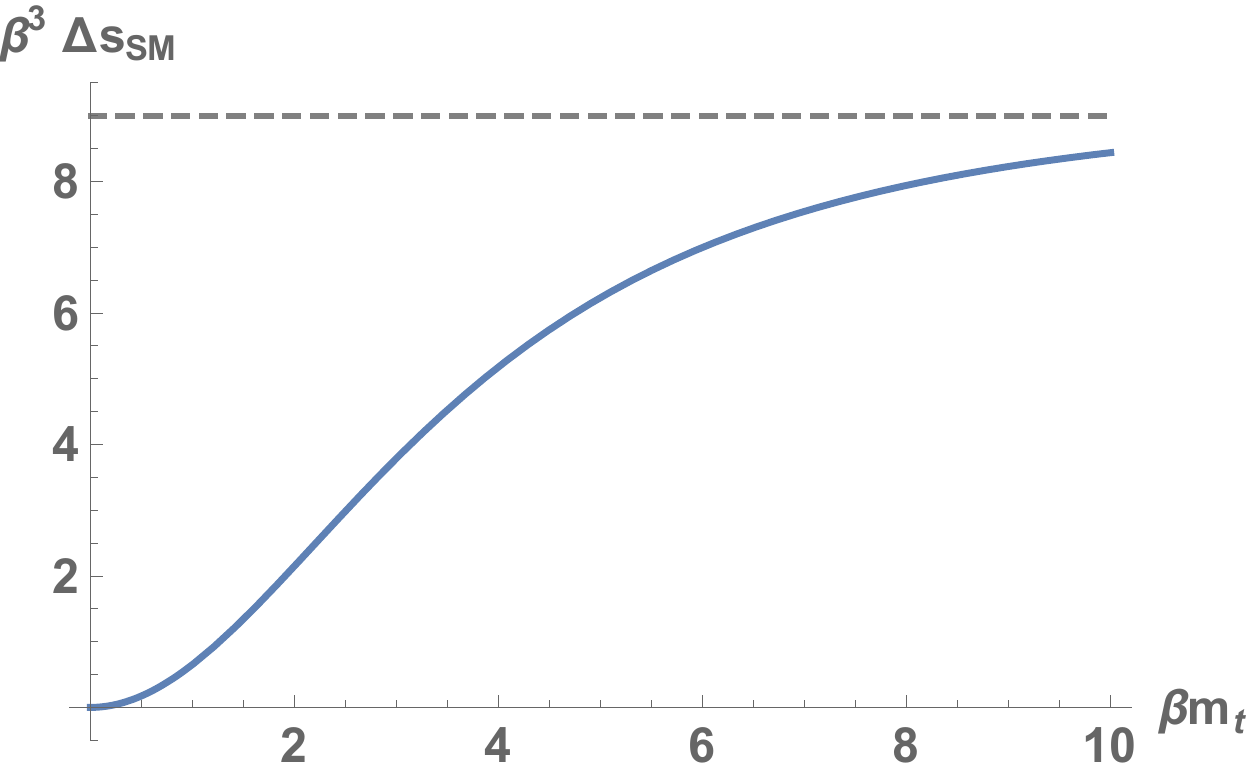}
\end{center}
\vskip-0.5cm
\caption{\small The change in the entropy density 
$\Delta s_{\rm SM}/T^3$ (solid blue) 
of the standard model plasma across a nucleated bubble as a function of
the ratio of the top mass (in the broken phase)  and the temperature, $\beta m_t=m_t(\phi_0)/T$.
The maximum amount by which the entropy density can change is $41\pi^2T^3/45$,
which is reached when $\beta m_t\rightarrow \infty$ (horizontal dashed).}
\label{figure entropy production SM}
\vskip-0.3cm
\end{figure}

To get an impression of how strong  the bubble force is 
in a theory with  the matter content of the standard model, 
in figure~\ref{figure entropy production SM} we plot the increase in the entropy density 
across the bubble wall 
as a function of $\beta m_t$ (when plotting figure~\ref{figure entropy production SM}
we made use of $\beta m_i=(m_i/m_t)\beta m_t$ with $i=H,W,Z$). When compared with 
the single scalar field in figure~\ref{entropy production}, we see that 
the entropy increase -- and thus also the force on the bubble -- is, as expected, much larger in the standard model than in the real scalar case. This is so because there are many more heavy degrees of freedom in the standard model. To be precise, {\it twenty two} in the standard model {\it vs}
{\it one} in the real scalar field.

\begin{figure}[h!]
\vskip-0.1cm
\begin{center}
\includegraphics[width=7.8cm]{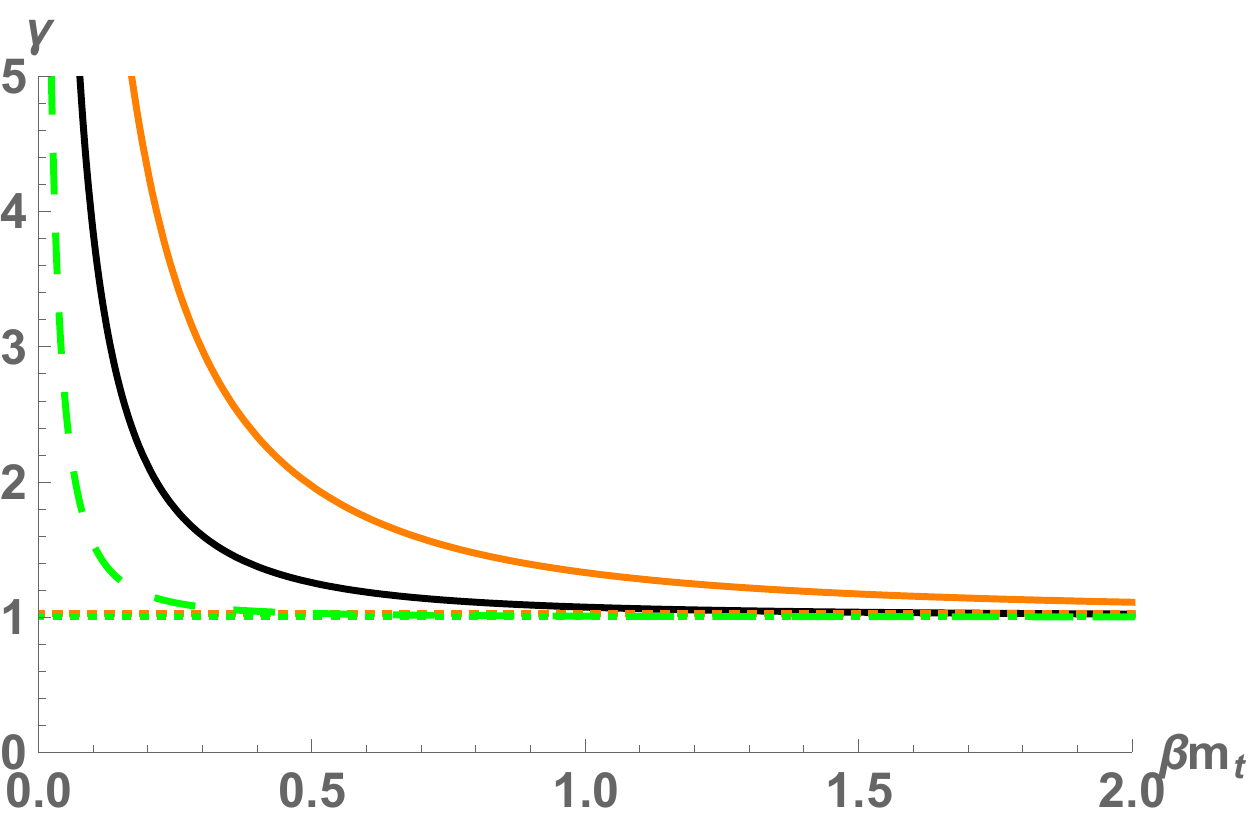}
\end{center}
\vskip-0.5cm
\caption{\small The bubble Lorentz factor $\gamma(v)$ for the standard model 
as a function of the top mass $m_t/T$ for $\Delta {\cal P}=-0.01/\beta^4$ 
(green dashed), $\Delta {\cal P}=-0.1/\beta^4$ (solid black) and 
$\Delta {\cal P}=-0.5/\beta^4$ (solid orange).}
\label{bubble speed gamma SM}
\vskip-0.3cm
\end{figure}

\begin{figure}[h!]
\vskip-0.cm
\begin{center}
\includegraphics[width=7.8cm]{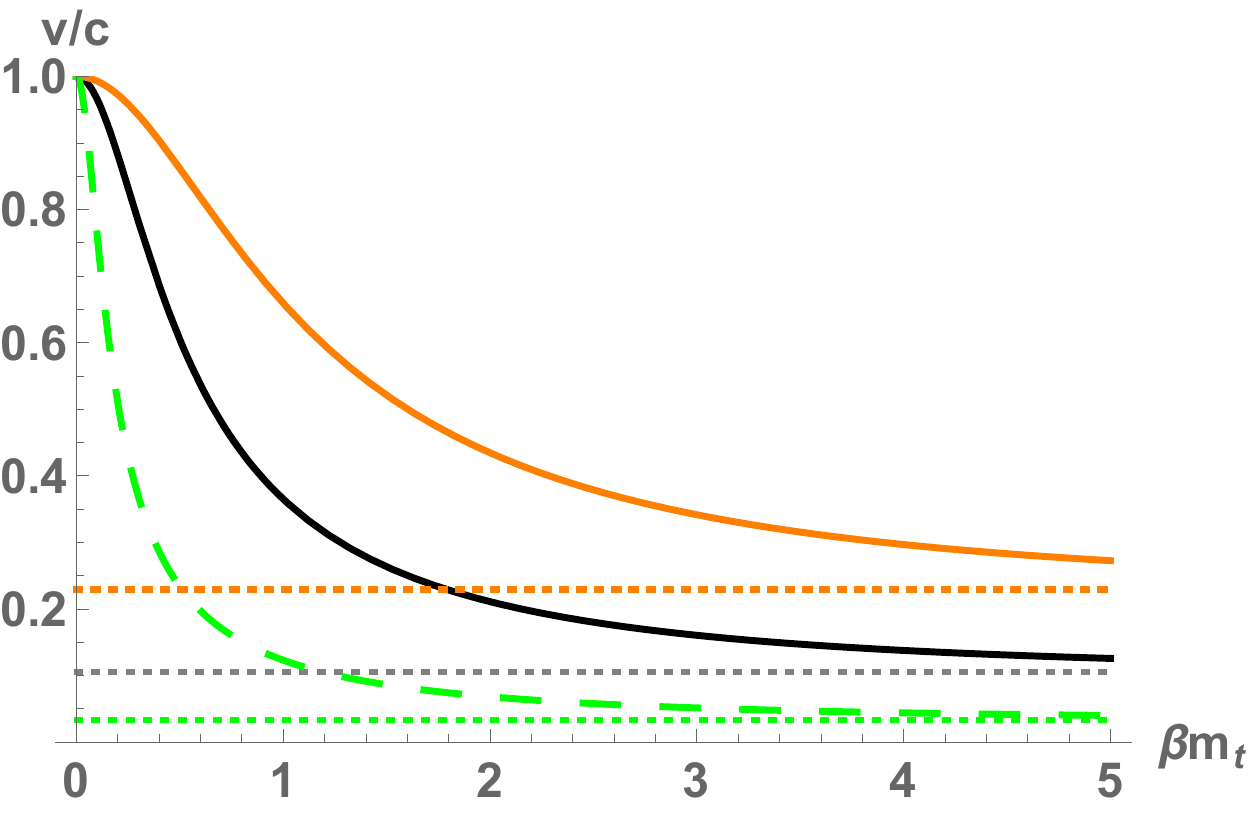}
\end{center}
\vskip-0.5cm
\caption{\small The bubble speed $v/c$ 
as a function of the top mass $m_t/T$ for the same choice of parameters
as in figure~\ref{bubble speed gamma SM}:
$\Delta {\cal P}=-0.01/\beta^4$ 
(green dashed), $\Delta {\cal P}=-0.1/\beta^4$ (solid black) and 
$\Delta {\cal P}=-0.5/\beta^4$ (solid orange).}
\label{bubble speed v SM}
\vskip-0.5cm
\end{figure}

In figures~\ref{bubble speed gamma SM} and~\ref{bubble speed v SM}
we show the bubble Lorentz factor and the corresponding expansion speed
for the standard model as a function of the top condensate in units of the temperature. 
When compared with the real scalar field in figures~\ref{bubble speed gamma}
and~\ref{bubble speed v},  we see that the bubbles become non-relativistic already 
for reasonably strong transitions for which $\beta m_t \simeq 1$. 
Given that the top Yukawa $y_t\simeq 1$,
this is equivalent to $\phi_0/T \simeq 1$. Recalling the Shaposhnikov's baryon washout criterion 
for the strength of the transition, $\phi_0/T\geq 1$, we infer that the bubbles 
at a strongly first order phase transition with the standard model matter content 
are typically subsonic, which broadly speaking agrees with the results of 
the more detailed microscopic analyses of Moore and Prokopec presented in 
Refs.~\cite{Moore:1995ua,Moore:1995si}.


\bigskip

{\bf Standard model with a singlet.} There are several types of extensions of the standard model 
in which the standard model is extended by a scalar field 
which is a singlet under the standard model gauge group. Examples include portal models~\cite{Hambye:2008bq} 
(an important class of which are conformal portal models, 
see {\it e.g.} Refs.~\cite{Hambye:2013dgv,Chataignier:2018aud,Chataignier:2018kay,Prokopec:2018tnq}
 and references therein)
and supersymmetric models, which include scalar singlet fields,
a notable example being the NMSSM. 

Quite generically both classes of models 
lead to strongly first order transitions. 
Even though in portal extensions the transition often proceeds in two stages
-- the nucleation along the scalar singlet direction is followed by a rolling
along the Higgs direction~\cite{Prokopec:2018tnq} -- studying how the additional field content affects the transition dynamics 
can still be useful. In this case, we ought to add to (\ref{change in entropy density: top}--\ref{change in entropy density: Z}) the contribution of the singlet,~\footnote{If the phase transition occurs along the scalar direction and then followed by rolling in the Higgs direction, then only the particles acquiring mass in the first stage are relevant for the pressure calculation.}
\begin{equation}
T\Delta s_s = N_s\left[\frac{2\pi^2}{45\beta^4} 
    \!-\! \frac{2}{3\pi^2\beta^5m_s}\left[\partial_z J_B(6,z)\right]_{z=\beta m_s}
         \!-\!\frac{m_s}{2\pi^2\beta^3}\left[\partial_z J_B(4,z)\right]_{z=\beta m_s}\right]
,\quad
\label{change in entropy density: singlet}
\end{equation}
where $m_s$ and  $N_s$ denote the singlet mass and its number of relativistic degrees of freedom, 
respectively (for example, $N_s=2$ if the singlet is a complex scalar). 

As a general remark, adding more massive degrees of freedom generally increases
the entropy production across the expanding bubbles, which in turn increases the force on 
the bubbles, thus slowing them down. 
 On the other hand, if the character of the phase transition is changed -- as it is for example in the aforementioned conformal models -- strong super-cooling can be present, resulting in a large latent heat release, which in turn accelerates the walls.
 It is also worth noting that, if there is more than 
 one scalar field in the theory, then multistage transitions are possible, see {\it e.g.}
 Ref.~\cite{Prokopec:2018tnq}.
 Even though the dynamics of such multistage transitions can be analyzed with 
Eq.~(\ref{conservation of Tmn:4}), such transitions are beyond the scope of this paper.

\section{Summary and discussion} 
\label{Summary and discussion}

 We derive a general {\it quantum}  field theoretic
 formula~(\ref{conservation of Tmn:4}) for the  terminal velocity of
 {\it expanding bubbles} of a first order phase transition. The formula has a simple physical interpretation.  If local thermal equilibrium is attained across the
 bubble interface, 
the friction force  acting on a bubble is proportional to 
the {\it entropy increase} across the bubble,
 with the proportionality constant being the Lorentz factor squared, $\gamma^2-1 =v^2/(1-v^2)$. Our formula is applicable to {\it quantum field theories} both {\it in} and {\it out-of equilibrium} and 
 can be applied at the one or higher-loop level. In this paper we show 
 how to apply~(\ref{conservation of Tmn:4})  at the 
 one-loop level in the toy model with one scalar field which exhibits spontaneous symmetry breaking
 (in section~\ref{Real scalar field}),
 as well as to the standard model and its simple extensions 
 (in section~\ref{Standard model and its extensions}). 
Our analysis applies to bubbles with a moderate Lorentz factor, $\gamma\lesssim 10$, and when scatterings are efficient,  
such that the local thermal equilibrium 
approximation 
applies and the relevant two-point functions are approximately thermal.
 A particular attention is devoted to how to obtain gauge independent results,
 the details of which are given in Appendix~A.
Our formula~(\ref{conservation of Tmn:4}) generalizes the previous known semi-classical formula of
Refs.~\cite{Moore:1995ua,Moore:1995si}, 
and reduces to it in the adiabatic one-loop approximation. However, our analysis reveals an important dependence of the force on the relativistic $\gamma$ factor of the wall which was not known before.

We apply our formula~(\ref{conservation of Tmn:4})
to two simple cases.  First we assume local thermal equilibrium.
Our results show that the force grows as the Lorentz factor-squared, $\propto(\gamma^2-1)$, and thus the bubbles quickly reach their terminal velocity. This allows to expect, even though our computation formally does not apply to $\gamma>10$, that runaway is not possible in this scenario. The other limit that we analyzed -- the ballistic approximation in which the particles do not thermalize efficiently -- displays behavior closer to the runaway scenario featuring friction force that saturates already for moderate $\gamma$ values.  These two limiting cases -- the local thermal equilibrium which overestimates the force and the ballistic case which underestimates it -- can be used to  estimate the true force acting on a bubble, which needs to be between the two extrema. Moreover, the force in more realistic scenarios can be estimated as perturbations around one of our two solutions.

While the analysis presented in this work is at places simplistic, 
it can capture the leading order contribution to the bubble force in a broad range of situations.
Nevertheless, there are situations in which our analysis is not accurate enough. 
For example, when a transition is not very 
strong, the force can be significantly altered by higher-loop effects which can induce 
a moderate, or even large, departure from equilibrium,
whose effects can be transported away from the bubble interfaces. 
This type of corrections, as well as their gauge 
dependence, should be carefully investigated and one should reassess how
the analysis presented here is affected when these effects are included.
Furthermore, the latent heat released by the transition can induce significant 
effects on the plasma, which can propagate and dissipate in the form of sound waves and turbulence,
both of which can heat up and induce a large scale motion of the plasma, thus affecting 
the force on the bubbles, which should also be taken into account. 
These are just some of the unaccounted-for effects, which can influence the dynamics of the transition 
and which are 
captured by our formula~(\ref{conservation of Tmn:4}). In order words, this paper provides 
an important step towards an accurate modeling of the dynamics of the electroweak phase transition, which can be of paramount importance for a quantitative understanding
of the gravitational wave production and 
baryogenesis at the electroweak scale.

\bigskip

\noindent{\bf Acknowledgments.}
We would like to thank Guy Moore for insightful comments which allowed to significantly improve this article, to Dietrich B{\"o}deker for critically reading the manuscript and to Yann Gouttenoire for a fruitful correspondence. This work was in part supported by the D-ITP consortium, a program of 
the Netherlands Organization for Scientific Research (NWO) that is 
funded by the Dutch Ministry of Education, Culture and Science (OCW).  B{\'S} acknowledges the support from the National Science Centre, Poland, through the SONATA project number 2018/31/D/ST2/03302 and through the HARMONIA project under contract
 UMO-2015/18/M/ST2/00518 (2016-2019).

\section*{Appendix: The energy-momentum tensor}

In this appendix we present some details of the calculations 
of the one-loop energy-momentum tensor 
for a massive fermionic field, massless and massive gauge fields and 
the Abelian Higgs model all
in thermal state in Minkowski space. The real scalar field is considered in the main 
text in section~\ref{Real scalar field}.
Most of the material covered in this Appendix can 
be found in the 
literature~\cite{Dolan:1973qd,Quiros:1994dr,Kapusta:2006pm,Bellac:2011kqa}, however not in a single source.

\bigskip

{\bf Dirac Fermion.} The action for a Dirac fermionic field $\Psi(x)$ 
that suffices for the one-loop 
calculation is of the form, 
\begin{equation}
S[\Psi] = \int d^Dx \sqrt{-g}{\cal L}_\Psi
\,,\quad 
  \mathcal{L}_\Psi = \frac{i}{2} \bar{\Psi} \gamma^\mu \nabla_\mu \Psi 
      - \frac{i}{2} \bar{\Psi} \overleftarrow{\nabla}_\mu \gamma^\mu \Psi - m \bar{\Psi} \Psi 
   \label{L_fermions}
\end{equation}
where $\bar{\Psi}=\Psi^\dagger \gamma^0$,
 $m=y\phi_0(x)$ is the field dependent mass, $y$ is a Yukawa coupling
  and $\nabla_\mu$ is the covariant 
derivative acting on a spinor field~\cite{Birrell:1982ix,Laenen:2020},
\begin{equation}
\nabla_\mu = \partial_\mu+\Gamma_\mu
\,,\qquad \Gamma_\mu 
 = \frac{1}{8}\left[\gamma^a,\gamma^b\right]e_a^\mu \nabla_\mu e_{\nu b}
\,.
\label{covariant derivative: fermion}
\end{equation}
Here $\Gamma_\mu$ is the spin(or) connection, 
 $e_a^\mu(x)$ is the tetrad field, 
which lifts tensors into the tangent space.
For example, $g_{\mu \nu}(x) e^\mu_a(x) e^\nu_b(x) = \eta_{ab}$
and $\sqrt{|g|}= |{\rm det}[e_{\nu b}]|\equiv |e|$, where 
$\eta_{ab}$ is the (flat) tangent space metric at a spacetime point $x^\mu$.
From~(\ref{L_fermions}) one easily gets the equation of motion 
for the fermionic Feynman propagator,
\begin{equation}
\sqrt{-g} \left(i\gamma^\mu\nabla_\mu-m\right)_{\alpha\gamma}iS^{\gamma\beta}(x;x')
   = i\delta_{\alpha}^{\beta}\delta^D(x\!-\!x')
\,,
\label{fermion propagator equation}
\end{equation}
where $\alpha,\beta,\gamma$ are spinor indices, and 
\begin{eqnarray}
iS^{\alpha\beta}(x;x')
 &=&{\rm Tr}\left[\hat \rho_{\rm in}
              T\left(\hat\Psi_\alpha(x)\hat{\bar\Psi}_\beta(x')\right)\right]
= - {\rm Tr}\left[\hat \rho_{\rm in}
              T\left(\hat{\bar\Psi}_\beta(x')\hat\Psi_\alpha(x)\right)\right]
\label{fermion propagator: def}\\
&\equiv&\!\!
\Theta(t\!-\!t'){\rm Tr}\left[\hat \rho_{\rm in}\hat\Psi_\alpha(x)\hat{\bar\Psi}_\beta(x')\right]
\!-\!\Theta(t'\!-\!t){\rm Tr}\left[\hat \rho_{\rm in}\hat{\bar\Psi}_\beta(x')\hat\Psi_\alpha(x)\right]
,
\nonumber
\end{eqnarray}
and $\hat \rho_{\rm in}$ is the fermionic density operator (in Heisenberg picture).

The Dirac matrices $\gamma^\mu(x)$ build a Clifford algebra, 
and obey the standard relations on spacetime 
\begin{equation}
 \left\{\gamma^\mu(x),\gamma^\nu(x)\right\}=-2g^{\mu\nu}(x)
\,,
\label{Dirac matrices: space time anticommutation}
\end{equation}
and on the tangent space, 
\begin{equation}
 \left\{\gamma^a,\gamma^b\right\}=-2\eta^{ab}
\,,
\label{Dirac matrices: tangent space anticommutation}
\end{equation}
respectively, such that $\gamma^\mu(x) = e^\mu_a(x) \gamma^a$,
where the Latin letters $a,b,c$ denote the flat tangent space indices.
The right hand side of~(\ref{fermion propagator equation}) follows  
from the canonical quantization relation,
\begin{equation}
 \left\{\hat\Psi_\alpha(t,\vec x),\hat\Psi^\dagger_\beta(t,\vec x^{\,\prime})\right\}
    = \delta_{\alpha\beta}\frac{\delta^{D-1}(\vec x-\vec x^{\,\prime})}{a^{D-1}}
\,,
\label{canonical quantization relation: fermions}
\end{equation}
which follows from the form of the canonical momentum of $\Psi$, $\Pi_\Psi=-ia^{D-1}\Psi^*$.
The factor $a^{D-1}$ in~(\ref{canonical quantization relation: fermions}) 
originates from specifying to a homogeneous cosmological
spacetime, in which the metric is diagonal with $\sqrt{-g}=a^D$ 
and $1/a$ comes from the tetrad 
$e^\mu_b=a^{-1}\delta^\mu_{\;b}$ that projects $\gamma^\mu(x)$ 
onto the tangent space according to, 
$\gamma^\mu=e^\mu_{\;a}\gamma^a$.

\begin{figure}[h!]
\vskip -0.3cm
\centering
\includegraphics[scale=0.65]{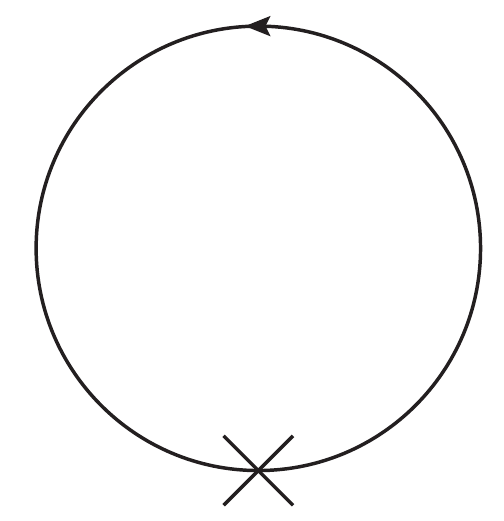}
\vskip -0.3cm
\caption{\small The Feynman diagram corresponding to the one-loop 
 stress-energy tensor for fermions. The cross indicates the fermionic stress-energy tensor 
insertion defined by~(\ref{T_fermions}).}
\label{fig:fermion_loop}
\vskip -0.2cm
\end{figure}

The stress-energy tensor is obtained by varying the action in Eq.~(\ref{L_fermions}) with respect to the 
tetrad field according to 
$T^\Psi_{\mu \nu} =-|e|^{-1}e_{a(\mu}\frac{\delta S}{\delta e^{\nu)}_a}$
and its expectation value reads, 
\begin{equation}
  \langle \hat T^\Psi_{\mu \nu}(x)\rangle 
   =  
\left\langle\!T\!\left[\hat{\bar{\Psi}}(x)\Big(\!-\frac{i}{2} \gamma_{(\mu} \partial_{\nu)} 
    \!+\!\frac{i}{2}\overleftarrow{\partial}_{(\mu} \gamma_{\nu)}\Big)\hat\Psi (x)
   \right]\right\rangle
   +\frac12 g_{\mu \nu}\left\langle\! T\!\left[\hat{\mathcal{L}}_\Psi(x)\right]\right\rangle
,\;\;
\label{T_fermions}
\end{equation}
which is covariantly conserved, 
$\partial^\mu \langle \hat T^\Psi_{\mu \nu}\rangle  = 0$.~\footnote{
For simplicity we assume that the background spacetime is expanding adiabatically slowly, 
such that time derivatives of the scale factor $a$ can be neglected. 
Then the way the expansion
enters the problem is through temperature's dependence on the scale factor,
whose precise form can be obtained from conservation of the entropy density $s$,  
$(d/dt)s\approx 0$.
If no large amounts of entropy are created by interactions, 
the temperature scales approximately inversely with the scale factor, {\it i.e.} 
$T\propto 1/a$.}
We wish to calculate~(\ref{T_fermions}) at the one-loop order,  
which corresponds to the Feynman diagram in figure~\ref{fig:fermion_loop},
where the solid oriented line is the free thermal fermionic propagator and the cross indicates 
the fermionic energy-momentum insertion, which is obtained by varying~(\ref{T_fermions}) 
with respect to $\Psi(y)$ and $\bar\Psi(y')$. 

In adiabatic regime (see section~\ref{The force on bubbles from the renormalized stress energy tensor}),  
when the effects of the expansion can be neglected, and
by making use of 
$(i\gamma^\mu\partial_\mu-m)(i\gamma^\nu\partial_\nu+m)
=(\partial^2-m^2)\mathbf{1}$, 
where $\partial^2=\eta^{\mu\nu}\partial_\mu\partial_\nu$,
$\partial_\mu\partial_\nu=\partial_\nu\partial_\mu$, and $\mathbf{1}$
is the unity matrix in the spinor space, 
one finds that the free thermal fermionic propagator~(\ref{fermion propagator: def})
can be expressed in terms of a `scalar' propagator $i\Delta_F(x;x')$,
\begin{equation}
 i S_{\alpha\beta}(x;x') = (i\gamma^\nu\partial_\nu+m)_{\alpha\beta} i\Delta_F(x;x')
\,,
\label{fermionic propagator: decomposition}
\end{equation}
where 
\begin{equation}
i\Delta_F(x;x') 
 = \frac{m^{D-2}}{(2\pi)^{D/2}}\frac{K_{\frac{D-2}{2}}\left(m\sqrt{\Delta x^2}\right)}
                                                        {(m\sqrt{\Delta x^2})^{\frac{D-2}{2}}}
\!-\! \int\frac{d^{D-1}p}{(2\pi)^{D-1}}
    \frac{{\rm e}^{i \vec p\cdot(\vec x\!-\!\vec x^{\prime})}}
                         {E_p}\frac{\cos[E_p(t-t^\prime)]}{{\rm e}^{\beta E_p}+1}
\,,
\label{fermionic propagator: decomposition 2}
\end{equation}
with $ E_p=\sqrt{p^2+m^2}$.
The principal difference between $i\Delta_F(x;x')$ and the scalar propagator $ i\Delta(x;x')$
in~(\ref{thermal scalar propagator}) is in the thermal part,
where the Bose-Einsten distribution function,
 $n_{\rm BE}=1/({\rm e}^{\beta E_p}-1)$ in Eq.~(\ref{thermal scalar propagator})
is replaced by the Fermi-Dirac distribution, $n_{\rm FD}=1/({\rm e}^{\beta E_p}+1)$ 
in~(\ref{fermionic propagator: decomposition 2}) and an overall minus sign
in the fermionic thermal part (which can be traced back 
to the anticommuting nature of the fermionic fields).
This affects, for example, the coincident fermionic propagator through,
\begin{equation}
i\Delta_F(x;x) 
 = \frac{m^{D-2}\Gamma\big(1-\frac{D}{2}\big)}{(4\pi)^{D/2}}
+\frac{1}{2\pi^2\beta^3m}\left[\partial_z J_F(4,z)\right]_{z=\beta m}
\,,
\label{thermal fermionic propagator: coincidence}
\end{equation}
where we introduced a {\it fermionic thermal integral},
\begin{equation}
 J_F(n,z) \equiv \int_0^\infty dx x^{n-2}\ln\left(1+{\rm e}^{-\sqrt{x^2+z^2}}\right)  
\,,
\label{JF n integral}
\end{equation}
and we should keep in mind that $\partial_z J_F(n,z)<0$.
Notice that we evaluated the thermal contribution
to the coincident fermionic propagator~(\ref{thermal fermionic propagator: coincidence}) in $D=4$.
The result~(\ref{thermal fermionic propagator: coincidence}) is to be compared with~(\ref{thermal scalar propagator: coincidence})
and~(\ref{JB n integral}), where the difference is in the form of 
the fermionic integral~(\ref{JF n integral}), but also in the sign of the 
coincident thermal propagator in~(\ref{thermal fermionic propagator: coincidence}). 
This sign difference can be traced back to the fact that the fermionic propagator 
is formally a one-loop quantity, and each fermionic loop contributes with a minus sign 
when compared with a bosonic loop.
 
We have now all the elements needed to evaluate the one-loop energy-momentum tensor~(\ref{T_fermions}). For that we firstly need the contribution of the mass term, 
\begin{equation}
 \langle T[\bar\Psi(x)\Psi(x)]\rangle = 
 \left(\!-\frac{m^{D-1}\Gamma\big(1-\frac{D}{2}\big)}{(4\pi)^{D/2}}
\!-\!\frac{1}{2\pi^2\beta^3}\left[\partial_z J_F(4,z)\right]_{z=\beta m}
\right)\!{\rm Tr}[\mathbf{1}]
\,,
\label{trace of the mass term: fermions}
\end{equation}
where ${\rm Tr}[\mathbf{1}]= 2^{D/2}$ is the number of fermionic degrees of freedom
in $D$ spacetime dimensions, which reduces to {\it four} in $D=4$ (two chiralities and 
two helicities). Next, 
\begin{equation}
\left\langle\!T\!\left[\hat{\bar{\Psi}}(x)\Big(\!-\frac{i}{2} \gamma_{(\mu} \partial_{\nu)} 
\Big)\hat\Psi (x)\right]\right\rangle 
= -\frac{i}{2}  \gamma^{\alpha\beta}_{(\mu} \partial^x_{\nu)}
\left\langle\!T^*\!\left[\hat{\bar{\Psi}}_\beta(x^\prime)\hat\Psi_\alpha (x)\right]\right\rangle
_{x^\prime\rightarrow x} 
\,,
\label{trace of the derivative term: fermions}
\end{equation}
where, as before when we considered the real scalar field
in section~\ref{Real scalar field}, 
the $T^*$ product indicates that the derivative
$\partial^x_{\nu}$ commutes with the time ordering operator $T^*$. 
With Eq.~(\ref{fermionic propagator: decomposition}) in mind, it is clear that 
only the term containing two $\gamma^\mu$'s contributes.
By recalling that, 
${\rm Tr}\left[i \gamma_{(\mu} \partial^x_{\nu)}
 i\gamma^\rho\partial^x_{\rho}\right] = \delta_{(\mu}^\rho\partial^x_{\nu)}\partial^x_{\rho}{\rm Tr}\left[\mathbf{1}\right]
= \partial^x_{\mu}\partial^x_{\nu}{\rm Tr}\left[\mathbf{1}\right]$, we can rewrite 
Eq.~(\ref{trace of the derivative term: fermions}) as,
\begin{eqnarray}
\left\langle\!T\!\left[\hat{\bar{\Psi}}(x)\Big(\!-\frac{i}{2}  \gamma_{(\mu} \partial_{\nu)} 
\Big)\hat\Psi (x)\right]\right\rangle 
&=&\frac{1}{2} 
\left[\partial^x_{\mu}\partial^x_{\nu}i\Delta_F(x;x') \right]_{x^\prime\rightarrow x}
\!{\rm Tr}[\mathbf{1}]
\nonumber\\
&=&\!-\frac{1}{2} \left[\partial^{x^\prime}_{\mu}\partial^x_{\nu}i\Delta_F(x;x') \right]_{x^\prime\rightarrow x}
\!{\rm Tr}[\mathbf{1}]
\,,
\label{trace of the derivative term: fermions:2}
\end{eqnarray}
where the last equality follows from the symmetry of the propagator under the exchange, 
$x\leftrightarrow x^\prime$. 
Analogous considerations show that the second term in~(\ref{T_fermions})
contributes equally, such that we have 
\begin{equation}
\left\langle\!T\!\left[\hat{\bar{\Psi}}(x)
   \Big(\!-\frac{i}{2}  \gamma_{(\mu} \partial_{\nu)} \!+\!\frac{i}{2}\overleftarrow{\partial}_{(\mu} \gamma_{\nu)}\Big)\hat\Psi (x)\right]\right\rangle
=-\left[\partial^\prime_{(\mu}\partial_{\nu)}i\Delta_F(x;x') \right]_{x^\prime\rightarrow x}
\!{\rm Tr}[\mathbf{1}]
,\;
\label{trace of the derivative term: fermions:3}
\end{equation}
where ${\rm Tr}[\mathbf{1}]=2^{D/2}$.
When Eqs.~(\ref{trace of the derivative term: fermions:3})
and~(\ref{trace of the mass term: fermions}) are compared with 
the analogous results for the real scalar field of section~\ref{Real scalar field}, we see
that -- up to the factor ${\rm Tr}[\mathbf{1}]=2^{D/2}$ which counts the number 
of degrees of freedom of a Dirac fermion -- 
the fermionic one loop energy-momentum tensor can be obtained 
from the real scalar one by replacing $i\Delta(x;x')$ by $i\Delta_F(x;x')$.
This then immediately implies that, the expectation value of 
the Lagrangian vanishes, {\it i.e.} $\langle T[\hat{\cal L}_\Psi]\rangle=0$
and that ({\it cf.} Eqs.~(\ref{Tmn: real scalar: vac}) and~(\ref{Tmn: real scalar: vac 2}))
\begin{eqnarray}
  \langle \hat T^\Psi_{\mu \nu}(x)\rangle 
   &=& -2^{D/2}
\left[\partial^\prime_{(\mu}\partial_{\nu)}i\Delta_f(x;x') \right]_{x^\prime\rightarrow x}
\,.
\label{T_fermions:2}
\end{eqnarray}
Just as in the case of the real scalar, the vacuum part of~(\ref{T_fermions:2}), which equals, 
\begin{eqnarray}
  \langle \hat T^\Psi_{\mu \nu}(x)\rangle_{\rm vac} 
   =-\eta_{\mu\nu} \frac{m^{D}\Gamma\big(\!-\!\frac{D}{2}\big)}{2(2\pi)^{D/2}}
\label{T_fermions: vacuum}
\end{eqnarray}
can be renormalized by the counterterm action~(\ref{counter term action: scalar}).
If the fermion mass $m$ is generated by a real scalar field condensate, 
\begin{equation}
 m(\phi_0) = y\phi_0
 \,,
 \label{field dependent mass: appendix}
\end{equation}
where $y$ is a Yukawa coupling and $\phi_0$
denotes a scalar condensate that may be 
adiabatically varying in space and time, then in 
the minimal subtraction scheme the counterterm coupling required 
to renormalize~(\ref{T_fermions: vacuum}) reads 
(see Eqs.~(\ref{counter term Tmn: scalar}--\ref{counter term lambda: scalar})), 
\begin{equation}
 \delta \lambda_\Psi = \frac{3y^4}{\pi^2}\frac{\mu^{D-4}}{D\!-\!4}
\,.
\label{counterterm coupling fermions}
\end{equation}
Upon adding the counterterm contribution to the energy-momentum tensor 
one obtains the sought-for renormalized 
one-loop energy-momentum tensor for the Dirac fermion, 
\begin{eqnarray}
\langle\hat T^\Psi_{\mu\nu}\rangle_{\rm ren} &=& \eta_{\mu\nu} \frac{m^4}{16\pi^2}\left[\ln\left(\frac{m^2}{2\pi\mu^2}\right)+\gamma_E-\frac32\right]
   -\frac{2\eta_{\mu\nu}}{3\pi^2\beta^5m}
                   \left[\partial_z J_F(6,z)\right]_{z=\beta m}
\quad
\nonumber\\
 &-& \!\delta_\mu^0\delta_\nu^0
        \left\{
        \frac{8}{3\pi^2\beta^5m}\left[\partial_z J_F(6,z)\right]_{z=\beta m}
         \!+\!\frac{2m}{\pi^2\beta^3}\left[\partial_z J_F(4,z)\right]_{z=\beta m}
        \right\}
\,.
 \quad
\label{renormalized energy-momentum tensor: fermion}
\end{eqnarray}
This result implies that the renormalized vacuum contribution to the energy density 
of a Dirac fermion is negative and four times
as large as that of a real scalar field. Its thermal contribution harbors four fermionic 
degrees of freedom which obey Fermi-Dirac statistics.

\bigskip

{\bf Abelian gauge field model.} 
The action for an Abelian gauge field reads, 
\begin{equation}
 S_{\rm EM}[A_\mu]=\!\int\! d^Dx\sqrt{-g}{\cal L}_{\rm EM}
\, ,\quad 
{\cal L}_{\rm EM} = -\frac14 g^{\mu\rho}g^{\nu\sigma}F_{\mu\nu}F_{\rho\sigma}
\,,
\label{EM action}
\end{equation}
where $F_{\mu\nu} = \partial_\mu A_\nu - \partial_\nu A_\mu$ is the field strength
and $A_\mu$ is an Abelian gauge field, a prime example of which is the {\it photon} of 
the quantum electrodynamics (QED).
The action~(\ref{EM action}) possesses an Abelian gauge symmetry, under which
the field transforms as, 
\begin{equation}
 A_\mu\rightarrow A_\mu + \partial_\mu \Lambda(x)
\,,
 \label{Abelian gauge transformation}
\end{equation}
where $\Lambda(x)$ is an arbitrary function of space and time. This means that 
$A_\mu$ contains redundant (unphysical) degrees of freedom which make the kinetic operator for $A_\mu$ derived from~(\ref{EM action}) non-invertible. 
Without going into the details of the gauge fixing procedure, this is resolved by 
adding a gauge fixing term to the action, which makes the kinetic operator invertible,
but does not change any physical quantity. A legitimate gauge fixing is the {\it Fermi
gauge}, which is convenient since it has one gauge parameter $\xi$ 
which can be used to control gauge dependence of the results.
The corresponding gauge fixing action and Lagrangian are, 
\begin{equation}
 S_{\rm Fermi}=\!\int\! d^Dx\sqrt{-g}{\cal L}_{\rm Fermi}
\, ,\quad 
{\cal L}_{\rm Fermi} = -\frac1{2\xi} \left(g^{\mu\nu}\nabla_\mu A_\nu\right)^2
\,.
\label{Fermi gauge}
\end{equation}
The canonical momentum of the theory is obtained by a variation of the total action,
 $S_{\rm tot} =S_{\rm EM}+S_{\rm Fermi}$ with respect to $\partial_0 A_\mu(x)$,
 and in Minkowski background it reads, 
 \begin{equation}
  \Pi_A^\nu(x) = \frac{\delta S_{\rm tot} }{\delta\partial_0 A_\nu(x) }
    = -\eta^{\mu\nu}\left(\partial_0 A_\mu - \partial_\mu A_0\right)
      + \frac{1}{\xi}\delta_0^\nu \eta^{\rho\sigma}\partial_\rho A_\sigma
\,,
\label{canonical momentum: Abelian gauge massless}
\end{equation}
such that {\it e.g.}  $\Pi_A^0 = \xi^{-1}\eta^{\rho\sigma}\partial_\rho A_\sigma$ 
does not vanish.
Due to the added gauge fixing term~(\ref{Fermi gauge}), 
all of the canonical momenta become dynamical,
such that the canonical commutation relation in the Fermi gauge is simple, 
\begin{equation}
 \left[\hat A_\mu(t,\vec x),\hat \Pi_A^\nu(t,\vec x^{\,\prime}\right] 
  = i\delta_\mu^{\;\nu}\delta^{D-1}(\vec x-\vec x^{\,\prime})
\,.
\label{canonical quantization: Abelian gauge field}
\end{equation}
The Feynman propagator equation in the Fermi gauge is therefore, 
\begin{equation}
\left(\partial^2\eta^{\mu\nu}-\left(1-\frac1{\xi}\right)\partial^\mu\partial^\nu\right)
  i\left[_\nu\Delta_\alpha\right](x;x^\prime)
     =i\delta^\mu_{\;\alpha}\delta^D(x\!-\!x^{\,\prime})
\,,
\label{Feynman propagator: Abelian gauge field in Fermi gauge}
\end{equation}
where 
\begin{equation}
  i\left[_\nu\Delta_\alpha\right](x;x^\prime)
     =\left\langle T\left[\hat A_\nu(x)\hat A_\alpha(x^{\,\prime})\right]\right\rangle
\,.
\label{Feynman propagator: Abelian gauge field in Fermi gauge: def}
\end{equation}
The solution of~(\ref{Feynman propagator: Abelian gauge field in Fermi gauge}) can be written as, 
\begin{equation}
 i\left[_\nu\Delta_\alpha\right](x;x^\prime)
 = \left[\eta_{\nu\alpha}
      -\left(1-\xi\right)\frac{\partial_\nu\partial_\alpha}{\partial^2}\right]
         i\Delta_0(x;x^{\,\prime})
\,,
\label{Feynman propagator: Abelian gauge field in Fermi gauge: solution}
\end{equation}
where $i\Delta_0(x;x^{\,\prime})$ is the massless limit of the thermal scalar 
propagator~(\ref{thermal scalar propagator}), which equals
(see the $n=0$ term of the second series in Eq.~(\ref{expansion of Knu})),
\begin{eqnarray}
         i\Delta_0(x;x^{\,\prime}) &=&\frac{\Gamma\left(\frac{D}{2}-1\right)}{4\pi^{D/2}}
\left(\frac{1}{\Delta x^2(x;x^\prime)}\right)^\frac{D-2}{2}
\label{massless scalar propagator: vacuum}\\
 &+&\!\! \frac{1}{4\pi^2}\!\left[\!-{\cal P}\frac{1}{-\Delta t^2+r^2}
    \! +\! \frac{\pi}{\beta r}\frac{{\rm e}^\frac{4\pi r}{\beta}-1}
{\Big(\!{\rm e}^\frac{2\pi (r+\Delta t)}{\beta}\!\!-\!1\Big)
 \Big(\!{\rm e}^\frac{2\pi (r-\Delta t)}{\beta}\!\!-\!1\Big)}
   \right] 
,\quad\;
\label{massless scalar propagator: thermal}
\end{eqnarray}
where $\Delta t = t-t^\prime$, $r=\|\vec x-\vec x^{\,\prime}\|$,
$\Delta x^2(x;x')$ is given in Eq.~(\ref{invariant distance}) 
and, for simplicity, we have evaluated 
the thermal part~(\ref{massless scalar propagator: thermal}) in $D=4$ (because it is finite in $D=4$ and therefore does not need 
to be regularized).
The first term in the second line~(\ref{massless scalar propagator: thermal})
is not there to cancel the principal part ($\cal P$) of the vacuum 
propagator~(\ref{massless scalar propagator: vacuum}) in $D=4$, 
but instead it is needed to get the thermal contribution to vanish 
in the zero temperature limit, $\beta\rightarrow \infty$. 

We are now ready to proceed with the calculation of the one-loop energy-momentum tensor,
which for the total photon action, consisting of 
Eq.~(\ref{EM action}) plus the gauge fixing part~(\ref{Fermi gauge}), reads,
%
%
%
%
\begin{eqnarray}
\langle \hat T_{\mu\nu}^{\rm EM}\rangle 
&=& \eta^{\alpha\rho}\left[\delta_{(\mu}^{\;\beta}\delta_{\nu)}^{\;\sigma}
-\frac14\eta_{\mu\nu}\eta^{\beta\sigma}\right]
       \big \langle T[\hat F_{\alpha\beta}\hat F_{\rho\sigma}]\big\rangle
-\frac{2}{\xi}\eta^{\alpha\beta}
  \Big \langle T\left[\hat A_{(\mu}\partial_{\nu)}\partial_\alpha \hat A_\beta\right]\Big\rangle
\nonumber\\
&&\hskip -1.8cm
+\,\frac1{\xi}\eta_{\mu\nu}\eta^{\alpha\beta}\eta^{\rho\sigma}
\Big \langle T\left[\hat A_\alpha\partial_\beta\partial_\rho \hat A_\sigma\right]\Big \rangle
+\frac1{2\xi}\eta_{\mu\nu}\eta^{\alpha\beta}\eta^{\rho\sigma}
 \Big\langle T\left[ (\partial_\alpha\hat A_\beta)(\partial_\rho \hat A_\sigma)\right]\Big\rangle
.\;\;
\label{energy momentum tensor: massless photon:2}
\end{eqnarray}
where we also took an expectation value of the operator-valued energy-momentum tensor.
Next, the derivatives in~(\ref{energy momentum tensor: massless photon:2}) can 
be taken out of the expectation values, provided one replaces the $T$ with the $T^*$ time ordering,
such that the energy-momentum 
tensor~(\ref{energy momentum tensor: massless photon:2}) can be recast as,
\begin{eqnarray}
\langle \hat T_{\mu\nu}^{\rm EM}(x)\rangle 
\!\!&=&\!\!\!\left[ 4\eta^{\alpha\rho}\delta_{(\mu}^{\;\beta}\delta_{\nu)}^{\;\sigma}
- \eta^{\alpha\rho}\eta_{\mu\nu}\eta^{\beta\sigma}\right]
  \partial^\prime_{\rho]}\partial_{[\alpha}
   \Big\langle T^*\left[\hat A_{\beta]}(x) \hat A_{[\sigma}(x^\prime)\right]\Big\rangle
            _{x^\prime\rightarrow x}
\label{energy momentum tensor: massless photon:3}\\
&&\hskip -2.5cm
+\,\frac{1}{\xi}\!\left[
 \!-2\eta^{\rho\sigma}\delta_{(\mu}^{\;\alpha}\delta_{\nu)}^{\;\beta}
           \partial^\prime_\alpha\partial^\prime_\rho
\!+\! \eta_{\mu\nu}\eta^{\alpha\beta}\eta^{\rho\sigma}
           \!\left(\!\partial^\prime_\alpha
                \!+\!\frac12\partial_\alpha\right)\!\partial^\prime_\rho
\right]\Big\langle\! T^*\!\left[\hat A_\beta(x) \hat A_\sigma(x^\prime)\!\right]\!\Big\rangle
            _{x^\prime\rightarrow x}\!\!
.\quad\;\;
\label{energy momentum tensor: massless photon:3b}
\end{eqnarray}
This expression can be simplified by making use of the tensor structure of 
the photon propagator~(\ref{Feynman propagator: Abelian gauge field in Fermi gauge: solution}),
\begin{eqnarray}
\langle \hat T_{\mu\nu}^{\rm EM}(x)\rangle 
\!\!&=&\!\!\left[\frac{D\!-\!3}{2}\eta_{\mu\nu}\partial^2
         -(D\!-\!2)\partial_{\mu}\partial_{\nu}\right]
  i\Delta_0(x;x^\prime)_{x^\prime\rightarrow x}
\label{energy momentum tensor: massless photon:4}\\
&+&\hskip 0cm
\!\!\left[\frac12\eta_{\mu\nu}\partial^2 - 2 \partial_\mu\partial_\nu
\right]  i\Delta_0(x;x^\prime)_{x^\prime\rightarrow x}
\,,\;\;\;
\label{energy momentum tensor: massless photon:4b}
\end{eqnarray}
 where we took account of the antisymmetry property of the derivatives, 
 $\partial^\prime_\alpha i\Delta_0(x;x^\prime)=-\partial_\alpha i\Delta_0(x;x^\prime)$.
The first line~(\ref{energy momentum tensor: massless photon:4})
originates from the gauge field 
action~(\ref{energy momentum tensor: massless photon:3}), while the second 
 line~(\ref{energy momentum tensor: massless photon:4b}) 
comes from the gauge fixing term 
contribution~(\ref{energy momentum tensor: massless photon:3b}). 
Notice that both contributions in~(\ref{energy momentum tensor: massless photon:4}--\ref{energy momentum tensor: massless photon:4b}) 
do not dependent on the gauge parameter $\xi$, and combine into, 
\begin{eqnarray}
\langle \hat T_{\mu\nu}^{\rm EM}(x)\rangle 
&=&\left[\frac{D\!-\!2}{2}\eta_{\mu\nu}\partial^2
         -D\partial_{\mu}\partial_{\nu}\right]
  i\Delta_0(x;x^\prime)_{x^\prime\rightarrow x}
\,.
\label{energy momentum tensor: massless photon:5}
\end{eqnarray}

It is now clear that in dimensional regularization the vacuum part of $ i\Delta_0(x;x^\prime)$
in~(\ref{massless scalar propagator: vacuum})
does not contribute to~(\ref{energy momentum tensor: massless photon:4}). 
However, the thermal part does contribute and the result is,
\begin{eqnarray}
\langle \hat T_{\mu\nu}^{\rm EM}(x)\rangle 
&\stackrel{?}{=}& 4\times\frac{\pi^2}{30\beta^4}\delta_\mu^0\delta_\nu^0
+4\times\frac{\pi^2}{90\beta^4}
   \left(\eta_{\mu\nu}+ \delta_\mu^0\delta_\nu^0\right)
\,,\quad
\label{EM energy momentum tensor massless photon: gauge dep}
\end{eqnarray}
where~\footnote{
The small argument expansion of~(\ref{massless scalar propagator: thermal}) is, 
$i\Delta^{\rm th}_0(x;x^\prime)
\sim\frac{1}{12\beta^2}-\frac{\pi^2}{180\beta^4}(r^2+3\Delta t^2)$.
}
%
\begin{eqnarray}
 \partial_j\partial_k i\Delta_0(x;x^\prime)_{x^\prime\rightarrow x} 
  &=& - \frac{\pi^2}{90\beta^4}\delta_{jk}
  \,,\quad
 \partial_0^2 i\Delta_0(x;x^\prime)_{x^\prime\rightarrow x} 
 =\! - \frac{\pi^2}{30\beta^4}
 \nonumber\\
    \partial_0\partial_j i\Delta_0(x;x^\prime)_{x^\prime\rightarrow x} 
  &=& 0
  \,,\qquad \qquad\quad\!
   \partial^2 i\Delta_0(x;x^\prime)_{x^\prime\rightarrow x} 
  = 0
\,.
\label{some useful equalities for the massless photon}
\end{eqnarray}
The result~(\ref{EM energy momentum tensor massless photon: gauge dep})
{\it cannot} be correct, since it suggests that the photon has {\it four} degrees of freedom,
instead of {\it two} of the physical photon (the two transverse polarizations with helicities,
$h=\pm\hbar $). The reason is the gauge fixing term~(\ref{Fermi gauge}) which 
made all four polarizations of the photon dynamical 
({\it cf.} Eq.~(\ref{canonical momentum: Abelian gauge massless})), thus  
explaining~(\ref{EM energy momentum tensor massless photon: gauge dep}).
Even though Eq.~(\ref{EM energy momentum tensor massless photon: gauge dep})
does not depend on the gauge parameter $\xi$, 
it is not correct because we did not take account of the contribution
from the {\it Faddeev-Popov ghosts}. It is well known that in the vacuum the contribution from 
ghosts in Abelian gauge theories vanishes, for more general states such as thermal states however,
ghosts do contribute and thus have to be taken into account. To show that, consider 
the Fadeed-Popov ghost action and Lagrangian associated with an Abelian gauge field, 
\begin{equation}
S_{\rm ghost} = \int d^D x\sqrt{-g}{\cal L}_{\rm ghost}
\,,\qquad
{\cal L}_{\rm ghost}= - g^{\mu\nu}(\partial_\mu \bar c)(\partial_\nu c)
\,,
\label{ghost lagrangian}
\end{equation}
 where $c=c(x)$ is a complex Grasmannian (anticommuting) scalar field and $\bar c=c^*(x)$.
 The corresponding canonical momenta are, 
 \begin{equation}
 \Pi_c=-\partial_0 \bar c\,\quad {\rm and}\quad \Pi_{\bar c}=\partial_0 c
 \,,
 \label{canonical momenta: Abelian ghosts}
\end{equation}
(the minus in $\Pi_c=\delta S_{\rm ghost} /\delta \partial_0c$ 
comes from the anticommuting of $\delta/\delta \partial_0c$ with $\bar c$) 
 which imply the following nonvanishing canonical quantization rules,~\footnote{
 The ghost quantization rules~(\ref{canonical commutation: ghosts}) accord
 with the fermionic one~(\ref{canonical quantization relation: fermions}), 
 if one recalls that $\Pi_\Psi(x) = -ia^{D-1} \Psi^*(x)$.
 } 
 \begin{equation}
 \left\{\hat c(t,\vec x),\hat \Pi_c(t,\vec x^{\,\prime})\right\} 
        =-i\delta^{D-1}(\vec x\!-\!\vec x^{\,\prime})
 \,,\quad 
 \left\{\hat {\bar c}(t,\vec x),\hat \Pi_{\bar c}(t,\vec x^{\,\prime})\right\} 
 = -i\delta ^{D-1}(\vec x\!-\!\vec x^{\,\prime})
\,,\quad
 \label{canonical commutation: ghosts}
 \end{equation}
from where one obtains the ghost propagator equation,
\begin{equation}
\partial^2 i\Delta_0^{\rm gh}(x;x^\prime) = i\delta^D(x\!-\!x^\prime)
\,,
\label{ghost equation of motion}
\end{equation}
with
\begin{eqnarray}
 i\Delta_0^{\rm gh}(x;x^\prime) &\equiv& \left\langle T\left[\hat c(x)\hat{\bar c}(x')\right]\right\rangle
    = -  \left\langle T\left[\hat{\bar c}(x')\hat{c}(x)\right]\right\rangle
\label{ghost propagator: definition:0}\\
  &=& \theta(t\!-\! t^\prime)\left\langle\hat  c(x)\hat{\bar c}(x')\right\rangle
 \!- \theta(t^\prime\!-\! t)\left\langle\hat{\bar c}(x')\hat{c}(x)\right\rangle
\,.\;
\label{ghost propagator: definition}
\end{eqnarray}
Had we defined the ghost propagator with the ghost field ordering as indicated 
after the second equality in~(\ref{ghost propagator: definition:0}), we would have obtained 
$-i\delta^D(x\!-\!x^\prime)$ on the right hand side of~(\ref{ghost equation of motion}). 
Ghosts are complex, bosonic, anticommuting fields, so their propagator 
equation is simply related to that of a massless scalar~(\ref{massless scalar propagator: thermal}),
$i\Delta_0^{\rm gh}(x;x^\prime) \leftrightarrow  i\Delta_0(x;x^\prime)$,~\footnote{
\label{footnote on ghosts}
One can consider the positive frequency thermal Wightman function for the ghost field, 
\begin{equation}
 i\Delta_{\rm gh}^+(x;x^\prime) = \left\langle\hat  c(x)\hat{\bar c}(x')\right\rangle
   = {\rm Tr}\left[\hat \rho_{\rm th}\hat  c(x)\hat{\bar c}(x')\right]
\,,\qquad \hat \rho_{\rm th} = \frac{{\rm e}^{-\beta \hat H_{\rm gh}}}
        {{\rm Tr}\big[{\rm e}^{-\beta \hat H_{\rm gh}}\big]}
\,,
\label{ghost positive frequency Wightman function}
\end{equation}
where $\hat \rho_{\rm th}$ is the thermal density matrix and $\hat H_{\rm gh}$ is the ghost Hamiltonian.
By taking the cyclic properly of the trace,
$c(t,\vec x) = {\rm e}^{-i\hat H_{\rm gh}}c(0,\vec x){\rm e}^{i\hat H_{\rm gh}}$
 and the non-commuting nature of the ghost fields, 
one obtains the following Kubo-Martin-Schwinger condition for the ghost, 
\begin{equation}
{\rm Tr}\left[\hat \rho_{\rm th}\hat  c(t,\vec x)\hat{\bar c}(t',\vec x')\right]
 = -{\rm Tr}\left[\hat \rho_{\rm th}\hat{\bar c}(t'+i\beta,\vec x')\hat  c(t,\vec x)\right]
\,,
\label{KMS for ghost}
\end{equation}
from where we infer, 
\begin{equation}
i\Delta^+(t,\vec x;t',\vec x')=-i\Delta^-(t,\vec x;t'+i\beta,\vec x')
\,,
\label{KMS for ghost:2}
\end{equation}
which reads in momentum space, 
$i\tilde\Delta^+(p^\alpha) = - {\rm e}^{-\beta p^0}i\tilde\Delta^-(p^\alpha) $.
This suggests that the thermal part of the ghost propagator 
should obey a Fermi-Dirac statistic.
}
%
\begin{eqnarray}
         i\Delta_0^{\rm gh}(x;x^{\,\prime}) &=&\frac{\Gamma\left(\frac{D}{2}-1\right)}{4\pi^{D/2}}
\left(\frac{1}{\Delta x^2(x;x^\prime)}\right)^\frac{D-2}{2}
\label{massless ghost propagator: vacuum}\\
 &+&\!\! \frac{1}{4\pi^2}\!\left[\!-{\cal P}
  \frac{1}{-\Delta t^2\!+\!r^2}
    \! +\! \frac{\pi}{\beta r}\frac{{\rm e}^\frac{4\pi r}{\beta}-1}
{\Big(\!{\rm e}^\frac{2\pi (r+\Delta t)}{\beta}\!\!-\!1\Big)
 \Big(\!{\rm e}^\frac{2\pi (r-\Delta t)}{\beta}\!\!-\!1\Big)}
   \right] \!
.\qquad
\label{massless ghost propagator: thermal}
\end{eqnarray}
The argument presented in footnote~\ref{footnote on ghosts} shows that, as a 
consequence of treating the ghosts as Grassmannian fields in thermal equilibrium, 
implies that they should obey a Fermi-Dirac statistic. 

The reason why ghosts are treated as anticommuting fields is that, upon integrating them out, 
one reproduces the correct Faddeev-Popov $O_{\rm FP}(x;x')$ determinant. This determinant 
is obtained by a functional variation of the gauge fixing condition with respect to the gauge parameter,
which ensures the correct field-dependent integration measure along gauge orbits, 
and which for the problem at hand reads, 
\begin{equation}
O_{\rm FP}(x;x') = \frac{\delta}{\delta \Lambda(x^\prime)}\nabla^\mu A_\mu(x)
                         =\Box_x \delta^D(x-x^\prime)
\,,
\label{Faddeev popov determinant}
\end{equation}
whose determinant, 
\begin{equation}
 {\rm det}\left[O_{\rm FP}(x;x') \right]={\rm det}\left[\Box_x \delta^D(x-x^\prime)\right]
\label{Faddeev Popov determinant:2}
\end{equation}
can be represented as a path integral over the anticommuting Faddeev-Popov complex ghost fields,
\begin{equation}
 {\rm det}\left[O_{\rm FP}(x;x') \right]=\int{\cal D}\bar c{\cal D}c
           {\rm exp}\left[{i\int d^Dx \sqrt{-g}\Big(-g^{\mu\nu}\partial_\mu \bar c\partial_\nu c\Big)}\right]
\,.
\label{Faddeev Popov determinant:3}
\end{equation}
On the other hand, Eq.~(\ref{Faddeev Popov determinant:2}) implies, 
\begin{equation}
{\rm det}[O_{\rm FP}(x;x')]=\exp\Big\{i\big[-i{\rm Tr}\ln\big({\cal O}_{\rm FP}(x;x')\big)\big]\Big\}
\,,
\label{Faddeev Popov determinant:3}
\end{equation}
such that its contribution to the effective action 
is of the form, 
\begin{equation}
\Gamma_{\rm FP}=-i{\rm Tr}\big[\ln\big({\cal O}_{\rm FP}(x;x')\big)\big]
\,,
\label{Faddeev Popov determinant:4}
\end{equation}
which equals {\it minus twice} that of a real scalar field, which contribute as, 
$\Gamma_{\phi}=(i/2){\rm Tr}\left[\ln\left({\cal O}_\phi(x;x')\right)\right]$.
This observation has led to the development of the ghost field formalism,
according to which ghosts are - just like fermions - anticommuting fields,
but they obey a Bose-Einstein statistic. However, as we argue in footnote~\ref{footnote on ghosts}, 
that is inconsistent with the notion that a thermal state is defined in terms of the thermal density operator.
Since ghosts fields are commonly viewed as `unphysical', 
the field practitioners have gotten used to this inconvenience, and declared by fiat that 
ghosts fields obey a Bose-Einstein statistic.~\footnote{An alternative strategy was proposed by 
Kobes, Semenoff and Weiss in Ref.~\cite{Kobes:1984vb}, where the authors posit that 
both the gauge and ghost sectors of gauge theories should be treated as particles in the vacuum state,
{\it i.e.} that the corresponding propagators should be constructed with boundary conditions imposed 
such that they possess only the vacuum part. While this may give correct answers for thermal states,
there is no guarantee that such a prescription will work in more general 
non-equilibrium situations. Furthermore, 
this prescription requires a separation of the gauge and ghost sectors from the `physical sector',
and that separation in general requires a use of nonlocal operators, 
and hence it is unnecessarily complicated and quite delicate. Moreover, since the principal goal
of quantization of constrained systems is to demonstrate that observables calculated in 
perturbation theory do not depend on the gauge fixing procedure or on the choice of 
the gauge part of the initial state, we find the proposal of Ref.~\cite{Kobes:1984vb} unsatisfactory. 
} 
Our take on this is that, once degrees of freedom are added
to the Hamiltonian, they are degrees of freedom and have to be dealt with as such.

 Here we take a different take on ghosts. Note that the Faddeev-Popov determinant can be 
also written as, 
\begin{eqnarray}
 {\rm det}\left[O_{\rm FP}(x;x') \right]  
 &=& \frac{1}{{\rm det}\left[\Box_x^{-1} \delta^D(x-x^\prime)\right]}
\nonumber\\
  &=& \int{\cal D}\bar \phi_{\rm gh}{\cal D}\phi_{\rm gh}
           {\rm exp}\left[{i\int d^Dx \sqrt{-g}\Big( \bar \phi_{\rm gh}\Box^{-1} \phi_{\rm gh}\Big)}\right]
 ,\qquad
\label{Faddeev Popov determinant:5}
\end{eqnarray}
where $\phi_{\rm gh}$ is now defined as a complex (commuting) scalar ghost field
and $\bar\phi_{\rm gh}=\phi_{\rm gh}^*$. The price to pay 
is the {\it non-local action}~(\ref{Faddeev Popov determinant:5}) 
which governs the dynamics of the ghosts and which, upon a partial integration, can be recast as, 
\begin{equation}
S_{\rm gh}= \!\int \!d^Dx \sqrt{\!-g}{\cal L}_{\rm gh}
,\quad
{\cal L}_{\rm gh}=\bar  \phi_{\rm gh}\frac{1}{\Box} \phi_{\rm gh}
 = 
-g^{\mu\nu}\! \left(\!\frac{\partial_\nu}{\Box}\bar  \phi_{\rm gh}\!\right)
    \left(\!\frac{\partial_\mu}{\Box} \phi_{\rm gh}\!\right) +{\rm bd.\, term}
 .
\label{scalar ghost action}
\end{equation}
%
The principal disadvantage of this action is that it is non-local. 
One should not be scared by non-local actions however, 
as they have successfully been dealt with in 
the literature on dark 
energy~\cite{Deser:2007jk,Woodard:2018gfj,Deser:2019lmm}.
One may wonder whether 
the nonlocality can be dealt with 
by introducing auxiliary fields, $\chi =\Box^{-1} \phi_{\rm gh}$, in terms of which 
the ghost action can be rewritten 
into an on-shell (weakly) equivalent  ($S_{\rm gh}^{\prime}\approx S_{\rm gh}$) form as,
 %
\begin{equation}
S_{\rm gh}^{\prime}= \int d^Dx \sqrt{-g}\left[-g^{\mu\nu}\partial_\nu \bar\chi\partial_\mu \chi
 +\bar \lambda\left(\chi \!-\!\Box^{-1} \phi_{\rm gh}\right)
 + \lambda\left(\bar\chi \!-\!\Box^{-1} \bar  \phi_{\rm gh}\right)
           \right]
 \,,
\label{scalar ghost action:2}
\end{equation}
where $\lambda=\lambda(x)$ and $\bar\lambda=\bar\lambda(x)$ are Lagrange multiplier fields. 
Upon solving for the constraint fields $\lambda$ and $\bar\lambda$, one gets the original action.
However,  varying with respect to $\chi$ and $\bar\chi$ yields,
\begin{equation}
 \Box \chi + \lambda = 0 \,,\qquad \Box\bar \chi + \bar \lambda = 0
 \,.
\label{scalar ghost action:3}
\end{equation}
Now $\lambda$ and $\bar\lambda$ can be eliminated in favor of $\chi$ and $\bar\chi$,
to obtain yet another on-shell equivalent ghost action,
\begin{equation}
S_{\rm gh}^{\prime\prime}
 = \int d^Dx \sqrt{-g}\left[g^{\mu\nu}\partial_\nu \bar\chi\partial_\mu \chi
 +\bar\chi  \phi_{\rm gh}+\bar  \phi_{\rm gh}\chi \right]
 \,,
\label{scalar ghost action:4}
\end{equation}
This action is local, and that is desirable. However the price we paid to get~(\ref{scalar ghost action:4}) 
is in that $\chi$ and $\bar\chi$ are ghost scalars, 
{\it i.e.} they have a negative kinetic term. As long as we do not include dynamical gravity,
this need not be fatal for the theory, and we can work safely with it.~\footnote{There are 
gauges in which some of the gauge field components have a ghost-like kinetic term.
However, because these belong to the gauge 
sector of the theory, that is considered not to be a problem.}
The action~(\ref{scalar ghost action:4})
is not problem free however. Indeed, upon varying~(\ref{scalar ghost action:4}) with respect to 
$\bar\chi$, $\chi$, $\bar \phi_{\rm gh}$ and $\phi_{\rm gh}$ we get, 
\begin{equation}
\Box \chi = \phi_{\rm gh}
\,,\quad \chi =0
\,,\quad
\Box \bar\chi = \bar\phi_{\rm gh}
\,,\quad \bar\chi =0
\,.
\label{scalar ghost action:5}
\end{equation}
Obviously,  these equations cannot be the correct equations for the ghost sector. The correct equations are 
obtained by acting $1/\Box $ on the first and third equation in~(\ref{scalar ghost action:5}), 
\begin{equation}
 \chi = \frac{1}{\Box}\phi_{\rm gh}
\,,\quad \chi =0
\,,\quad
\bar\chi = \frac{1}{\Box}\bar\phi_{\rm gh}
\,,\quad \bar\chi =0
\,,
\label{scalar ghost action:6}
\end{equation}
which are equivalent to the equations obtained from the original 
action~(\ref{Faddeev Popov determinant:5}--\ref{scalar ghost action}). 
The above exercise is instructive, as it teaches us that one ought to be extra careful when making use of 
the procedure which is known to be valid for constrained systems and whose dynamics is described
by local Hamiltonians.

In order to avoid such pitfalls, we shall work here with 
the non-local version of the theory~(\ref{scalar ghost action}), which is feasible at 
the one-loop level and that is what we do next.  Varying the action~(\ref{scalar ghost action})
with respect to $\bar\phi_{\rm gh}$ and $\bar\phi_{\rm gh}$ gives, 
\begin{equation}
 \frac{1}{\partial^2} i\Delta_{\rm gh}(x;x^\prime) = i\delta^D(x\!-\!x^\prime)
\,,
\label{EOM for a ghost scalar}
\end{equation}
where 
\begin{equation}
 i\Delta_{\rm gh}(x;x^\prime) 
  = \Big\langle T\left[\hat\phi_{\rm gh}(x)\hat{\bar\phi}_{\rm gh}(x^\prime)\right]\Big\rangle
 \label{scalar ghost propagator: definition}
\end{equation}
with 
\begin{equation}
T\left[\hat\phi_{\rm gh}(x)\hat{\bar\phi}_{\rm gh}(x^\prime)\right]
 = \theta(t\!-\!t^\prime)\hat\phi_{\rm gh}(x)\hat{\bar\phi}_{\rm gh}(x^\prime)
 +\theta(t^\prime\!-\!t)\hat{\bar\phi}_{\rm gh}(x^\prime)\hat\phi_{\rm gh}(x)
\,,
 \label{T ordering ghosts}
 \end{equation}
and for simplicity we assumed in~(\ref{EOM for a ghost scalar}) a Minkowski background. 
Eq.~(\ref{T ordering ghosts}) implies that the ghosts $\phi_{\rm gh}$ and $\bar\phi_{\rm gh}$ are 
bosonic fields which obey a Bose-Einstein statistic, and the solution 
with thermal boundary conditions can be written as (see footnote~\ref{gauge field propagator split} below),
\begin{equation}
 i\Delta_{\rm gh}(x;x^\prime) =\partial^2 i\delta^D(x\!-\!x^\prime)
  +\left[\partial^4 i\Delta_M^{\rm th}\left(x;x^\prime\right)\right]_{M^2\rightarrow 0}
  =\left[\partial^4 i\Delta_M\left(x;x^\prime\right)\right]_{M^2\rightarrow 0}
.
\label{EOM for a ghost scalar: solution}
\end{equation}
where $i\Delta_M^{\rm th}\left(x;x^\prime\right)$ denotes the 
thermal part of the massive scalar 
propagator~(\ref{thermal scalar propagator}). 
In Eq.~(\ref{EOM for a ghost scalar: solution}) we shifted the poles of the 
massless propagator, $(k^0)^2=\|\vec k\|^2$ by $\delta M^2=M^2$,
which regulates the would-be singular behavior of its thermal part. 
In what follows, it will become clear how to use~(\ref{EOM for a ghost scalar: solution})
in practical calculations.

The next step is the one-loop energy-momentum tensor from the ghost fields,
$T^{\rm gh}_{\mu\nu}=-(2/\sqrt{-g})(\delta S_{\rm ghost}/\delta g^{\mu\nu})$,
whose expectation value is,  
 \begin{eqnarray}
 \langle\hat T^{\rm gh}_{\mu\nu}\rangle
   &=& -2\left\langle T\left[\left(\frac{\partial_{(\mu}}{\Box}\bar\phi_{\rm gh}\right)
       \left(\frac{\partial_{\nu)}}{\Box}\phi_{\rm gh}\right)\right]\right\rangle
       + g_{\mu\nu}\left\langle T\left[\left(\frac{1}{\Box}\bar\phi_{\rm gh}\right)\phi_{\rm gh}
       \right]\right\rangle
\nonumber\\
&&\hskip -1.5cm
+\; g_{\mu\nu}g^{\alpha\beta}\left\langle T\left[\left(\frac{\partial_{\alpha}}{\Box}\bar\phi_{\rm gh}
                                              \right)
       \left(\frac{\partial_{\beta}}{\Box}\phi_{\rm gh}\right)\right]\right\rangle
    + g_{\mu\nu}\left\langle T\left[\bar\phi_{\rm gh}\left(\frac{1}{\Box}\phi_{\rm gh}\right)
       \right]\right\rangle
,
\qquad
\label{Tmn massless ghosts}
\end{eqnarray}
which, upon extracting the derivatives, can be recast as,
 \begin{eqnarray}
 \langle\hat T^{\rm gh}_{\mu\nu}\rangle
   \!&=&\!\! -2\frac{\partial^\prime_{(\mu}\partial_{\nu)}}{{\partial^\prime}^2\partial^2}
       \left\langle T^*\!\left[\bar\phi_{\rm gh}(x^\prime)\phi_{\rm gh}(x)\right]\right\rangle
                _{x^\prime\rightarrow x}
       \!\!+\! \frac{g_{\mu\nu}}{{\partial^\prime}^2}
              \left\langle T^*\!\left[\bar\phi_{\rm gh}(x^\prime)\phi_{\rm gh}(x)\right]\right\rangle
                _{x^\prime\rightarrow x}
\nonumber\\
&&\hskip -1.5cm
+\; g_{\mu\nu}g^{\alpha\beta}\frac{\partial^\prime_\alpha\partial_\beta}{{\partial^\prime}^2\partial^2}
 \left\langle T^*\!\left[\bar\phi_{\rm gh}(x^\prime)\phi_{\rm gh}(x)\right]\right\rangle
                _{x^\prime\rightarrow x}
    \!+\!\frac{g_{\mu\nu}}{\partial^2}
 \left\langle T^*\!\left[\bar\phi_{\rm gh}(x^\prime)\phi_{\rm gh}(x)\right]\right\rangle
                _{x^\prime\rightarrow x}
,\nonumber\\
\qquad
\label{Tmn massless ghosts:2}
\end{eqnarray}
where we set, $\Box\rightarrow \partial^2$ and $\Box^\prime\rightarrow {\partial^\prime}^2$. 
We can now use the ghost propagator~(\ref{EOM for a ghost scalar: solution})
to show that the vacuum part of the energy momentum tensor~(\ref{Tmn massless ghosts})
vanishes in dimensional regularization. The thermal part, on the other hand, gives a non-trivial 
contribution. Indeed, by taking account of, 
\begin{equation}
\frac{1}{\partial^2}i\Delta_M^{\rm th}(x;x^\prime)_{x^\prime\rightarrow x} 
=\frac{1}{{\partial^\prime}^2}i\Delta_M^{\rm th}(x;x^\prime)_{x^\prime\rightarrow x}
=\frac{1}{M^2}i\Delta_M^{\rm th}(x;x^\prime)_{x^\prime\rightarrow x}
\,,
\nonumber
\end{equation}
one sees that only the first term in Eq.~(\ref{Tmn massless ghosts}) contributes (on-shell), 
\begin{eqnarray}
 \langle\hat T^{\rm gh}_{\mu\nu}\rangle
   \!&=&\!\! -2\partial^\prime_{(\mu}\partial_{\nu)}i\Delta_0^{\rm th}(x;x^\prime)
                _{x^\prime\rightarrow x}
\,,
\qquad
\label{Tmn massless ghosts:3}
\end{eqnarray}
which equals {\it minus twice} the contribution of a {\it massless scalar} field
({\it cf.} Eqs.~(\ref{some useful equalities for the massless photon})
and~(\ref{massless ghost propagator: thermal})), resulting in,  
\begin{eqnarray}
\langle \hat T_{\mu\nu}^{\rm gh}(x)\rangle 
\!\!&=&\! -2\times\frac{\pi^2}{30\beta^4}\delta_\mu^0\delta_\nu^0
-2\times\frac{\pi^2}{90\beta^4}
   \left(\eta_{\mu\nu}+ \delta_\mu^0\delta_\nu^0\right)
\,.\quad
\label{ghost energy momentum tensor massless photon}
\end{eqnarray}
Upon adding this ghost contribution 
to~(\ref{EM energy momentum tensor massless photon: gauge dep})
one finally obtains the gauge independent and physically correct 
one-loop energy-momentum tensor for the photon in thermal equilibrium,
\begin{eqnarray}
\langle \hat T_{\mu\nu}^{\rm EM}(x)\rangle 
\!\!&=&\! 2\times\frac{\pi^2}{30\beta^4}\delta_\mu^0\delta_\nu^0
+2\times\frac{\pi^2}{90\beta^4}
   \left(\eta_{\mu\nu}+ \delta_\mu^0\delta_\nu^0\right)
\,.\quad
\label{EM energy momentum tensor massless photon: gauge independent}
\end{eqnarray}
\begin{figure}[t]
\vskip -0.5cm
\centering
\includegraphics[scale=0.6]{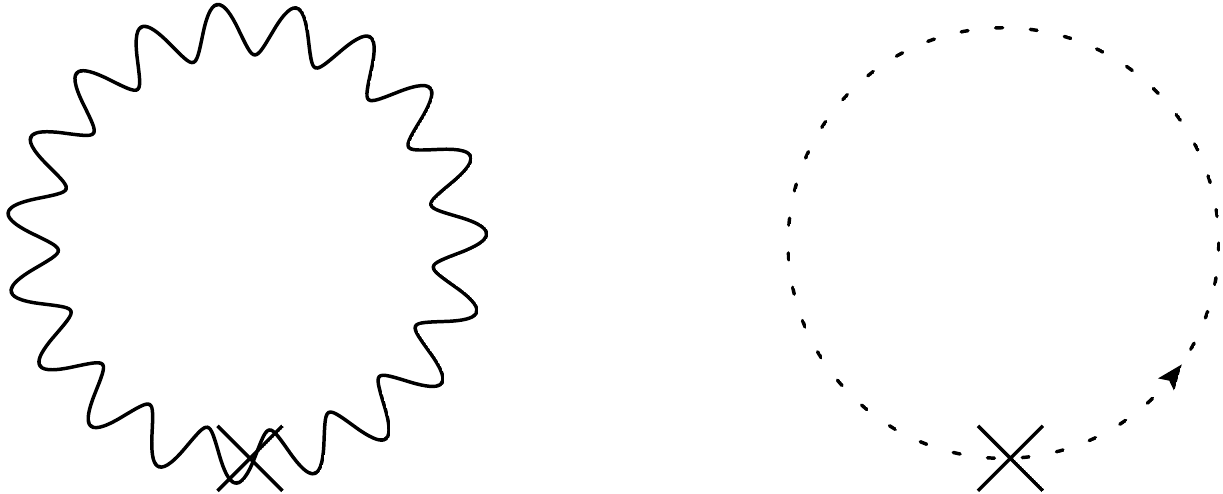}
\caption{The one-loop Feynman diagrams contributing to the one-loop energy-momentum tensor 
of a massless gauge field. The photon contribution is the diagram with wiggly lines 
and the ghost diagram is dashed and oriented.}
\label{fig:photon_ghosts}
\end{figure}
The Feynman diagrams contributing to the one-loop energy-momentum tensor
of a massless gauge field (the photon) are shown in figure~\ref{fig:photon_ghosts}, 
where both the photon loop (wiggly) and the ghost loop (dashed) are shown.
From the result~(\ref{EM energy momentum tensor massless photon: gauge independent}), 
one can read off the energy density and pressure of an ideal gas of photons, 
$\rho=\pi^2 T^4/15$, ${\cal P} =\pi^2 T^4/45$, such that ${\cal P}=(1/3) \rho$,
which are the correct results for an ultrarelativistic plasma with two degrees of freedom. 
The corresponding entropy density is, $s= k_B ({\cal P}+\rho)/T= k_B4\pi^2 T^3/45$. 
The important lesson to learn from this calculation is that, 
even though we obtained a result in which any dependence on the gauge parameter $\xi$ 
dropped out even without including the ghosts,
the result was in a subtle way incorrect, and only when we included 
the ghosts' contribution we obtained the physically correct result.

\bigskip

{\bf Massive gauge field.} We include in this appendix a massive Proca gauge field 
$A_\mu$, since it appears in the calculation of the energy-momentum tensor 
of the Abelian Higgs model and consequently of the standard model. 
The Proca action is given by,
\begin{equation}
 S_{\rm Proca}[A_\mu] = \!\int\! d^Dx\sqrt{-g}{\cal L}_{\rm Proca}
 ,\quad 
{\cal L}_{\rm Proca} = -\frac14 g^{\mu\rho}g^{\nu\sigma}F_{\mu\nu}F_{\rho\sigma}
   \!-\! \frac{1}2 M^2g^{\mu\nu}A_\mu A_\nu
\,,
\label{Proca action}
\end{equation}
where $F_{\mu\nu} = \partial_\mu A_\nu - \partial_\nu A_\mu$ is the field strength 
and $M$ is its mass. The Proca field has $D-1$ dynamical degrees of freedom and 
no gauge symmetry, such that the calculation of the energy-momentum tensor 
is rather simple since it is not complicated by gauge symmetry. 
From~(\ref{Proca action}) one easily obtains the equation of motion (in Minkowski space) for the Proca field,
\begin{equation}
\left((\partial^2-M^2)\eta^{\mu\nu}-\partial^\mu\partial^\nu\right)\hat A_\nu(x) = 0
\,.
\label{Proca field equation of motion}
\end{equation}
Upon acting $\partial_\mu$ on~(\ref{Proca field equation of motion}) 
we see that the Lorentz condition, $\eta^{\mu\nu}\partial_\mu \hat A_\nu=0$
is automatically imposed by the equation of motion, which kills one degree of freedom.
Because of this, it is natural to require that the Feynman propagator obeys the equation
of motion,~\footnote{The precise meaning of the operator, 
$G(x;x^\prime)\equiv (1/\partial^2)\delta^D(x-x^\prime)$, is revealed upon acting with
$\partial^2$ on both sides of the equation, which gives  $\partial^2G(x;x^\prime)=\delta^D(x-x^\prime)$,
which means that  $iG(x;x^\prime)=i\Delta_0(x;x^\prime)$ is the massless scalar propagator, 
whose precise form is subject to boundary conditions. Since here we are interested in
the thermal propagator on Minkowski space, $iG(x;x^\prime)$ is the thermal propagator for 
a massless scalar and its spacetime dependence is given in Eqs.~(\ref{massless ghost propagator: vacuum}--\ref{massless ghost propagator: thermal}).
}
\begin{equation}
\left((\partial^2-M^2)\eta^{\mu\nu}-\partial^\mu\partial^\nu\right)
 i\left[_\nu\Delta_\alpha\right](x;x^\prime) 
 = \left(\delta^\mu_{\;\alpha} -\frac{\partial^\mu\partial_{\alpha}}{\partial^2}\right)
  i\delta^D(x-x^\prime) 
\,,
\label{Feynman propagator equation: Proca field}
\end{equation}
such that the Proca propagator is transverse on both legs, 
\begin{equation}
\partial_x^\nu i\left[_\nu\Delta_\alpha\right](x;x^\prime) 
 = 0 = \partial_{x^\prime}^\alpha i\left[_\nu\Delta_\alpha\right](x;x^\prime) 
\,.
\label{Feynman propagator equation: Proca field: transversality}
\end{equation}
In the language of gauge theories this can be formally viewed as the Landau gauge 
of the massive gauge field propagator of the Abelian Higgs model, 
whose detailed analysis can be found below in this Appendix.
 The transversality 
 property~(\ref{Feynman propagator equation: Proca field: transversality}) dictates 
 the form of the Proca propagator, 
\begin{equation}
 i\left[_\nu\Delta_\alpha\right](x;x^\prime)  
  = \left(\eta_{\nu\alpha} - \frac{\partial_\nu\partial_\alpha}{\partial^2}\right)
                     i\Delta_M(x;x')
\label{Proca propagator: solution}
\end{equation}
where $i\Delta_M(x;x')$ is the real scalar propagator~(\ref{thermal scalar propagator})
whose mass is $m=M$.

We are now ready to calculate the energy-momentum tensor of the Proca field.
Varying the action~(\ref{Proca action}) with respect to the inverse metric gives, 
\begin{eqnarray}
\left\langle\hat T_{\mu\nu}^{\rm Proca}\right\rangle 
 &=&  \eta^{\alpha\rho}\left[\delta_{(\mu}^{\;\beta}\delta_{\nu)}^{\;\sigma}
-\frac14\eta_{\mu\nu}\eta^{\beta\sigma}\right]
       \big \langle T[\hat F_{\alpha\beta}\hat F_{\rho\sigma}]\big\rangle
\nonumber\\
&+&M^2\left[\delta_{(\mu}^{\;\alpha}\delta_{\nu)}^{\;\beta}
-\frac12\eta_{\mu\nu}\eta^{\alpha\beta}\right]
\big \langle T[\hat A_{\alpha}\hat A_{\beta}]\big\rangle
\,.
\label{energy momentum: Proca field}
\end{eqnarray}
Upon pulling the derivatives out of~(\ref{energy momentum: Proca field}) and 
making use of the transverse form of the Proca propagator~(\ref{Proca propagator: solution})
one obtains ({\it cf.} Eq.~(\ref{energy momentum tensor: massless photon:4})), 
\begin{eqnarray}
\left\langle\hat T_{\mu\nu}^{\rm Proca}(x)\right\rangle 
&=&  \left[\frac{D\!-\!3}{2}\eta_{\mu\nu}\partial^2
         \!-\!(D\!-\!2)\partial_{\mu}\partial_{\nu}\right]\!
  i\Delta_M(x;x^\prime)_{x^\prime\rightarrow x}
\nonumber\\
&-&\!\!\!M^2
\left(\frac{D\!-\!3}{2}\eta_{\mu\nu}\!+\!\frac{\partial_\mu\partial_\nu}{\partial^2}\right)\!
 i\Delta_M(x;x^\prime)_{x^\prime\rightarrow x}
\,.\;
\label{energy momentum: Proca field:2}
\end{eqnarray}
The vacuum contribution can be extracted from Eq.~(\ref{thermal scalar propagator})
and the integer series of~(\ref{expansion of Knu}),
\begin{eqnarray}
\left\langle\hat T_{\mu\nu}^{\rm Proca}\right\rangle_{\rm vac} 
&=&  
(D\!-\!1)\frac{M^D}{2(4\pi)^{D/2}}\Gamma\left(\!-\frac{D}{2}\right)
    \eta_{\mu\nu}
      \nonumber\\
      &=& -3\times \frac{M^4}{32\pi^2}\left[\frac{\mu^{D-4}}{D\!-\!4}
       +\frac12\ln\left(\frac{M^2}{4\pi \mu^2}\right)+\frac{\gamma_E}{2}-\frac{5}{12}\right]
         \eta_{\mu\nu}
\quad
\nonumber\\
&+& {\cal O}(D\!-\!4)
\,.
\label{energy momentum: Proca field:3}
\end{eqnarray}
The thermal contribution is a homogeneous solution that satisfies the on-shell 
condition, $(\partial^2-M^2)i\Delta_{M}^{\rm th}(x;x^\prime)=0$, such that can be evaluated 
\begin{eqnarray}
\left\langle\hat T_{\mu\nu}^{\rm Proca}(x)\right\rangle_{\rm th} 
&=&  
 -(D\!-\!1)\partial_{\mu}\partial_{\nu}
  i\Delta_{M}^{\,\rm th}(x;x^\prime)_{x^\prime\rightarrow x}
\,,\;
\label{energy momentum: Proca field:4}
\end{eqnarray}
where we made use of, 
$(1/\partial^2)i\Delta_{M}^{\,\rm th}(x;x^\prime)=(1/M^2)i\Delta_{M}^{\,\rm th}(x;x^\prime)$.
Comparing this with~(\ref{part of Tmn: real scalar}) we conclude that 
a massive Proca field has $D-1$ degrees of freedom ({\it three} degrees in $D=4$), 
which is in agreement with the expectation. Namely, {\it one}
out of the $D$ degrees of freedom of 
the massive vector field $A_\mu$ gets removed by the transversality 
condition~(\ref{Feynman propagator equation: Proca field: transversality}). 

The vacuum contribution~(\ref{energy momentum: Proca field:3})
diverges and thus must be renormalized. If $M$ is a tree-level mass,
it is easy to see that the counterterm action is that a cosmological 
constant,~\footnote{If $M$ is generated by a scalar field condensate, then 
the correct counterterm is the scalar self-coupling term.} 
\begin{equation}
 S_{\rm ct}^{\rm Proca} = \delta \left(\frac{\Lambda}{16\pi G}\right)\int d^Dx\sqrt{-g}
\,,
\label{counterterm Proca}
\end{equation}
whose energy-momentum tensor is, 
\begin{equation}
 (T_{\mu\nu})_{\rm ct}^{\rm Proca} = \delta \left(\frac{\Lambda}{16\pi G}\right)
                 \eta_{\mu\nu}
\,.
\label{counterterm Proca: Tmn}
\end{equation}
By comparing~(\ref{counterterm Proca: Tmn}) with~(\ref{energy momentum: Proca field:3})
one can see that the minimal subtraction demands, 
\begin{equation}
 \delta \left(\frac{\Lambda}{16\pi G}\right) 
 = \frac{3M^4}{32\pi^2}\frac{\mu^{D-4}}{D\!-\!4}
\,.
\label{counterterm Proca: Tmn: 2}
\end{equation}
When~(\ref{counterterm Proca: Tmn}) is added 
to~(\ref{energy momentum: Proca field:3}) one obtains the renormalized vacuum 
contribution to the energy-momentum tensor. Taking account of 
the thermal contribution as well (which is three times that of the real scalar field 
given in~(\ref{renormalized energy-momentum tensor: scalar})), one obtains 
the renormalized one-loop energy-momentum tensor of a massive vector field
in the thermal state,
\begin{eqnarray}
\left\langle\hat T_{\mu\nu}^{\rm Proca}\right\rangle_{\rm ren} 
&=& -\frac{3M^4}{64\pi^2}\left[
       \ln\left(\frac{M^2}{4\pi \mu^2}\right)\!+\!\gamma_E
                 \!-\!\frac{5}{6}\right]
   +\frac{\eta_{\mu\nu}}{2\pi^2\beta^5M}
                   \left[\partial_z J_B(6,z)\right]_{z=\beta M}
\quad
\nonumber\\
 && \hskip -1.1cm
 +\,\delta_\mu^0\delta_\nu^0
        \left\{
        \frac{2}{\pi^2\beta^5M}\left[\partial_z J_B(6,z)\right]_{z=\beta M}
         \!+\!\frac{3M}{2\pi^2\beta^3}\left[\partial_z J_B(4,z)\right]_{z=\beta M}
        \right\}
.
 \;\;
\label{energy momentum: Proca field:final}
\end{eqnarray}

\bigskip\bigskip
 
{\bf Abelian Higgs model.} This model, also known as 
{\it scalar quantum electrodynamics} (SQED), consists of one complex scalar $\Phi$ 
and one Abelian gauge field $A_\mu$ and its action is given by, 
\begin{eqnarray}
S_{\rm SQED}[\Phi,A_\mu] &=& \int d^Dx \sqrt{-g}\mathcal{L}_{\rm SQED}(\Phi,A_\mu)
\,,
 \label{L_abelian_Higgs:1}
\\
&&\hskip -3.4cm
  \mathcal{L}_{\rm SQED} 
   = -\frac{1}{4}g^{\mu\rho}g^{\nu\sigma} F_{\mu \nu} F_{\rho\sigma}
      \!-\! g^{\mu\nu} (D_\mu \Phi)^*(D_\mu \Phi)  \!-\! \mu^2 \Phi^\dagger \Phi 
        \!-\! \lambda_\Phi (\Phi^*\Phi)^2
\,,\quad
\label{L_abelian_Higgs:2}
\end{eqnarray}
where the gauge-covariant derivative is defined as,
\begin{equation}
D_\mu \Phi = \left(\partial_\mu + ig A_\mu \right) \Phi
\,, 
\label{covariant derivative}
\end{equation}
where $g$ is the gauge coupling constant, $\mu^2$ is the scalar mass parameter
 and $\lambda_\Phi$ its quartic self-coupling.  The model possesses an Abelian gauge symmetry. This means that the Lagrangian~(\ref{L_abelian_Higgs:1})
is invariant under the following local field transformations,
\begin{equation}
 \Phi \rightarrow {\rm e}^{-ig\Lambda(x)}\Phi
 \,,\qquad 
 A_\mu  \rightarrow A_\mu + \partial_\mu \Lambda(x)
\,.
 \label{gauge transformations}
\end{equation}
 For example, the covariant derivative~(\ref{covariant derivative}) transforms multiplicatively
 under~(\ref{gauge transformations}),
\begin{equation}
 D_\mu \Phi  \rightarrow{\rm e}^{-ig\Lambda} D_\mu \Phi
\,.
\end{equation}

 When the mass parameter $\mu^2>0$ in~(\ref{L_abelian_Higgs:2}), then  
 $\mu$ is the tree level scalar mass. In this {\it symmetric} phase, 
 the model~(\ref{L_abelian_Higgs:1}--\ref{L_abelian_Higgs:2}) contains one massive 
 complex scalar with mass $m=\mu$ and one massless gauge field. The corresponding 
  contributions to the one-loop energy-momentum tensor have already been calculated:
the energy-momentum tensor of a complex scalar field is two times that of a real 
scalar given in Eq.~(\ref{renormalized energy-momentum tensor: scalar}) with mass 
$m\rightarrow \mu$ and the energy-momentum tensor of
a massless gauge field is given in 
Eq.~(\ref{EM energy momentum tensor massless photon: gauge independent}).  
 
\bigskip 
 
\begin{figure}[h!]
\vskip -0.5cm
\centering
\includegraphics[scale=0.75]{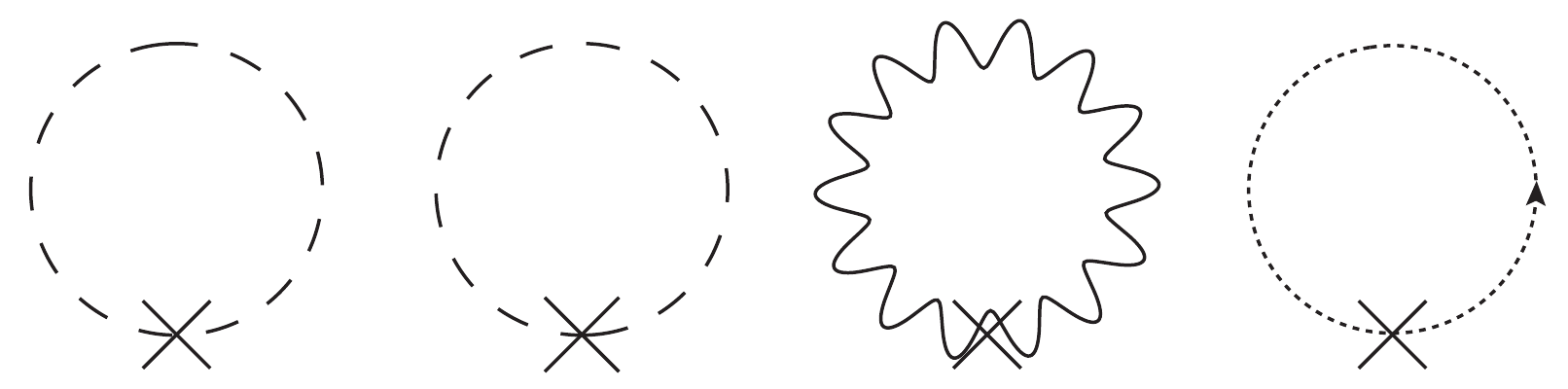}
\caption{The Feynman diagrams contributing to the one-loop energy-momentum
 tensor for the Abelian Higgs model. There are four separate contributions. The 
 Higgs field and the Goldstone boson diagrams are denoted by long dashed lines, 
 the massive gauge boson is the wiggly diagram and the ghost contribution is the short-dashed
 diagram.}
\label{fig:massive_gauge}
\end{figure}

 When, on the other hand, 
$\mu^2<0$, the scalar field $\Phi$ exhibits a {\it spontaneous symmetry breaking}
and acquires a condensate by the famous Brout-Englert-Higgs (BEH) mechanism. 
To study the mechanism, it is convenient to decompose $\Phi$ as,
\begin{equation}
  \Phi = \frac{1}{\sqrt{2}}(\varphi_1+i\varphi_2+ \phi_0)
\,,
\label{scalar decomposition}
\end{equation}
where we take the field condensate $\phi_0=\langle\hat\Phi\rangle$ 
to be real (this can be always achieved by a suitable global gauge 
transformation~(\ref{gauge transformations})). 
The gauge field $A_\mu$ is assumed not to develop a condensate. 
Then $\varphi_1$ is the Higgs field and $\varphi_2$ is the Goldstone boson 
of the symmetry `broken' by the condensation.~\footnote{In fact, since 
the Ward identities  generated by the symmetry transformation~(\ref{gauge transformations}) are respected both in the symmetric and in the condensate phase,
the gauge symmetry is never really broken. Nevertheless, because the scalar condensate 
generates a mass for the gauge field, and a massive Proca theory does not
possess a gauge symmetry, the condensate phase is often -- but not correctly 
-- referred to as 
the broken phase.}
Upon inserting~(\ref{scalar decomposition}) into~(\ref{L_abelian_Higgs:2})
we get (up to boundary terms) for the quadratic part of the Lagrangian,
\begin{eqnarray} 
	\mathcal{L}^{(2)} 
&=& -\frac{1}{4}\eta^{\mu\rho}\eta^{\nu\sigma} F_{\mu \nu} F_{\rho\sigma} 
- \frac{1}{2}\eta^{\mu\nu}(\partial_\mu \varphi_1 \partial_\nu \varphi_1 
      + \partial_\mu \varphi_2 \partial_\nu \varphi_2)  
\nonumber\\ &&
      - \frac{1}{2} M^2\eta^{\mu\nu} A_\mu A_\nu
      + M \eta^{\mu\nu}\partial_\mu A_\nu \varphi_2
        - \frac{1}{2} m_H^2 \varphi_1^2 + \frac{m_H^2 \phi_0^2}{8}
 \label{L_abelian_Higgs_decomp.}
\end{eqnarray}
where we have defined $M^2 = (g\phi_0)^2$ and $m_H^2 = 2\lambda \phi_0^2$
and for simplicity calculated $\mathcal{L}^{(2)}$ on Minkowski background,
on which $g_{\mu\nu}\rightarrow \eta_{\mu\nu}$. Note that 
the scalar and gauge perturbations couple in 
the gauge~(\ref{scalar decomposition}) {\it via}
 the term $M \partial_\mu A^\mu \varphi_2$.
The fields decouple in the `t Hooft gauge~\cite{tHooft:1971qjg}, 
\begin{equation}
 {\cal L}_{\rm tHooft}= -\frac{1}{2\xi} (g^{\mu\nu}\nabla_\nu A_\mu + \xi M  \varphi_2)^2
 \label{tHooft gauge fixing term}
\end{equation}
in which the total quadratic gauge fixed Lagrangian reads 
(with $g_{\mu\nu}\rightarrow \eta_{\mu\nu}$),
\begin{eqnarray}
	  \mathcal{L}^{(2)}_{\rm tot} 
	  &= &-\frac{1}{4}\eta^{\mu\rho}\eta^{\nu\sigma} F_{\mu \nu}F_{\rho \sigma} 
	  - \frac{1}{2}\eta^{\mu\nu}(\partial_\mu \varphi_1 \partial_\nu \varphi_1 
	   + \partial_\mu \varphi_2 \partial_\mu \varphi_2) 
	   - \frac{1}{2} M^2\eta^{\mu\nu} A_\mu A_\mu
\nonumber\\
&& 
         - \frac{1}{2} m_H^2 \varphi_1^2 
         - \frac{1}{2}\xi M^2 \varphi^2_2 
          - \frac{1}{2\xi} (\partial_\mu A^\mu)^2  
           + \mathcal{L}_\text{ghosts}
           + \frac{m_H^2 \phi_0^2}{8}
 \,,
 \label{Abelian Higgs: total lagrangian}
\end{eqnarray}
where we also included the Faddeev-Popov ghost Lagrangian, which can be obtained 
from the Faddeev-Popov determinant, which is obtained by considering how 
the gauge fixing condition varies with respect to infinitesimal gauge transformations,
$\Lambda(x)\rightarrow \theta(x)$, 
\begin{eqnarray}
{\cal D}_{\rm FP} &=& {\rm det} \left[\frac{\delta}{\delta \theta(x^\prime)}
    \big[\Box \!-\! \xi M g(\phi_0+\varphi_1)\big]\theta(x) \right]
\nonumber\\
&=& {\rm det} \left[\left(\Box\!-\! M_c^2 \!-\! \xi g M \varphi_1\right)\delta^D(x-x^\prime)\right]
\,,
\label{Faddeev Popov determinant massive ghost}
\end{eqnarray}
where $M_c^2=\xi M^2$  and we made use of~(\ref{tHooft gauge fixing term}), 
$\delta_\theta A_\mu=\partial_\mu\theta$ and  
$\delta_\theta\varphi_2= - g\theta(\phi_0+\varphi_1)$. 
From~(\ref{Faddeev Popov determinant massive ghost}) it follows that 
the corresponding Lagrangian can be written in terms of Grassmannian ghost fields as,
\begin{equation}
  \mathcal{L}_\text{ghosts}
   = -g^{\mu\nu}(\partial_\mu \bar{c})( \partial_\nu c) 
    - \xi M^2 \bar{c} c - \xi g M \bar{c} c \varphi_1
\,,
\label{ghost lagrangian with interaction}
\end{equation}
such that in the `t Hooft gauge the ghosts and the Abelian scalar interact.
The quadratic Lagrangian~(\ref{Abelian Higgs: total lagrangian})
splits nicely into the contributions from different fields, 
\begin{equation}
  \mathcal{L}^{(2)}_{\rm tot} 
   = \mathcal{L}^{(2)}_{\varphi_1}  + \mathcal{L}^{(2)}_{\varphi_2} 
  + \mathcal{L}^{(2)}_{A_\mu} + \mathcal{L}^{(2)}_\text{ghosts}
  + \frac{m_H^2 \phi_0^2}{8}
\,,
\label{AHM: split Lagrangian}
\end{equation}
where $ \mathcal{L}^{(2)}_\text{ghosts}$ is the quadratic part of~(\ref{ghost lagrangian with interaction}).
That means that the model consists of two massive scalars with masses 
$m_1^2=m_H^2=2\lambda\phi_0^2$
and $m_2^2=\xi M^2=\xi (g\phi_0)^2$, one massive gauge field with 
$M^2 = g^2\phi_0^2$ 
and a massive ghost with $M_{\rm c}^2= \xi M^2$, where we assume 
$\xi\geq 0$ for stability. All of these particles contribute to the energy-momentum tensor, 
and the corresponding Feynman diagrams are shown in figure~\ref{fig:massive_gauge}.
 In section~\ref{Real scalar field} we analyzed a massive scalar field,
and in this appendix a massless gauge field, a massless ghost and a massive gauge field, 
but the massive gauge field was a Proca field which does not possess any gauge symmetry. 

\medskip

This means that we still ought to analyze the massive gauge field and the massive ghost. Let us start 
with the massive gauge field. From~(\ref{Abelian Higgs: total lagrangian}) we can 
read off the propagator equation of motion, 
\begin{equation}
\left((\partial^2\!-\!M^2)\eta^{\mu\nu}-\left(1\!-\!\frac1{\xi}\right)\partial^\mu\partial^\nu\right)
  i\left[_\nu\Delta_\alpha\right](x;x^\prime)
     =i\delta^\mu_{\;\alpha}\delta^D(x\!-\!x^{\,\prime})
\,,
\label{Feynman propagator: massive Abelian gauge field in Fermi gauge}
\end{equation}
which, in the massless limit $M\rightarrow 0$, reduces to that of a massless gauge 
field in Fermi gauge~(\ref{Feynman propagator: Abelian gauge field in Fermi gauge}).
A formal solution of~(\ref{Feynman propagator: massive Abelian gauge field in Fermi gauge})
 can be written as
({\it cf.} Eq.~(\ref{Feynman propagator: Abelian gauge field in Fermi gauge: solution})),
\begin{equation}
 i\left[_\nu\Delta_\alpha\right](x;x^\prime)
 = \left[\eta_{\nu\alpha}
      -\left(1-\xi\right)\frac{\partial_\nu\partial_\alpha}{\partial^2-\xi M^2}\right]
         i\Delta_M(x;x^{\,\prime})
\,,
\label{Feynman propagator: massive Abelian gauge field in Fermi gauge: solution}
\end{equation}
where $i\Delta_M(x;x^{\,\prime})$ is the massive scalar propagator~(\ref{thermal scalar propagator})
of mass $M$. 
The solution~(\ref{Feynman propagator: Abelian gauge field in Fermi gauge: solution}) can be 
recast in a more convenient form (for $\xi\neq 1$ and $\xi\geq 0$) 
as,~\footnote{\label{gauge field propagator split}
When writing
the solution~(\ref{Feynman propagator: massive Abelian gauge field in Fermi gauge: solution:2}) 
we made use of the fact that the particular solution of a sourced differential equation of the form,
\begin{equation}
 (\partial^2 - \xi M^2)G(x;x^\prime) = i\Delta_M(x;x^\prime)
 \,,
 \label{sourced propagator equation}
\end{equation}
can be written as, 
\begin{equation}
 G(x;x^\prime) = \frac{1}{(1\!-\!\xi)M^2}
    \left[ i\Delta_M(x;x^\prime)- i\Delta_{\sqrt{\xi}M}(x;x^\prime)\right]
,\quad (\xi\neq 1)
\,.
\label{sourced propagator equation: solution}
\end{equation}
In the singular (Feynman) gauge when $\xi=1$ the solution of~(\ref{sourced propagator equation})
 is a parametric derivative, 
\begin{equation}
 G(x;x^\prime) = \frac{\partial}{\partial M^2} i\Delta_M(x;x^\prime)\,,\quad (\xi=1)
\,.
  \label{sourced propagator equation: solution: sing}
\end{equation}
We leave the analysis of this case as an exercise to the reader.
}
\begin{equation}
 i\left[_\nu\Delta_\alpha\right](x;x^\prime)
 =\left(\eta_{\nu\alpha}-\frac{\partial_\nu\partial_\alpha}{M^2}\right)  i\Delta_M(x;x^{\,\prime})
 +\frac{\partial_\nu\partial_\alpha}{M^2} i\Delta_{\sqrt{\xi}M}(x;x^{\,\prime}) 
\,.
\label{Feynman propagator: massive Abelian gauge field in Fermi gauge: solution:2}
\end{equation}
This form of the solution is convenient since 
the propagator~(\ref{Feynman propagator: massive Abelian gauge field in Fermi gauge: solution:2}) 
splits naturally into a gauge independent transverse part 
and a gauge dependent longitudinal part. 

We are now ready to calculate the corresponding energy-momentum tensor.
By varying Eq.~(\ref{Abelian Higgs: total lagrangian}) with respect to $g^{\mu\nu}$ one gets
({\it cf.} Eqs.~(\ref{energy momentum tensor: massless photon:3}--\ref{energy momentum tensor: massless photon:3b}) and~(\ref{energy momentum: Proca field})), 
\begin{eqnarray}
\langle \hat T_{\mu\nu}^{\rm gauge}\rangle 
\!\!&=&\!\!\!\left[ 4\eta^{\alpha\rho}\delta_{(\mu}^{\;\beta}\delta_{\nu)}^{\;\sigma}
\!-\! \eta^{\alpha\rho}\eta_{\mu\nu}\eta^{\beta\sigma}\right]
  \partial^\prime_{\rho]}\partial_{[\alpha}
   \Big\langle T^*\!\left[\hat A_{\beta]}(x) \hat A_{[\sigma}(x^\prime)\right]\Big\rangle
            _{x^\prime\rightarrow x}
\label{energy momentum tensor: massive gauge field:a}\\
&&\hskip -2.2cm
+\,\frac{1}{\xi}\!\left[
 \!-2\eta^{\rho\sigma}\delta_{(\mu}^{\;\alpha}\delta_{\nu)}^{\;\beta}
           \partial^\prime_\alpha\partial^\prime_\rho
\!+\! \eta_{\mu\nu}\eta^{\alpha\beta}\eta^{\rho\sigma}
           \!\left(\!\partial^\prime_\alpha
                \!+\!\frac12\partial_\alpha\right)\!\partial^\prime_\rho
\right]\!\Big\langle\! T^*\!\left[\hat A_\beta(x) \hat A_\sigma(x^\prime)\!\right]\!\Big\rangle
            _{x^\prime\rightarrow x}\!
\qquad
\label{energy momentum tensor: massive gauge field:b}
\\
&&\hskip -2.2cm
+\,M^2\left[\delta_{(\mu}^{\;\alpha}\delta_{\nu)}^{\;\beta}
-\frac12\eta_{\mu\nu}\eta^{\alpha\beta}\right]
\big \langle T[\hat A_{\alpha}\hat A_{\beta}]\big\rangle
\,.
\label{energy momentum tensor: massive gauge field:c}
\end{eqnarray}
Acting on the transverse and longitudinal 
tensor structures of 
the propagator~(\ref{Feynman propagator: massive Abelian gauge field in Fermi gauge: solution:2})
yields, 
\begin{eqnarray}
\langle \hat T_{\mu\nu}^{\rm gauge}\rangle 
\!\!&=&\!\left[\frac{D\!-\!3}{2}\eta_{\mu\nu}(\partial^2-M^2)-(D\!-\!1)\partial_\mu\partial_\nu\right]
  i\Delta_{M}(x;x^\prime)_{x^\prime\rightarrow x}
\label{energy momentum tensor: massive gauge field:2a}\\
&&\hskip 0cm -\partial_\mu\partial_\nu
  i\Delta_{\sqrt{\xi}M}(x;x^\prime)_{x^\prime\rightarrow x}
\,.
\qquad
\label{energy momentum tensor: massive gauge field:2b}
\end{eqnarray}
By making use of~(\ref{thermal scalar propagator}) and~(\ref{expansion of Knu})
we can evaluate the vacuum part of~(\ref{energy momentum tensor: massive gauge field:2a}--\ref{energy momentum tensor: massive gauge field:2b}),
\begin{eqnarray}
\langle \hat T_{\mu\nu}^{\rm gauge}\rangle_{\rm vac}
\!=(D\!-\!1) \frac{M^D}{2(4\pi)^{D/2}}\Gamma\left(\!-\frac{D}{2}\right)\eta_{\mu\nu}
\!+\!  \frac{(\sqrt{\xi}M)^D}{2(4\pi)^{D/2}}\Gamma\left(\!-\frac{D}{2}\right)\eta_{\mu\nu}
\,,
\label{energy momentum tensor: massive gauge field:3}
\end{eqnarray}
which corresponds to $(D-1)$ massive scalar degrees of freedom with mass $M$ and 
{\it one} massive gauge degree of freedom with mass $\sqrt{\xi}M$.
Since the thermal contribution is on-shell, we see that when 
$(\partial^2-M^2)$ acts in~(\ref{energy momentum tensor: massive gauge field:2a}) 
it yields zero and we have, 
\begin{eqnarray}
\langle \hat T_{\mu\nu}^{\rm gauge}\rangle_{\rm th}
\!\!&=&3\partial^\prime_{(\mu}\partial_{\nu)}
  i\Delta_{M}^{\,\rm th}(x;x^\prime)_{x^\prime\rightarrow x}
+\partial^\prime_{(\mu}\partial_{\nu)}
  i\Delta_{\sqrt{\xi}M}^{\rm th}(x;x^\prime)_{x^\prime\rightarrow x}
\,.
\qquad
\end{eqnarray}
This can be evaluated by taking account of the thermal contributions
of a real scalar field in Eqs.~(\ref{thermal scalar propagator: coincidence})
and~(\ref{renormalized energy-momentum tensor: scalar}),
\begin{eqnarray}
\langle \hat T_{\mu\nu}^{\rm gauge}\rangle_{\rm th}
&=& 
 \eta_{\mu\nu}\frac{1}{2\pi^2\beta^5M}
                   \left[\partial_z J_B(6,z)\right]_{z=\beta M}
\quad
\label{renormalized energy-momentum tensor: massive gauge vector 1}
\\
 &&\hskip -2.5cm
 +\,\delta_\mu^0\delta_\nu^0
        \left\{
        \frac{2}{\pi^2\beta^5M}\left[\partial_z J_B(6,z)\right]_{z=\beta M}
         \!+\!\frac{3M}{2\pi^2\beta^3}\left[\partial_z J_B(4,z)\right]_{z=\beta M}
        \right\}
 \quad
\nonumber\\
&&\hskip -2.5cm
 +\,\eta_{\mu\nu}\frac{1}{6\pi^2\beta^5\sqrt{\xi}M}
                   \left[\partial_z J_B(6,z)\right]_{z=\beta \sqrt{\xi}M}
\quad\;\;
\label{renormalized energy-momentum tensor: massive gauge vector 2}
\\
 &&\hskip -2.5cm
 +\, \delta_\mu^0\delta_\nu^0
        \left\{
        \frac{2}{3\pi^2\beta^5\sqrt{\xi}M}\left[\partial_z J_B(6,z)\right]_{z=\beta \sqrt{\xi}M}
         \!+\!\frac{\sqrt{\xi}M}{2\pi^2\beta^3}\left[\partial_z J_B(4,z)\right]_{z=\beta \sqrt{\xi}M}
        \right\}
\,.
\nonumber
\end{eqnarray}
The first two lines~(\ref{renormalized energy-momentum tensor: massive gauge vector 1})
 are generated by the transverse part of the massive vector while the latter
two~(\ref{renormalized energy-momentum tensor: massive gauge vector 2})
originate from the longitudinal (gauge dependent) part. 
In conclusion, we have found out that the massive gauge field in covariant `t Hooft gauge 
contributes as $(D-1)$ massive degrees of freedom with mass $M$ and 
one gauge degree of freedom with mass $\sqrt{\xi}M$.

\bigskip

Next we analyze the {\it massive ghost} field. 
One way to proceed is to use the Grassmannian ghost fields whose 
the Lagrangian is Eq.~(\ref{ghost lagrangian with interaction}).~\footnote{In the analysis 
of the massless Abelian gauge field above, we observed that, if one treats massless ghosts as 
Grasmannian fields and define their thermal state in a standard way by 
means of the thermal density operator, 
one finds that ghosts obey the Fermi-Dirac statistic, which leads to invalid results. 
In order to arrive at the correct result one ought to {\it postulate} that the ghosts 
obey a Bose-Einstein statistic,
such that the ghost propagator is that of a massive scalar field~(\ref{thermal scalar propagator}), 
\begin{equation}
 i\Delta^{\rm gh}_{M_c}(x;x^\prime) 
 = \left\langle T\left[\hat{\bar c}(x^\prime)\hat c(x)\right]\right\rangle
= i\Delta_{M_c}(x;x^\prime)
\,.
\label{massive ghost propagator Grassmannian}
\end{equation}
The corresponding energy-momentum tensor is then, 
\begin{equation}
\langle \hat T^{\rm gh}_{\mu\nu}(x)\rangle 
 =  2 \left\langle T\left[\partial_{(\mu}\hat{\bar c}(x)\partial_{\nu)}\hat c(x)\right]\right\rangle
 =-2\partial_{(\mu}\partial^\prime_{\nu)}  i\Delta^{\rm gh}_{M_c}(x;x^\prime)_{x^\prime\rightarrow x}
 \,,
 \label{massive ghost Tmn Grassmannian}
 \end{equation}
implying that the ghost energy-momentum tensor is minus 
twice that of a massive scalar field,
whose vacuum and thermal contributions can be written as~(\ref{Tmn: massive ghost: vac:B})
and~(\ref{energy-momentum tensor: massive ghost thermal part:B}), respectively.
} 
Here we use a nonlocal prescription, already discussed in some detail 
above, {\it cf.} Eqs.~(\ref{Faddeev Popov determinant:5}) and~(\ref{scalar ghost action}) .

From the inverse of the Faddeev-Popov determinant~(\ref{Faddeev Popov determinant massive ghost}) (see Eq.~(\ref{Faddeev Popov determinant:5})), we can infer 
the quadratic part of the non-local ghost action,  
\begin{equation}
S_{\rm gh}=\int d^D \sqrt{-g} {\cal L}_{\rm gh}
\,,\qquad 
{\cal L}_{\rm gh} = \bar\phi_{\rm gh}\frac{1}{\Box -M_c^2}\phi_{\rm gh}
\,.
\label{massive scalar ghost action and lagrangian}
\end{equation}
where $\phi_{\rm gh}$ and $\bar\phi_{\rm gh}$ are 
scalar ghost fields. Eq.~(\ref{massive scalar ghost action and lagrangian}) implies
 the following equation of motion for the massive scalar ghost propagator, 
\begin{equation}
\frac{1}{\partial^2 \!-\!M_c^2}i\Delta_{M_c}^{\rm gh}(x;x^\prime) = i\delta^D(x\!-\!x^\prime)
\label{massive scalar ghost eom}
\end{equation}
and its solution can be written formally as ({\it cf.} Eq.~(\ref{EOM for a ghost scalar: solution})), 
\begin{eqnarray}
 i\Delta_{M_c}^{\rm gh}(x;x^\prime) 
  &=& (\partial^2-M_c^2) i\delta^D(x\!-\!x^\prime)
  +\left[(\partial^2 -M_c^2)^2i\Delta_{M}^{\rm th}\left(x;x^\prime\right)\right]
                _{M^2\rightarrow M_c^2}
\nonumber\\
 &=& \left[(\partial^2-M_c^2)({\partial^\prime}^2-M_c^2)i\Delta_{M}\left(x;x^\prime\right)\right]
                _{M^2\rightarrow M_c^2}
\,,
\label{massive scalar ghost solution}
\end{eqnarray}
where $i\Delta_{M}$ denotes the massive scalar propagator~(\ref{thermal scalar propagator}).
Just as in the massless ghost case discussed above
(see Eqs.~(\ref{EOM for a ghost scalar}--\ref{ghost energy momentum tensor massless photon})), 
in order to 
regulate the would-be singular behavior of the solution~(\ref{massive scalar ghost solution})
we shifted the poles by $\delta M^2=M^2-M_c^2$ in~(\ref{massive scalar ghost solution}).
That is enough to regulate the one-loop the energy-momentum tensor of the 
massive scalar ghosts, and that is what we do in what follows.

The corresponding one-loop energy-momentum tensor is then ({\it cf.} Eq.~(\ref{Tmn massless ghosts:2})), 
 \begin{eqnarray}
 \langle\hat T^{\rm gh}_{\mu\nu}\rangle
   \!&=&\!\! -2\frac{\partial^\prime_{(\mu}\partial_{\nu)}}{({\partial^\prime}^2-M_c^2)(\partial^2-M_c^2)}
       \left\langle T^*\!\left[\bar\phi_{\rm gh}(x^\prime)\phi_{\rm gh}(x)\right]\right\rangle
                _{x^\prime\rightarrow x}
\nonumber\\
&+&
 \frac{g_{\mu\nu}}{{\partial^\prime}^2-M_c^2}
              \left\langle T^*\!\left[\bar\phi_{\rm gh}(x^\prime)\phi_{\rm gh}(x)\right]\right\rangle
                _{x^\prime\rightarrow x}
\nonumber\\
&+& g_{\mu\nu}\frac{g^{\alpha\beta}\partial^\prime_\alpha\partial_\beta+M_c^2}
                           {({\partial^\prime}^2-M_c^2)(\partial^2-M_c^2)}
 \left\langle T^*\!\left[\bar\phi_{\rm gh}(x^\prime)\phi_{\rm gh}(x)\right]\right\rangle
                _{x^\prime\rightarrow x}
  \nonumber\\
&+&\frac{g_{\mu\nu}}{\partial^2-M_c^2}
 \left\langle T^*\!\left[\bar\phi_{\rm gh}(x^\prime)\phi_{\rm gh}(x)\right]\right\rangle
                _{x^\prime\rightarrow x}
\,.
\label{Tmn massive scalar ghost}
\end{eqnarray}
Now, by making use of~(\ref{massive scalar ghost solution}) and of,
\begin{equation}
 \frac{1}{({\partial^\prime}^2-M_c^2)(\partial^2-M_c^2)}i\Delta_{M_c}^{\rm gh}(x;x^\prime)
    = i\Delta_{M_c}(x;x^\prime)
\,,
\label{simplifying the ghost Tmn}
\end{equation}
where $ i\Delta_{M_c}(x;x^\prime)$ is the massive scalar 
propagator~(\ref{Tmn massive scalar ghost}), 
one sees that --
just as in the massless ghost case~(\ref{Tmn massless ghosts:3})
 -- only the first term in~(\ref{Tmn massive scalar ghost}) contributes,
 and which can be simplified as, 
 \begin{eqnarray}
 \langle\hat T^{\rm gh}_{\mu\nu}\rangle
   \!&=&\!\! -2\partial^\prime_{(\mu}\partial_{\nu)} i\Delta_{M_c}(x;x^\prime)
\,.
\qquad
\label{Tmn massive scalar ghost:2}
\end{eqnarray}
When this is evaluated one obtains minus twice the contribution of the
real massive scalar field~(\ref{renormalized energy-momentum tensor: scalar}) 
with $m^2=M_c^2$. Therefore, we have for the vacuum ghost contribution,
\begin{equation}
\langle \hat T^{\rm gh}_{\mu\nu}\rangle _{\rm vac}
 =-\eta_{\mu\nu} \frac{(\sqrt{\xi}M)^{D}}{(4\pi)^{D/2}}\Gamma\Big(\!-\frac{D}{2}\Big)
 \,,
 \label{Tmn: massive ghost: vac:B}
 \end{equation}
and the thermal contribution of the massive scalar ghost reads
({\it cf.} Eq.~(\ref{renormalized energy-momentum tensor: scalar})), 
\begin{eqnarray}
\langle \hat T^{\rm gh}_{\mu\nu}\rangle _{\rm th} 
&=&\!-\,\frac{\eta_{\mu\nu}}{3\pi^2\beta^5\sqrt{\xi}M}
                   \left[\partial_z J_B(6,z)\right]_{z=\beta \sqrt{\xi}M}
\label{energy-momentum tensor: massive ghost thermal part:B}
\\
&&\hskip -2cm
 -\,\delta_\mu^0\delta_\nu^0
        \left\{
        \frac{4}{3\pi^2\beta^5\sqrt{\xi}M}\left[\partial_z J_B(6,z)\right]_{z=\beta \sqrt{\xi}M}
         \!+\!\frac{\sqrt{\xi}M}{\pi^2\beta^3}\left[\partial_z J_B(4,z)\right]_{z=\beta \sqrt{\xi}M}
        \right\}
\,.
 \quad
 \nonumber
\end{eqnarray}
This concludes our analysis of the massive ghost.

\bigskip

We now have all the pieces we need to calculate the one-loop energy-momentum tensor of 
the Abelian Higgs model in its Higgs phase. 
Let us first consider the {\it vacuum} part. Adding the vacuum contribution 
of the two massive scalars, one with mass $m_H$ and the other with mass $\sqrt{\xi}M$
to that of the massive gauge field~(\ref{energy momentum tensor: massive gauge field:3})
 and the massive ghost~(\ref{Tmn: massive ghost: vac:B}) results in, 
\begin{eqnarray}
\langle \hat T_{\mu\nu}^{\rm AH}\rangle_{\rm vac}
&=&\frac{m_H^D}{2(4\pi)^{D/2}}\Gamma\left(\!-\frac{D}{2}\right)\eta_{\mu\nu}
+(D\!-\!1) \frac{M^D}{2(4\pi)^{D/2}}\Gamma\left(\!-\frac{D}{2}\right)\eta_{\mu\nu}
\,.\quad\;\;\;
\label{energy momentum tensor: massive gauge field: vacuum total}
\end{eqnarray}
The gauge dependent contributions from the second scalar and 
the massive vector 
are canceled by the ghost contribution. As expected, the energy 
 momentum tensor~(\ref{energy momentum tensor: massive gauge field: vacuum total})
 diverges in $D=4$ and it ought to be renormalized. 
The counterterm action is clearly ({\it cf.} Eqs.~(\ref{L_abelian_Higgs:2})), 
\begin{equation}
S_{\rm ct}^{\rm AH}= \int d^Dx \sqrt{-g}\left(-\delta\lambda_\Phi (\Phi^* \Phi)^2\right)
\rightarrow \int d^Dx \sqrt{-g}\left(-\frac14 \delta\lambda_\Phi \phi_0^4\right)
\,.
 \label{ghost counterterm action}
\end{equation}
The corresponding energy-momentum tensor is 
$T_{\mu\nu}^{\rm ct, AH}=-\frac14\delta\lambda_\Phi \phi_0^4\eta_{\mu\nu}$,
such that the following minimal subtraction choice,
\begin{equation}
\delta\lambda_\Phi  =- \frac{4\lambda_\Phi^2\!+\!3g^4}{8\pi^2}\frac{\mu^{D-4}}{D\!-\!4}
\,,
\end{equation}
removes the divergent part of~(\ref{Tmn: massive ghost: vac:B}) 
({\it cf.} Eq.~(\ref{Tmn: real scalar: vac 2})). 
Therefore, the sought-for renormalized one-loop thermal energy-momentum tensor 
of the Abelian Higgs model in the condensate phase is, 
\begin{eqnarray}
\langle \hat T^{\rm AH}_{\mu\nu}\rangle _{\rm ren} 
\!&=&\! - \,\frac{m_H^4}{64\pi^2}\!\left[\ln\left(\frac{m_H^2}{4\pi\mu^2}\right)
\!+\!\gamma_E\!-\!\frac32\!-\!\frac{4\pi^2}{\lambda_\Phi}\right]\!\eta_{\mu\nu}
\label{energy momentum Abelian Higgs: total}
\\
&-& \frac{3M^4}{64\pi^2}\left[\ln\left(\frac{M^2}{4\pi\mu^2}\right)
\!+\!\gamma_E\!-\!\frac56\right]\!\eta_{\mu\nu}
\nonumber
\\
 &+&\!\!\!\eta_{\mu\nu}\frac{1}{6\pi^2\beta^5m_H}
                   \left[\partial_z J_B(6,z)\right]_{z=\beta m_H}
\quad
\nonumber
\\
 &+&\!\!\!\delta_\mu^0\delta_\nu^0
        \left\{
        \frac{2}{3\pi^2\beta^5m_H}\left[\partial_z J_B(6,z)\right]_{z=\beta m_H}
         \!+\!\frac{m_H}{2\pi^2\beta^3}\left[\partial_z J_B(4,z)\right]_{z=\beta m_H}
        \right\}
\quad
\nonumber\\
 &+&\!\!\!\eta_{\mu\nu}\frac{1}{2\pi^2\beta^5M}
                   \left[\partial_z J_B(6,z)\right]_{z=\beta M}
\nonumber
\\
 &+&\!\!\!\delta_\mu^0\delta_\nu^0
        \left\{
        \frac{2}{\pi^2\beta^5M}\left[\partial_z J_B(6,z)\right]_{z=\beta M}
         \!+\!\frac{3M}{2\pi^2\beta^3}\left[\partial_z J_B(4,z)\right]_{z=\beta M}
        \right\}
\,,
\nonumber
\end{eqnarray}
where we also included the relevant thermal 
contributions~(\ref{renormalized energy-momentum tensor: scalar})
(with $m\rightarrow m_H$)
as well as Eq.~(\ref{renormalized energy-momentum tensor: massive gauge vector 1}).  The results of this Appendix are used in 
the main part of the paper, in particular in 
section~\ref{Standard model and its extensions}.

\end{document}